\documentclass[amssymb, amsmath, aps, pra, letterpaper]{revtex4}
\usepackage{color}
\usepackage{graphicx}% Include figure files
\usepackage{bm}% bold math
\newlength{\strikewidth} 
\newlength{\strikelength} 
\begin{document}

\title{Modeling of Nanoscale Devices}

\vspace{0.1in}

\author{M.~P.~Anantram$^1$, M.~S.~Lundstrom$^2$, and D.~E.~Nikonov$^3$}

\author{}

\author{}

\vspace{0.4in}

\affiliation{$^1$ 
Electrical and Computer Engineering Department,\\
University of Waterloo, \\
Ontario, Canada N2L 3G1}
\affiliation{$^2$ Department of Electrical and Computer Engineering,\\
Purdue University, West Lafayette, IN 49097, USA}
\affiliation{$^3$ Intel Corporation, Mail Stop: SC1-05\\
Santa Clara, CA 95052, USA}
\maketitle

\section{\label{section:introduction} Introduction}

Semiconductor devices operate by controlling the flow of electrons and holes 
through a device, and our understanding of charge carrier transport has both 
benefited from and driven the development of semiconductor devices.  
When William Shockley wrote "Electrons and Holes in Semiconductors" \cite{Sho50}, 
semiconductor physics was at the frontier of research in condensed matter physics.  
Over the years, the essential concepts were clarified and simplified into the 
working knowledge of device engineers.  The treatment of electrons and holes 
as semiclassical particles with an effective mass was usually adequate.  
Electronic devices were made of materials (e.g. silicon, gallium arsenide, etc.) 
with properties (e.g. bandgap, effective mass, etc.) that could be looked up.  
For most devices, the engineer's drift-diffusion equation provided a simple, 
but adequate description of carrier transport.  Today things are changing.  
Device dimensions have shrunk to the nanoscale.  The properties of materials 
can be engineered by intentional strain and size effects due to quantum confinement.  
Devices contain a countable number of dopants and are sensitive to structure 
at the atomistic scale.  In addition to familiar devices like the MOSFET, 
which have been scaled to nanometer dimensions, new devices built from carbon 
nanotubes, semiconductor nanowires, and organic molecules are being explored.  
Device engineers will need to learn to think about devices differently.
To describe carrier transport in nanoscale devices, engineers must learn 
how to think about charge carriers as quantum mechanical entities rather 
than as semiclassical particles, and they must learn how to think at 
the atomistic scale rather than at a continuum one.  Our purpose 
in this chapter is to provide engineers with an introduction to the 
non-equilibrium Green's function (NEGF) approach, which provides a powerful 
conceptual tool and a practical analysis method to treat small electronic 
devices quantum mechanically and atomistically.  We first review the basis 
for the traditional, semiclassical description of carriers that has served 
device engineers for more than 50 years in section \ref{section:semi-clas-trans-diff}.  
We then describe why this traditional approach loses validity at the nanoscale.  
Next, we describe semiclassical ballistic transport in section \ref{section:semi-clas-ballistic}
and the Landauer-Buttiker approach to phase coherent quantum transport 
in section \ref{section:quantum}. Realistic devices include interactions 
that break quantum mechanical phase and also cause energy relaxation. 
As a result, transport in nanodevices are between diffusive and phase coherent. 
We introduce the non equilbrium Green's function (NEGF) approach, which can be 
used to model devices all the way from ballistic to diffusive limits in 
section \ref{section:quantum-gf}. This is followed by a summary of 
equations that are used to model a large class of layered structures 
such as nanotransistors, carbon nanotubes and nanowires in section \ref{section:ngf-summary}. 
An application of the NEGF method in the ballistic and scattering 
limits to silicon nanotransistors is discussed in sections  \ref{section:2D-ballistic-mosfet} 
and \ref{section:mode-space} respectively. We conclude with a summary in section 
\ref{section:Discussion}. The Dyson's equations and algorithms to solve for 
the Green's functions of layered structures are presented in appendices B and C. 
These appendices can be left out by a reader whose aim is to gain a basic 
understanding of the NEGF approach to device modeling.

\section{\label{section:semi-clas-trans-diff}Semiclassical Transport: Diffusive}

\begin{figure}
  \begin{center}
    \leavevmode
\includegraphics[height=7cm,angle=-0]{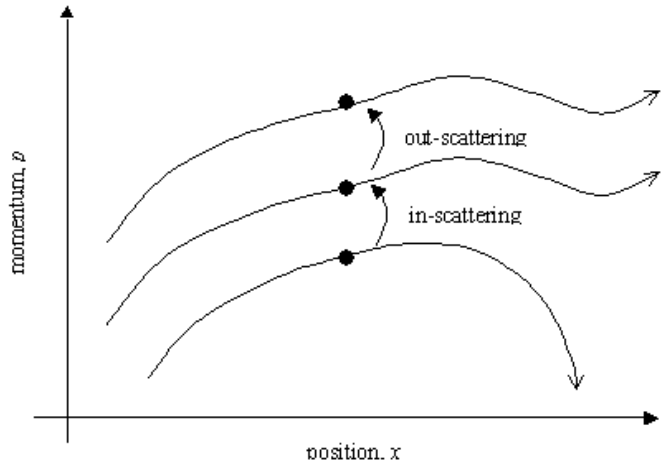}
  \end{center}
\caption{Carrier trajectories in phase space showing free flights along a trajectory 
interrupted by scattering events that begin another free flight.}
\label{fig:scatt1}
\end{figure}

Electrical engineers have commonly treated electrons as semiclassical particles that  
move through a device under the influence of an electric field and random scattering  
potentials.  As sketched in Fig. 1, electrons move along a trajectory in phase space 
(position and momentum space).  In momentum space, the equation of motion looks like 
Newton's Law for a classical particle
\begin{eqnarray}
\frac{d\vec{\hbar k}}{dt} = -\nabla_r E_C(\vec{r},t) \mbox{ ,} \label{eq:newton}
\end{eqnarray}
where $\vec{k}$ is the crystal momentum and $E_C$  is the bottom of the conduction band.  
In position space, the equation of motion is
\begin{eqnarray}
\frac{d\vec{r}}{dt} = \frac{1}{\hbar} \nabla_k E(\vec{k}) \mbox{ ,} \label{eq:newton-k}
\end{eqnarray}
where $E(k)$ describes the bandstructure of the semiconductor.  The right hand side of 
eqn. (\ref{eq:newton-k}) is simply the velocity of the semiclassical particle, and in 
the simplest case it is just $\hbar k/m^\ast$.  By solving eqns. (\ref{eq:newton}) 
and (\ref{eq:newton-k}), we trace the trajectory of a carrier in phase space as shown in Fig. \ref{fig:scatt1}.

Equations (\ref{eq:newton}) and (\ref{eq:newton-k}) describe the ballistic transport of 
semiclassical carriers.  In practice, carriers frequently scatter from various perturbing 
potentials (defects, ionized impurities, lattice vibrations, etc.).  The result is that 
carriers hop from one trajectory in phase space to another as shown in Fig. 1.  
The average distance between scattering events, the mean-free-path, $l$, has (until recently) 
been much smaller than the critical dimensions of a device.  Carriers undergo a random 
walk through a device with a small bias in one direction imposed by the electric field.  
To describe this scattering dominated (so-called diffusive) transport, we should add a 
random force ($F_S(\vec{r},t)$) to the right hand side of eqn. (1)
\begin{eqnarray}
\frac{d\vec{\hbar k}}{dt} = -\nabla_r E_C(\vec{r},t) + F_S(\vec{r},t) \mbox{ ,} \label{eq:newton-diff}
\end{eqnarray}

It is relatively easy to solve eqns. (\ref{eq:newton}) and (\ref{eq:newton-diff}) numerically.  
One solves the equations of motion, eqns. (1) and (2) to move a particle through phase space.  
Random numbers are chosen to mimic the scattering process and occasionally kick a carrier to 
another trajectory.  By averaging the results for a large number of simulated trajectories, 
these so-called Monte Carlo techniques provide a rigorous, though computationally demanding 
description of carrier transport in devices as described in [Jac89].

Device engineers are primarily interested in average quantities such as the average electron 
density, current density, etc. (There are some exceptions; noise is important too.)  
Instead of simulating a large number of particles, we can ask:  What is the probability 
that a state at position, $\vec{r}$, with momentum $\hbar \vec{k}$, is occupied at time $t$? 
The answer is given by the distribution function, $f(\vec{r},\vec{k},t)$,
 which can be computed by averaging the results of a large number of simulated trajectories.  
Alternatively, we can adopt a collective viewpoint instead of the individual particle viewpoint 
and formulate an equation for $f(\vec{r},\vec{k},t)$. The result is known as the Boltzmann 
Transport Equation (BTE),
\begin{eqnarray}
\frac{\partial f}{\partial t} + \vec{v}\cdot \nabla_r f - \frac{q\vec{E}}{\hbar k} \nabla_k f = 
\hat{C} f \label{eq:BTE}
\end{eqnarray}
where $\vec{E}$ is the electric field, and $\hat{C}f$ describes the effects of scattering.  
In equilibrium  $f(\vec{r},\vec{k},t)$ is simply the Fermi function, but in general, 
we need to solve eqn. (\ref{eq:BTE}) to find f.  Once  $f(\vec{r},\vec{k},t)$ is known, 
quantities of interest to the device engineer are readily found.  For example, to find 
the average electron density in a volume $\Omega$ centered at position, $\vec{r}$ , 
we simply add up the probability that all of the states in $\Omega$ are occupied and divide by the volume,
\begin{eqnarray}
n(\vec{r},t) = \frac{1}{\Omega} \sum_k f(\vec{r},\vec{k},t)\mbox{ .} \label{eq:B-n}
\end{eqnarray}
Similarly, we find, 
\begin{eqnarray}
\mbox{Current Density} && \vec{J}(\vec{r},t) = \frac{1}{\Omega} \sum_k (-q) \vec{v} f(\vec{r},\vec{k},t)\mbox{ ,}
           \label{eq:B-J} \\
\mbox{Kinetic Energy Density} && W(\vec{r},t) = \frac{1}{\Omega} \sum_k E(\vec{k})  f(\vec{r},\vec{k},t)\mbox{ ,}
           \label{eq:B-W} \\
\mbox{Energy Current Density} && \vec{J}_E(\vec{r},t) = \frac{1}{\Omega} \sum_k E(\vec{k})  
\vec{v} f(\vec{r},\vec{k},t)\mbox{ ,}         \label{eq:B-JE}
\end{eqnarray}
where $q>0$ is the absolute value of the electron charge.
This approach provides a clear and fairly rigorous description of semiclassical carrier 
transport, but solving the six dimensional BTE is enormously difficult.  One might ask 
if we can't just find a way to solve directly for the quantities of interest in eqns. 
(\ref{eq:B-n}) - (\ref{eq:B-JE}).  The answer is yes, but some simplifying assumptions are necessary.

Device engineers commonly describe carrier transport by few low order moments of the 
Boltzmann transport equation, eqn. (\ref{eq:BTE}).  A mathematical prescription for 
generating moment equations exists, but to formulate them in a tractable manner, 
numerous simplifying assumptions are required \cite{Lun00}.  Moment equations provide 
a phenomenological description of transport that gives insight and quantitative results when properly calibrated. 

The equation for the zeroth moment of $f(\vec{r},\vec{k},t)$ gives the well known 
continuity equation for the electron density, $n(\vec{r},t)$,
\begin{eqnarray}
\frac{\partial n(\vec{r},t)}{\partial t} = -\nabla_r F_n + G_n - R_n \mbox{ ,} \label{eq:BTE-n}
\end{eqnarray} 
where $F_n$ is the electron flux, 
\begin{eqnarray}
J_n = -q F_n \mbox{ ,}
\end{eqnarray}
$G_n$ the electron generation rate, and $R_n$ the 
electron recombination rate. Equation (\ref{eq:BTE-n}) states that the electron 
density at a location increases with time if there is a net flux of electrons 
into the region (as described by the first term, minus the divergence of the 
electron flux) or if carriers are being generated there.  Recombination causes 
the electron density to decrease with time.  Any physical quantity must obey a 
conservation law like eqn. (\ref{eq:BTE-n}).

The equation for the first moment of $f(\vec{r},\vec{k},t)$ gives the equation 
for average current density (eqn. (\ref{eq:B-J})) projection on the $x$-axis,
\begin{eqnarray}
\frac{\partial J_{nx}}{\partial t} = \frac{2q}{m} \frac{d W_{xx}}{dx} 
+ \frac{nq^2}{m^\ast} E_x - \frac{J_{nx}}{\tau_m} \mbox{ .} \label{eq:BTE-Jnx}
\end{eqnarray}
Each term on the right hand side of eqn. (\ref{eq:BTE-Jnx}) is analogous 
to the corresponding terms in eqn. (\ref{eq:BTE-n}).  The current typically 
changes slowly on the scale of the momentum relaxation time, $\tau_m$, 
(typically a sub-picosecond time) so the time derivative can be ignored and eqn. (\ref{eq:BTE-Jnx}) solved for
\begin{eqnarray}
J_{nx} = nq\mu_nE_x + \frac{2}{3} \mu_n \frac{d W_{xx}}{dx} \mbox{ ,} \label{eq:Jnx}
\end{eqnarray}
where
\begin{eqnarray}
\mu_n = \frac{q \tau_m}{m^\ast}
\end{eqnarray}
is the electron mobility, and we have assumed equipartition of energy so that $W_{xx}=W/3$, 
where W is the total kinetic energy density. This assumption can be justified when there 
is a lot of isotropic scattering, which randomizes the carrier velocity. Eqn. (\ref{eq:Jnx}) 
is a drift-diffusion equation; it says that electrons drift in electric fields and diffuse 
down kinetic energy gradients. Near equilibrium,
\begin{eqnarray}
W=\frac{3}{2} nk_BT \mbox{ ,}
\end{eqnarray}
so when $T$ is uniform, eqn. (\ref{eq:Jnx}) becomes
\begin{eqnarray}
J_{nx} = nq \mu_n E_x + k_BT \mu_n \frac{dn}{dx} = nq \mu_n E_x + qD_n \frac{dn}{dx} \mbox{,} \label{eq:Jnx2}
\end{eqnarray}
the drift-diffusion equation. By inserting eqn. (\ref{eq:Jnx2}) in the electron 
continuity equation, eqn. (\ref{eq:BTE-n}), we get an equation for the electron 
density that can be solved for the electron density within a device. This is the 
traditional and still most common approach for describing transport in semiconductor devices \cite{Pie87}.

Since most devices contain regions with high electric fields, the assumption that $W=3 n k_BT/2$ 
is not usually a good one. The carrier energy enters directly into the second term of the 
transport equation, eqn. (\ref{eq:Jnx}), but it also enters indirectly because the mobility
is energy dependent. To treat transport more rigoursly, we need an equation for the electron energy.

The second moment of $f(\vec{r},\vec{k},t)$ gives the carrier energy density, $W(\vec{r},t)$ 
according to eqn (\ref{eq:B-JE}). The second moment of the BTE gives a continuity 
equation for the energy density \cite{Lun00}
\begin{eqnarray}
\frac{\partial W}{\partial t} = - \frac{dJ_W}{dx} + J_{nx} E_x - \frac{W-W_0}{\tau_E} \mbox{ ,} \label{eq:B-JW}
\end{eqnarray}
where $W_0$ is the equilibrium energy density and $\tau_E$ the energy relaxation time. 
Note that the energy relaxation time is generally longer than the momentum relaxation 
time because phonon energies are small so that it takes several scattering events to 
thermalize an energetic carrier but only one to randomize its momentum.

To solve eqn. (\ref{eq:B-JW}), we need to specify the energy current. The third moment 
of $f(\vec{r},\vec{k},t)$ gives the carrier energy flux, $J_W(\vec{r},t)$ according to 
eqn. (\ref{eq:B-JE}). The third moment of the BTE gives a continuity equation for the energy flux,
\begin{eqnarray}
J_W = W \mu_E E_x + \frac{d(D_EW)}{dx} \mbox{ ,} \label{eq:JW}
\end{eqnarray}
where $\mu_E$ and $D_E$ are appropriate energy transport mobility and diffusion coefficient \cite{Lun00}.

\begin{figure}
  \begin{center}
    \leavevmode
\includegraphics[height=7cm,angle=-0]{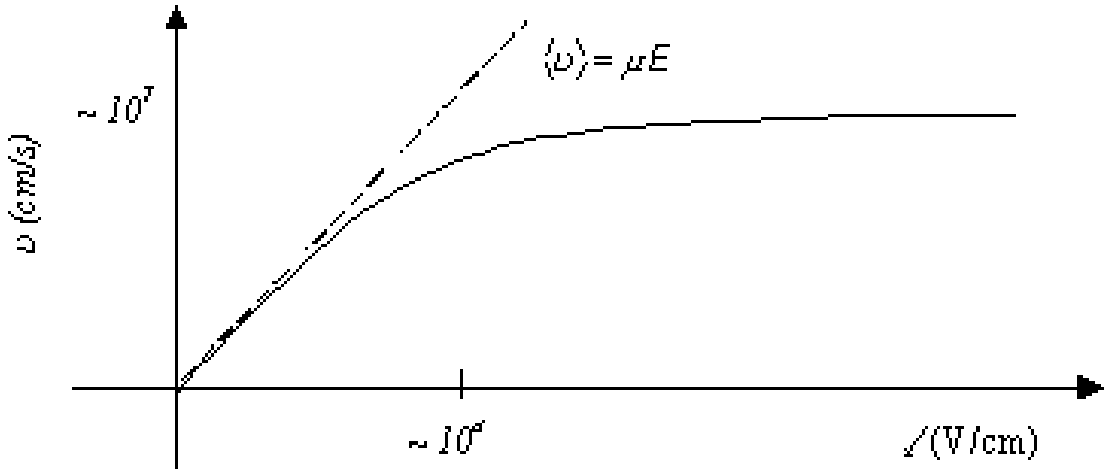}
  \end{center}
\caption{The average velocity vs. electric field for electrons in bulk silicon at room temperature.}
\label{fig:vel}
\end{figure}

Equations (\ref{eq:BTE-n}), (\ref{eq:Jnx2}), (\ref{eq:B-JW}) and (\ref{eq:JW}) can now be 
solved self-consistently to simulate carrier transport. Fig. (\ref{fig:vel}) sketches the 
result for uniformly doped, bulk silicon with a constant electric field. At low electric 
fields, $W \sim 3 nk_BT/2$, and we find that $<v_x> \sim \mu_n E_x$. For electric fields 
above  $\sim 10^4 V/cm$, the kinetic energy increases, which  increases the rate of 
scattering and lowers the mobility so that at high fields the velcoity saturates at 
$\sim 10^7 cm/s$. In a bulk semiconductor, there is a one-to-one relation between 
the magnitude of the electric field and the kinetic energy, so the mobility and 
diffusion coefficient can be parametrized as known functions of the local electric 
field. The result is that for bulk semiconductors or for large devices in which the 
electric field changes slowly, there is no need to solve all four equations; we need 
to solve the carrier continuity and drift-diffusion equations with field-dependent parameters.

\begin{figure}
  \begin{center}
    \leavevmode
\includegraphics[height=9cm,angle=-0]{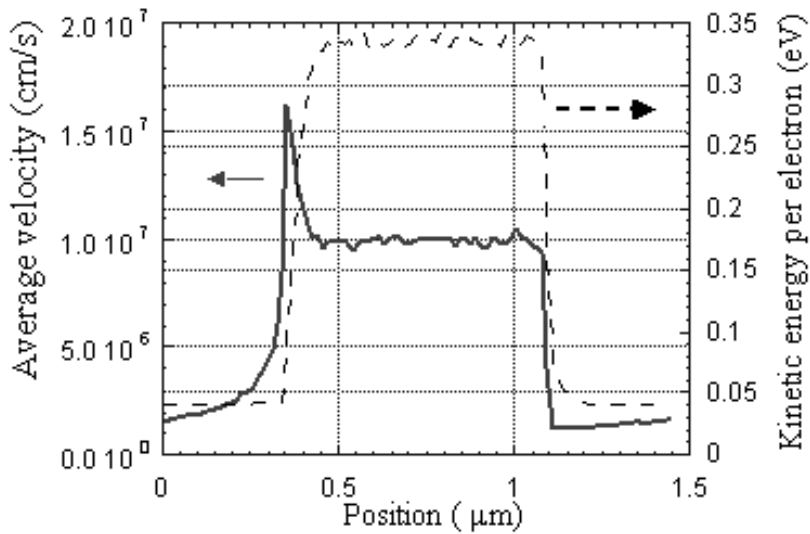}
  \end{center}
\caption{The average steady-state velocity and kinetic energy vs. postion for electrons injected 
into a short slab of silicon with low-high-low electric field profile.}
\label{fig:overshoot}
\end{figure}

Electric fields above $10^4 V/cm$ are common in nanoscale devices. This is certainly 
high enough to cause velocity saturation in the bulk, but in a short, high field region, 
transients occur. Fig. \ref{fig:overshoot} illustrates what happens for a hypothetical 
situation in which the electric field abrubtly jumps from a low value to a high value 
and then back to a low value again. Electrons injected from the low field region are 
accelerated by the high electric field, but energy relaxation times are longer than 
momentum relaxation times, so the eneregy is slow to respond. The result is that 
the mobility is initially high (even though the electric field is high), so the 
velocity can be higher than the saturated value shown in Fig. \ref{fig:vel}. 
As the kinetic energy increases, however, scattering increases, the mobility 
drops, and the velocity eventually decreases to $\sim 10^7 cm/s$, the saturated 
velocity for electrons in bulk silicon. The spatial width of the transient is 
roughly 100 nm; modern devices frequently have dimensions on this order, 
and strong velocity overshoot should be expected.

The example shown in Fig. \ref{fig:overshoot} demonstrates that it is better to think of the mobility 
and diffusion coefficient as functions of the local 
kinetic energy rather than the local electric field.  What this means is that eqns. (\ref{eq:BTE-n}), 
(\ref{eq:Jnx2}), (\ref{eq:B-JW}) and (\ref{eq:JW}) should all be solved self-consistently to simulate 
carrier transport in small devices.  Device simulation programs commonly permit two options:  
1) the solution of eqns. (\ref{eq:BTE-n}) and (\ref{eq:Jnx2}) self-consistently with the Poisson 
equation using mobilities and diffusion coefficients that depend on the local electric field 
(the so-called drift-diffusion approach) or 2) the solution of eqns. 
(\ref{eq:BTE-n}), (\ref{eq:Jnx2}), (\ref{eq:B-JW}) and (\ref{eq:JW}) 
self-consistently with the Poisson equation using mobilities and diffusion coefficients 
that depend on the local kinetic energy (so-called energy transport or hydrodynamic approaches).  
Solving the four equations self-consistently is more of a computational burden, 
but it is necessary for when the electric field changes rapidly on the scale of 
a mean-free-path for scattering.  Actually, numerous simplifying assumptions are also 
necessary to even write the current and energy flux equations as eqns. 
(\ref{eq:Jnx2}) and (\ref{eq:JW}) \cite{Lun00}.  
The most rigorous (and computationally demanding) simulations (so-called Monte Carlo simulations) 
go back to the individual particle picture and track carriers trajectories according to eqns. 
(\ref{eq:newton-k}) and (\ref{eq:newton-diff}).  
Drift-diffusion and energy transport approaches for treating carrier transport in semiconductor 
devices have two things in common; the first is the assumption that carriers can be treated as 
semiclassical particles and the second is the assumption that there is a lot of scattering.  
Both of these assumptions are losing validity as devices shrink.

\section{\label{section:semi-clas-ballistic}Semiclassical Transport: Ballistic}

\begin{figure}
  \begin{center}
    \leavevmode
\includegraphics[height=13cm,angle=-0]{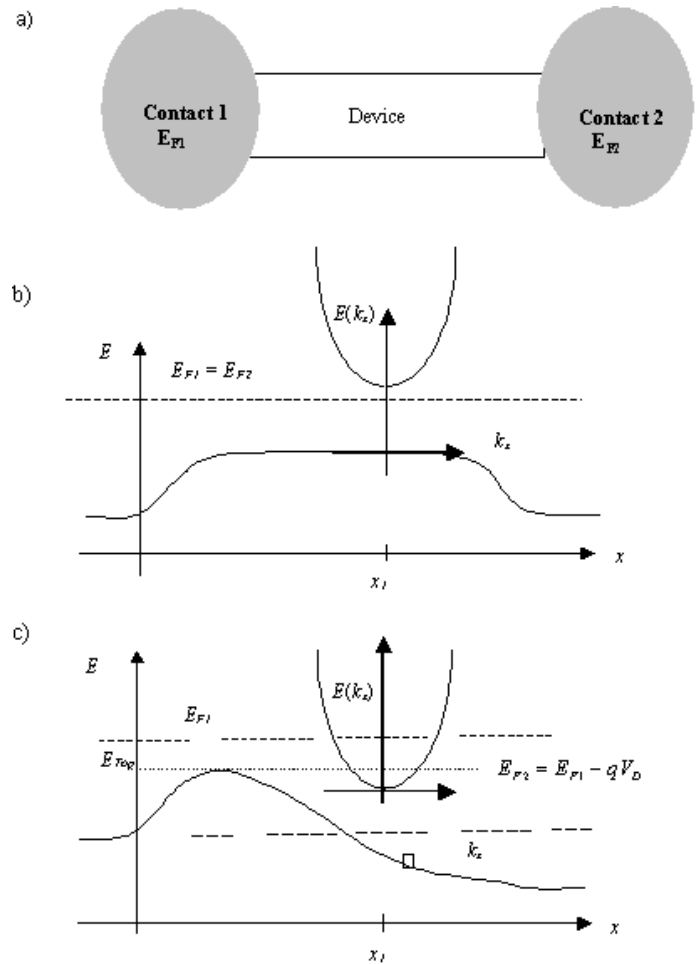}
  \end{center}
\caption{Sketch of a ballistic device with two contacts that function as reservoirs 
of thermal equilibrium carriers. (a) Device and the two contacts. (b) Energy 
band diagram under equilibrium conditions $(V_D = 0)$.  (c) Energy band diagram 
under bias ($V_D > 0$).}
\label{fig:ballistic-Boltzmann}
\end{figure}

Consider the "device" sketched in Fig. \ref{fig:ballistic-Boltzmann}(a), 
which consists of a ballistic region attached to two contacts. The left contact 
(source) injects a thermal equilibrium flux of carriers into the device; some 
carriers reflect from the potential barriers within the device, and the rest 
transmit across and enter the right contact (drain). A similar statement 
applies to the drain contact.  The source and drain contacts are assumed 
to be perfect absorbers, which means that carriers impinging them from 
the device travel without reflecting back into the device. To compute 
the electron density, current, average velocity, etc. within the device, 
we have two choices. The first choice treats the carriers as semiclassical 
particles and the Boltzmann equation is solved to obtain the distribution 
function $f(\vec{r},\vec{k},t)$ as discussed in the previous section. 
The second choice treats the carriers quantum mechanically as discussed 
in the next section. In this section, we will use a semiclassical description 
in which the local density-of-states within the device is just that of a 
bulk semiconductor, but shifted up or down by the local electrostatic potential.  
This approximation works well when the electrostatic potential does not vary too 
rapidly, so that quantum effects can be ignored.
To find how the k-states within the ballistic device are occupied, we solve 
the Boltzmann Transport Equation, eqn. (4).  Because the device is ballistic, 
there is no scattering, and $\hat{C}f=0$.  It can be shown \cite{Lun00} that the 
solution to the BTE with  $\hat{C}f=0$ is any function of the electron's total energy,
\begin{eqnarray}
E = E_C(x) + E(k)
\end{eqnarray}
where, $E_C(x)$ is the conduction band minimum versus position and E(k) is the band 
structure for the conduction band. We know that under equilibrium conditions sketched 
in Fig. 4(b), the proper function of total energy is the Fermi function,
\begin{eqnarray}
f(E) = \frac{1}{1+exp(\frac{E-E_{F}}{k_BT})} \mbox{ ,}
\end{eqnarray}
where the Fermi level, $E_F$, and temperature, $T$ are constant in equilibrium.

Now consider the situation in Fig. 4c where a drain bias has been applied 
to the ballistic device.  Although two thermal equilibrium fluxes are injected 
into the device, it is now very far from equilibrium.  Since scattering is what 
drives the system to equilibrium, the ballistic device is as far from equilibrium
 as it can be.  Nevertheless, for the ballistic device, the relevant steady-state 
Boltzmann equation is the same equation as in equilibrium.  The solution is again 
a function of the carrier's total kinetic energy.  At the contacts, we know that 
the solution is a Fermi function, which specifies the functional dependence 
on energy. For the ballistic device, therefore, the probability that 
a k-state is occupied is given by an equilibrium Fermi function. 
The only difficulty is that we have two Fermi levels, so we need to decide which one to use.

Return again to Fig. 4c and consider how to fill the states at $x = x_1$.  
We know that the probability that a k-state is occupied is given by a 
Fermi function, so we only need to decide which Fermi level to use for each k-state. 
For the positive k-states with energy above $E_{TOP}$ , the top of the energy barrier, 
the states can only have been occupied by injection from the source, so the appropriate 
Fermi level to use is the source Fermi level.  Similarly, negative k-states with 
energy above $E_{TOP}$ can only be occupied by injection from the drain, 
so the appropriate Fermi level to use is the drain Fermi level.  
Finally, for k-states below $E_{TOP}$, both positive and negative 
velocity states are populated according to the drain Fermi level. 
The negative velocity k-states are populated directly by injection 
from the drain and the positive k-states are populated when negative 
velocity carriers reflect from the potential barrier.

Ballistic transport can be viewed as a special kind of equilibrium.  
Each k-state is in equilibrium with the contact from which it was populated.  
Using this reasoning, one can compute the distribution function and any moment 
of it (e.g. carrier density, carrier velocity, etc.) at any location within the 
device.  Figure 5 shows that computed distribution function in a ballistic 
nanoscale MOSFET under high gate and drain bias \cite{Rhe02}.  A strong ballistic 
peak develops as carriers are injected from the source are accelerated in the 
high electric field near the drain.  Each k-state is in equilibrium with one of 
the two contacts, but the overall carrier distribution is very different from 
the equilibrium Fermi-Dirac distribution.  When scattering dominates, carriers 
quickly lose their "memory" of which contact they were injected from, but for 
ballistic transport there are two separate streams of carriers; one injected 
from the source and one from the drain.

To evaluate the electron density vs. position within the ballistic device, we should 
compute a sum like eqn. (\ref{eq:B-n}), but we must do two sums. one for the states 
filled from the left contact and another for the states filled from the right contact,
\begin{eqnarray}
n(x) = \sum_{k_L} \Theta_L(x) f_L(E) + \sum_{k_R} \Theta_R(x) f_R(E) \mbox{ ,} 
\label{eq:semi-ball-n}
\end{eqnarray}
where $f_L$ and $f_R$ are the equilibrium Fermi functions of contacts L and R and 
$\Theta_{L,R}(x)$ is a function that selects out the k-states at position, x, 
that can be filled by contact L or R according to the procedure summarized in Fig. 4.  
It is often convenient to do the integrals in energy space rather than in k-space 
in which case, eqn. (\ref{eq:semi-ball-n}) becomes
\begin{eqnarray}
n(x) = \int dE \left[ LDOS_{L} (x,E) f_L(E) +  \int LDOS_{R} (x,E) f_R(E) \right]
\label{eq:semi-ball-n2}
\end{eqnarray}
where $LDOS_{L,R}(x,E)$  is the local density of states at energy E, fillable from 
contact L or R.  For diffusive transport, we deal with a single density-of-states 
and fill it according to a since quasi-Fermi level, but for ballistic devices, 
the density of states separates into parts fillable from each contact.

The current flowing from source to drain (drain to source) contact is simply the 
transmission probability $T(E)$, times the Fermi function of the source (drain) 
contact. The net current flowing in the device is then,
\begin{eqnarray}
I = \frac{e}{h} 2 \int dE T(E) \left[f_L(E) - f_R(E)\right] \mbox{ .}\label{eq:semi-ball-current}
\end{eqnarray}
For the semiclassical example of Fig. \ref{fig:ballistic-Boltzmann}, 
T(E)=0 for $E<E_{TOP}$ and T(E)=1 for $E>E_{TOP}$.

\section{\label{section:quantum} Phase Coherent Quantum Transport: The Landauer-Buttiker Formalism}

\begin{figure}
  \begin{center}
    \leavevmode
\includegraphics[height=7cm,angle=-0]{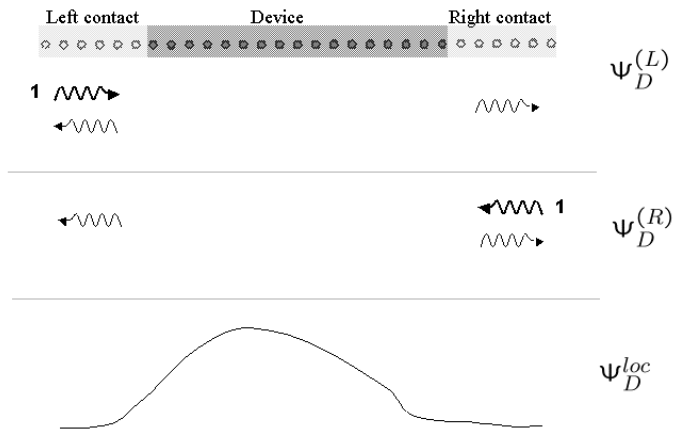}
  \end{center}
\caption{All wave funcitons in the device can be represented as left incident 
($\Psi_D^{(L)}$), right incident ($\Psi_D^{(R)}$) or localized states ($\Psi_D^{(loc)}$).}
\label{fig:wavefn}
\end{figure}

Quantum mechanically the electron is a wave and the wave function 
$\Psi(\vec{r})$ is obtained by solving Schrodinger's equation,
\begin{eqnarray}
\left[-\frac{\hbar^2}{2m} \nabla^2 + V(\vec{r}) \right] \Psi_n(\vec{r}) = E_n \Psi_n(\vec{r}) \mbox{  ,}
\end{eqnarray}
where $E_n$ is the energy. Consider a device connected to two contacts as shown 
in Fig. \ref{fig:wavefn}, where the contacts are assumed to have a constant 
electrostatic potential. In a manner identical to the discussion of semiclassical modeling in the 
previous section, where we considered corpuscular electrons incident from the 
left and right contacts, in quantum mechanical modeling, we need to consider 
electron waves incident from the left and right contacts. The electron wave 
function in device region D can be thought to arise from:
\begin{itemize}
\item Waves incident from left lead of the form $e^{ikx}$, which have 
transmitted and reflected components  $te^{ik^\prime x}$ and $r e^{ik x}$ 
in the right and left leads respectively. The wave function in the device 
region $D$ due to this wave is represented by $\Psi_D^{(L)}$.
\item Waves incident from right lead of the form $e^{-ikx}$, which have 
transmitted and reflected components $t e^{-ik^\prime x}$ and $r e^{ik x}$ 
in the right and left leads respectively. The wave function in the device 
region $D$ due to this wave is represented by $\Psi_D^{(R)}$.
\item States localized in Device region represented by $\Psi_D^{(loc)}$. 
Localized and quasi-localized states are filled up by scattering due to
 electron-phonon and electron-electron interaction. We will assume here 
that localized states are absent.

\end{itemize}

The Landauer-Buttiker approach expresses the expectation value of an
operator in terms of the left and right incident electrons from the contacts
and their distribution functions.
The expectation value of operator $\hat{Q}$ is:
\begin{eqnarray}
Q = \sum_{k,s} \left[ <\Psi^{(L)}_D| \hat{Q} |\Psi^{(L)}_D> f_L(E) + 
           <\Psi^{(R)}_D| \hat{Q} |\Psi^{(R)}_D> f_R(E) \right] \mbox{ .}
\label{eq:expec1}
\end{eqnarray}
Here the summation is performed over the momentum, $k$, and the spin, $s$, states.
Since in this chapter we do not consider any spin-depnednt phenomena, 
the spin summation trnaslates into a factor of 2. We will distinguish this factor
by keeping in front of sums and integrals.
Eq. (\ref{eq:expec1}) has contributions from two physically different
sources. The first term corresponds to contribution from waves
incident from the left contact ($\Psi^{(L)}_D$) at energy $E$, 
weighted by the Fermi
factor of the left contact ($f_L$). The second term corresponds to 
waves incident from the right contact ($\Psi^{(R)}_D$) weighted by the
Fermi factor of the right contact ($f_R$). More generally, if region D
is connected to a third contact G, then the expectation
value of operator $\hat{Q}$ is,
\begin{eqnarray}
Q = \sum_{k,s} \left[ <\Psi^{(L)}_D| Q |\Psi^{(L)}_D> f_L(E) + 
               <\Psi^{(R)}_D| Q |\Psi^{(R)}_D> f_R(E) +
               <\Psi^{(G)}_D| Q |\Psi^{(G)}_D> f_G(E) \right] \mbox{ ,}      
\label{eq:expectation}
\end{eqnarray}
where $\Psi^{(G)}_D$ corresponds to the wave function in the Device due to waves
incident from contact $G$ and $f_G$ is the Fermi factor of contact G.
Using eqn. (\ref{eq:expec1}), the contribution to electron ($n$) and current (J) 
densities at $x$ in the device region D are given by,
\begin{eqnarray}
n(x) =  \sum_{k,s} \left[ |\Psi^{(L)}_D(x)|^2 f_L(E) + |\Psi^{(R)}_D(x)|^2 f_R(E) \right]
                   \label{eq:expectation_q}
\end{eqnarray}
and
\begin{eqnarray}
J(x) = \sum_{k,s} \frac{e\hbar}{2mi} 
   \left[{\Psi^{(L)}_D(x)}^\dagger \frac{d\Psi^{(L)}_D(X)}{dx} f_L(E) +
         {\Psi^{(R)}_D(x)}^\dagger \frac{d\Psi^{(R)}_D(x)}{dx} f_R(E)
              - c.c\right] \mbox{ . } 
      \label{eq:expectation_J}
\end{eqnarray}
The quantum mechanical density of states at energy $E$ due to waves incident 
from the left ($LDOS_L$) and right ($LDOS_R$) are:
\begin{eqnarray}
LDOS_L (x,E) &=& 2 \sum_{\mbox{k states with energy E incident from the left contact}}  |\Psi^{(L)}_D(x)|^2 \\
LDOS_R (x,E) &=& 2 \sum_{\mbox{k states with energy E incident from the right contact}}  |\Psi^{(R)}_D(x)|^2 \mbox{ .}
\end{eqnarray}
Then the electron density can be written in the same form as eqn. (\ref{eq:semi-ball-n2}),
\begin{eqnarray}
n(x) = \int LDOS_{L} (x,E) f_L(E) dE +  \int LDOS_{R} (x,E) f_R(E) dE \mbox{ ,}
\end{eqnarray}
except that the expressions for the local density of states are different.
Similarly, eqn. (\ref{eq:expectation_J}) can be expressed in a form identical 
to eqn. (\ref{eq:semi-ball-current}). The above formalism can be extended to calculate noise (shot and Johnson-Nyquist) 
in nanodevices [Buttiker92].
Device modeling in the phase coherent limit involves solving Schrodinger's 
equation to obtain the electron density self-consistently with Poisson's equation.

\section{\label{section:quantum-gf} Quantum Transport with Scattering: The need for Green's functions}

The description in the previous section is valid only in the phase coherent limit.
The terminology "phase coherent" refers to a deterministic evolution of both 
the amplitude and phase of $\Psi_n(\vec{r})$ as given by Schrodinger's equation. 
The quantum mechanical wave function evolves phase coherently only in 
the presence of rigid scatterers, a common example of which is the electrostatic 
potential felt by an electron in the device. The wave function of an electron 
loses phase coherence due to scatterers which have an internal degree of freedom 
such as phonons. Phase incoherent scattering involves irreversible loss of phase 
information to phonon degrees of freedom. 
Naturally, including loss of phase information is  important 
when device dimensions become comparable to the scattering lengths due to 
phonons and other phase breaking mechanisms. Accurate modeling of nanodevices 
should have the ability to capture:
\begin{itemize}
\item Interference effects
\item Quantum mechanical tunneling
\item Discrete energy levels due to confinement in 2D and 3D device
geometries
\item Scattering mechanisms (electron-phonon, electron-electron).
\end{itemize}
The first three effects can be modeled by solving Schodinger's equation in a 
rigid potential as discussed in section \ref{section:quantum}. 
While in the semiclassical device modeling, the Boltzmann equation accounts 
for the energy and momentum relaxation due to scattering mechanisms, in quantum mechanical device modeling, 
the NEGF approach is necessary to account for energy, momentum and quantum 
mechanical phase relaxation. 

In the remainder of this section, we explain the NEGF approach in the phase 
coherent limit by starting
from Schrodinger's equation \cite{Dat96}.  We will start by an explanation of the 
tight binding Hamiltonian and relate this Hamiltonian to a device with open 
boundary conditions (section \ref{sec:tbh}). The open boundary conditions 
lead to an infinite dimensional matrix. We will describe a procedure to 
fold the effect of the open boundaries into the finite device region in 
section \ref{section:eliminate}. This will allow us to deal with small 
matrices where the open boundaries are modeled by
{\it contact self energies}. The Green's functions, self energies and 
their relationship to current and electron density are derived in
 section \ref{sect:GF}. Then in section \ref{sec:self-energy}, we 
extend the discussion in section \ref{sect:GF} to one include 
electron-phonon interaction, which is where the NEGF approach is 
really essential.

\subsection{\label{sec:tbh}Tight binding Hamiltonian for a one dimensional device}

Consider a system described by a set of one dimensional grid / lattice 
points with uniform spacing $a$. Further assume that only nearest 
neighbor grid points are coupled. A spatially {\it uniform system} with
a constant potential has the Hamiltonian:
\begin{eqnarray}
(E-H) \Psi = 0
\rightarrow
\left(
\begin{array}{cccccccccc}
\bullet & \bullet & \bullet &            &            &            &
    &         &         &    \\
        & \bullet & \bullet &   \bullet  &            &            &
    &         &         &    \\
        &         &     -t  & E-\epsilon &    -t      &            &
    &         &         &    \\
        &         &         &            &    -t      & E-\epsilon &     -t
    &         &         &    \\
        &         &         &            &            &    -t      & 
E-\epsilon & -t      &         &    \\
        &         &         &            &            &            &   
\bullet  & \bullet & \bullet &    \\
        &         &         &            &            &            &
    & \bullet & \bullet & \bullet
\end{array}
\right)
\left(
\begin{array}{c}
\bullet \\
\bullet\\
\Psi_{-1}\\
\Psi_0\\
\Psi_{+1}\\
\bullet\\
\bullet
\end{array}
\right)=0
\end{eqnarray}
\normalsize
or
\begin{eqnarray}
-t \Psi_{q-1} + (E - \epsilon) \Psi_q - t \Psi_{q+1} = 0 \mbox{,} 
\label{eq:tb-un1}
\end{eqnarray}
where, $E$ is the energy and $\Psi_q$ is the wave function at grid
point $q$. The Hamiltonian matrix is tridiagonal because of nearest
neighbor interaction.
The diagonal and off-diagonal elements of the Hamiltonian $\epsilon$ 
and $t$ represent the potential and interaction between nearest 
neighbor grid points $q$ and $q+1$ respectively.

The solution of eqn. (\ref{eq:tb-un1}) can be easily verified 
using Bloch theorem to be,
\begin{eqnarray}
E &=& \epsilon + 2t cos(ka)  \label{eq:E-tb-un1} \\
\Psi_q &=& e^{ikqa}          \label{eq:Psi-tb-un1}   \mbox{ ,}
\end{eqnarray}
and the group velocity is,
\begin{eqnarray}
v = \frac{1}{\hbar} \frac{\partial E}{\partial k} 
  =  -\frac{2at}{\hbar} sin(ka)  \mbox{ .}
\end{eqnarray}

The uniform tight binding Hamiltonian in eqn. (\ref{eq:tb-un1}) can be
extended to a {\it general nearest neighbor tight binding} Hamiltonian 
given by,
\begin{eqnarray}
t_{q,q-1} \Psi_{q-1} + (E - \epsilon_{q}) \Psi_q + t_{q,q+1} \Psi_{q+1} = 0
\mbox{,} \label{eq:tb-nn}
\end{eqnarray}
where, $\epsilon_q$ is the on-site potential at grid point $q$ and
$t_{q,q+1}$ is the Hamiltonian element connecting grid points $q$ and
$q+1$. $t_{q+1,q}=t_{q,q+1}^\dagger$ because the Hamiltonian is Hermitian.

\begin{center}
\fcolorbox{red}{white}{
\begin{minipage}{\textwidth}
In the special case of the discretized Schrodinger equation on a uniform
grid,
\begin{eqnarray}
t = t_{q,q+1} = t_{q+1,q} = - \frac{\hbar^2}{2ma^2} \;\; \mbox{ and }
\;\; \epsilon_q &=& V_q + \frac{\hbar^2}{ma^2} \mbox{ ,} \label{eq:eps}
\end{eqnarray}
where $a$ is the grid spacing and $V_q$ is the electrostatic potential
at grid point $q$.
\end{minipage}
}
\end{center}

\begin{figure}
  \begin{center}
    \leavevmode
\includegraphics[width=15cm,angle=-0]{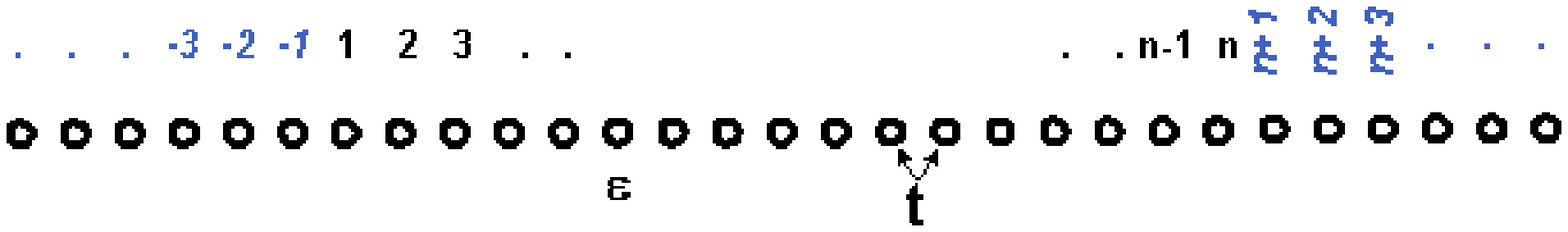}
  \end{center}
\caption{A one dimensional device connected to two semi-infinite leads. 
While the potential in the leads are held fixed, the potential in the 
device can spatially vary.}
\label{fig:1D-chain}
\end{figure}

\subsection{\label{section:eliminate} 
Eliminating the Left and Right semi-infinite leads}

A typical nanodevice  can be conceptually divided into three regions
(Fig. \ref{fig:1D-chain}):
\begin{itemize}
\item Left semi-infinite lead (L) with a constant potential $\epsilon_l$
\item Device (D) with and arbitrary potential, and 
\item Right semi-infinite lead (R) with a constant potential $\epsilon_r$
\end{itemize}
The potenial of the left (right) lead $\epsilon_l$ ($\epsilon_r$) and the 
hopping parameter $t_l$ ($t_r$) are assumed to be constant, which signifies 
that the leads are highly conducting and uniform. 
Then the Hamiltonian of the device and leads, eqn. (\ref{eq:tb-nn}), is an 
infinite dimensional matrix that can be expanded as,
\begin{eqnarray}
\bullet \nonumber \\
\bullet \nonumber \\
    -t_{l} \Psi_{l -1} + (E-\epsilon_{l}) \Psi_{l 0} - t_{l,d} \Psi_{1}  &=&
0 \label{eq:1D1} \\
  -t_{d,l} \Psi_{l 0} + (E-\epsilon_{1}) \Psi_{1}   - t_{1,2} \Psi_{2}  &=&
0 \label{eq:1D2} \\
  -t_{1,2} \Psi_{1}   + (E-\epsilon_{2}) \Psi_{2}   - t_{2,3} \Psi_{3}  &=&
0 \label{eq:1D5} \\
-t_{n-1,n} \Psi_{n-1} + (E-\epsilon_{n}) \Psi_{n}   - t_{dr}  \Psi_{rn+1} &=&
0 \label{eq:1D3} \\
  -t_{r,d} \Psi_{n}   + (E-\epsilon_{r}) \Psi_{rn+1}  - t_{r}   \Psi_{rn+2} &=&
0 \label{eq:1D4} \\
  \bullet \nonumber \\
\bullet \nonumber
\end{eqnarray}
where, the bullets represent the semi-infinite Left and Right leads. 
The subscript $lm$ ($rm$) refer to grid point $m$ in the Left
(Right) lead. However, to find the electron density in eqn. (\ref{eq:expectation_q}), 
the wave function is only required at the device grid points. We will now discuss a 
procedure to fold the influence
of the left and right semi-infinite leads into the device region.

\noindent
{\it Terminating the semi-infinite Left and Right leads:}
The wave function in the contacts due to waves incident from the
left lead are,
\begin{eqnarray}
\Psi_{ln} &=& (e^{+ik_ln} + s_{ll} e^{-ik_ln} ) u_{ln} \;\;\;\;\;
\mbox{ in region L}         \label{eq:wavefn1}\\
\Psi_{rn} &=&  s_{rl} e^{ik_r n} u_{rn}
\;\;\;\;\;\;\;\;\;\;\;\;\;\;\;\;\;\;\;\;\;\;
\mbox{ in region R}\mbox{, } \label{eq:wavefn2}
\end{eqnarray}
and the corresponding eigen values are [eqn. (\ref{eq:E-tb-un1})],
\begin{eqnarray}
E-\epsilon_l = 2 t_l cos(k_la) = t_l (e^{ik_la} + e^{-ik_la})
\label{eq:eigenvall1} \mbox{ ,}
\end{eqnarray}
and similarly for the right contact, with the indices $r$.
$s_{ll}$ and $s_{rl}$ are the reflection and transmission amplitudes.
Substituting Eqs. (\ref{eq:wavefn1}) and (\ref{eq:eigenvall1}) in eqn.
(\ref{eq:1D1}) yields,
\begin{eqnarray}
&&s_{ll}  = t_l^{-1} (-t_l  + t_{ld} \Psi_1) \label{eq:1D13} 
\mbox{ .}
\end{eqnarray}
Substituting Eqs. (\ref{eq:wavefn1}) and (\ref{eq:1D13}) in 
eqn. (\ref{eq:1D2}), we obtain,
\begin{eqnarray}
(E-\epsilon_{1} - t_{d,l} e^{+ik_la} t_l^{-1} t_{l,d}) \Psi_{1} +
t_{1,2} \Psi_{2} = -2i t_{d,l} sin(k_la)     \label{eq:1D21}
\end{eqnarray}
Eq. (\ref{eq:1D21}) is a modification of Schrodinger's equation 
centered at grid point $1$ of the Device (eqn. (\ref{eq:1D2})) to 
\underline{include} the influence of the entire semi-infinite Left lead.
Similarly, substituting eqn. (\ref{eq:wavefn2}) and
$E-\epsilon_r = 2 t_r cos(k_ra)$
 in eqn. (\ref{eq:1D4}) we get,
\begin{eqnarray}
&& s_{rl}  = t_r^{-1} t_{d,r} \Psi_n    \label{eq:1D43} \mbox{ .}
\end{eqnarray}
Now, substituting Eqs. (\ref{eq:wavefn2}) and (\ref{eq:1D43}) in 
eqn. (\ref{eq:1D3}), we can terminate the right semi infinite region
to yield,
\begin{eqnarray}
-t_{n-1,n} \Psi_{n-1} + (E-\epsilon_{n} - t_{d,r} e^{ik_r} t_r^{-1}
t_{r,d}) 
\Psi_{n} &=& 0 \mbox{ .}    \label{eq:1D31}
\end{eqnarray}
Eq. (\ref{eq:1D31}) is a modification of Schrodinger's equation centered at 
grid point $n$ of the Device (eqn. (\ref{eq:1D3})) to include the influence 
of the entire semi-infinite Right lead.

\begin{center}
\fcolorbox{red}{white}{
\begin{minipage}{\textwidth}
The influence of the semi-infinite Left and Right leads have been folded into 
grid points $1$ and $n$
of the device, for waves incident from the Left lead
[Eqs. (\ref{eq:1D21}) and (\ref{eq:1D31})]. Now the wave function in the 
device due to waves incident from the left lead can be obtained by solving the 
following $n$ dimensional matrix instead of the infinite dimensional matrices 
in eqns. (\ref{eq:1D1}) - (\ref{eq:1D4}):
\begin{eqnarray}
A \Psi_D^{(L)} = i_L \mbox{,} \label{eq:A1}
\end{eqnarray}
where, A is a square matrix of dimension n, $\Psi_D^{(L)}$ and $i_L$ 
are n by 1 vectors. $i_L$ is the source function at $(k,E)$ due to the
Left lead. Matrix A is,
\begin{eqnarray}
A  = EI - H_D - \Sigma_{lead} \mbox{} \label{eq:A-define}
\end{eqnarray}
and the non zero elements of $\Sigma_{lead}$ are,
\begin{eqnarray}
{\Sigma_{lead}}_{_{1,1}} &=&  t_{d,l} e^{+ik_la} t_l^{-1} t_{l,d} = \Sigma_L
\label{eq:sigmaLr} \\ 
{\Sigma_{lead}}_{_{n,n}} &=& t_{d,r} e^{+ik_ra} t_r^{-1} t_{r,d} = \Sigma_R 
                                  \mbox{ .} \label{eq:sigmaRr}
\end{eqnarray}
$\Sigma_L$ and $\Sigma_R$ are called the {\bf self-energies},
and they represent the influence of the semi-infinite Left and 
Right leads on the Device respectively.
The real part of self-energy shifts the on-site potential at grid
point 1 from $\epsilon_1$ to $\epsilon_1 + \mbox{Re}(\Sigma_L)$.
The imaginary part of self energy multiplied by $-2$
is the scattering rate of electrons
from grid point 1 of Device to Left lead (scattering rate $={- \mbox{2Im}[\Sigma_L]}$), in the weak coupling limit.

In a manner identical to
the derivation of eqn. (\ref{eq:A1}), for
waves incident from right contact, the wave function in the device 
($\Psi_D^{(R)}$) can be obtained by solving:
\begin{eqnarray}
A \Psi_D^{(R)} = i_R \mbox{,} \label{eq:A9}
\end{eqnarray}
where $i_R$ is the source function due to the Right
semi infinite contact.
The only non zero elements of $A$, $i_L$ and $i_R$ are:
\begin{eqnarray}
A(1,1)&=& E-\epsilon_{1} - \Sigma_L  \mbox{ and }
A(n,n)= E-\epsilon_{n} - \Sigma_R \label{eq:A4} \label{eq:A2}  \\
A(i,i)&=& E-\epsilon_{i} \mbox{ , } A(i,i+1)= -t_{i,i+1} \mbox{ and }
A(i+1,i)= -t_{i,i+1}^\dagger \label{eq:A3} \\
i_L(1) &=& - 2i t_{d,l} sin(k_la)  \mbox{ and } \label{eq:A7} \\
i_R(n) &=& -2i t_{d,r} sin(k_ra)  \mbox{ .} \label{eq:A10}
\end{eqnarray}
\end{minipage}
}
\end{center}

\subsection{\label{sect:GF} Electron and current densities expressed in terms of Green's functions}

The  Green's function corresponding to Schrodinger's equation
($[E-H]\Psi = 0$) for the device and contacts is,
\begin{eqnarray}
[E-H+i\eta] G = I \mbox{ ,}  \label{eq:genGr}
\end{eqnarray}
where $\eta$ is an infinitesimally small positive number which pushes
the poles of $G$ to the lower half plane in complex energy, and $H$
is the Hamiltonian.
The Green's function of device region D with the influence of
the contacts included is,
\begin{eqnarray}
A G = I \mbox{,} \label{eq:Gr0}
\end{eqnarray}
where
\begin{eqnarray}
A = EI - H_D - \Sigma_{lead}(E)
\end{eqnarray}
is an $n$ dimensional matrix defined in eqns. (\ref{eq:A2}) and (\ref{eq:A3}).
The formal derivation of eqn. (\ref{eq:Gr0}) is given in the appendix \ref{sect:appendix1}.

Using the definition for $G$ in Eqs. (\ref{eq:A1}) and 
(\ref{eq:A9}), the wave function in region D due to waves incident 
from Left and Right contacts can be written as, 
\begin{eqnarray}
\Psi_D^{(L)} &=& G i_L \;\;\;\;\; \mbox{ and } \label{eq:Psil} \\
\Psi_D^{(R)} &=& G i_R \mbox{.} \label{eq:Psir}
\end{eqnarray}
As $i_L$ and $i_R$ are non zero only at grid points $1$ and $n$, the full $G$ 
matrix is not necessary to find the wave function in the device; Only the two columns $G(:,1)$ and 
$G(:,n)$ are necessary.

The {\bf electron density} at grid point $q$ can now be written using eqns. (\ref{eq:Psil}) 
and (\ref{eq:Psir}) in eqn. (\ref{eq:expectation_q}) as
\begin{eqnarray}
n_q &=&  \sum_{k,s} {G}_{q,1} i_L i_L^\dagger {G^\dagger}_{1,q} f_L + 
                {G}_{q,n} i_R i_R^\dagger {G^\dagger}_{n,q} f_R
                     \label{eq:expectation_q1} \\
       &=&  \sum_{k,s} {G}_{q,1} [4 t_{d,l} sin^2(k_la) t_{l,d} f_L] {G^\dagger}_{1,q} +
        {G}_{q,n} [4 t_{d,r} sin^2(k_ra) t_{r,d} f_R]  {G^\dagger}_{n,q}
                      \mbox{, } \label{eq:expectation_q2} 
\end{eqnarray}
where $G^\dagger$ is the Hermitean conjugate of the Green's function.
The summation over $k$ can be converted to an integral over $E$ by,
\begin{eqnarray}
\sum_k \rightarrow \int \frac{dE}{2 \pi} |\frac{dk}{dE}| 
\label{eq:k-to-E}
\mbox{ .}
\end{eqnarray}
Using eqn. (\ref{eq:k-to-E}) and $|\frac{dE}{dk}| = 2a|t||sin(k_l a)|$,
eqn.  (\ref{eq:expectation_q2}) becomes,
\begin{eqnarray}
n_q &=&  2 \int \frac{dE}{2\pi} \left[G_{q,1}(E) \Sigma_L^{in}(E) G^\dagger_{1,q}(E)
          + G_{q,n}(E) \Sigma_R^{in}(E) G^\dagger_{n,q}(E) \right]
          \cdot \frac{1}{a}
\mbox{, } \label{eq:expectation_q3}
\end{eqnarray}
where,
\begin{eqnarray}
\Sigma_L^{in}(E) &=&  2 t_{d,l} \frac{1}{|t|} |sin(k_la)| t_{l,d} f_L (E)  
\label{eq:sigmaL<} \;\;\;\;\;\;\mbox{ and }\\
\Sigma_R^{in}(E) &=&  2 t_{d,r} \frac{1}{|t|} |sin(k_ra)| t_{r,d} f_R (E) 
\mbox{ .}  \label{eq:sigmaR<}
\end{eqnarray}
$k_l$ and $k_r$ at energy $E$ are determined by $E = \epsilon_l - 2t_l cos(k_la)$ and 
$E = \epsilon_r - 2t_r cos(k_ra)$
(eqn. (\ref{eq:E-tb-un1}))
respectively. It can be seen from Eqs. (\ref{eq:sigmaLr}), (\ref{eq:sigmaRr}),
(\ref{eq:sigmaL<}) and (\ref{eq:sigmaR<}):

\begin{eqnarray}
\Sigma_L^{in} (E) &=&  - 2 Im[\Sigma_L (E)] f_L(E) 
                                      \label{eq:sigmaL<-r} \\
\Sigma_R^{in} (E) &=&  - 2 Im[\Sigma_R (E)] f_R(E)  \mbox{ .}  
                                      \label{eq:sigmaR<-r}
\end{eqnarray}

\begin{center}
\fcolorbox{red}{white}{
\begin{minipage}{\textwidth}
The electron density [eqn. (\ref{eq:expectation_q3})] can then be 
written as,
\begin{eqnarray}
n_q = 2 \int \frac{dE}{2\pi} G(E) \Sigma^{in}_{lead}(E) G^\dagger(E)|_{q,q}
\cdot \frac{1}{a}
                      \mbox{, } \label{eq:expectation_q31}
\end{eqnarray}
where, the non zero elements of $\Sigma^{in}_{lead}$ are
\begin{eqnarray}
{\Sigma^{in}_{lead}}_{_{1,1}} (E) &=& \Sigma_L^{in} (E)  
    \;\mbox{ and }\
      \label{eq:Sigma<lead-1}\\
{\Sigma^{in}_{lead}}_{_{n,n}} (E) &=& \Sigma_R^{in} (E)
      \label{eq:Sigma<lead-n}
\mbox{.}
\end{eqnarray}
$\Sigma_L^{in}$ and $\Sigma_R^{in}$ defined above in Eqs. (\ref{eq:sigmaL<-r}) and (\ref{eq:sigmaR<-r}) 
are called the {\bf in-scattering self-energies} due to contacts. These self-energies physically 
represent in-scattering of electrons from the semi-infinite leads to the device and so 
play an important role in determining the charge occupancy in the device. They depend 
on the Fermi-Dirac factor / occupancy in the contacts $f_L$ and $f_R$, and the strength 
of coupling between contacts and device, $Im[\Sigma_L (E)]$ and $Im[\Sigma_R (E)]$.

It is easy to see using eqns. (\ref{eq:expectation_q31}) and (\ref{eq:Sigma<lead-n}) 
that the electron density can also be written as,
\begin{eqnarray}
n_q =  \int {dE} [LDOS_L(q,E) f_L(E) + LDOS_R(q,E) f_R(E) ]
                      \mbox{, } \label{eq:expectation_q311}
\end{eqnarray}
where, $LDOS_L(q,E)$ ($LDOS_R(q,E)$) are the density of states due to waves incident 
from the left (right) contact, at grid point $q$, and
\begin{eqnarray}
LDOS_L (q,E) &=& \frac{ G_{q,1} \Gamma_L G^\dagger_{1,q} }{\pi} \cdot \frac{1}{a} \\
LDOS_R (q,E) &=& \frac{ G_{q,n} \Gamma_R G^\dagger_{n,q} }{\pi} \cdot \frac{1}{a} \mbox{ ,}
\end{eqnarray}
where,
\begin{eqnarray}
\Gamma_L (E) &=& -2 Im[\Sigma_L (E)] \\
\Gamma_R (E) &=& -2 Im[\Sigma_R (E)] \mbox{ .}
\end{eqnarray}
Note that eqn. (\ref{eq:expectation_q311}) is identical to eqn. (\ref{eq:semi-ball-n2}) 
for semiclassical ballistic transport.
\end{minipage}
}
\end{center}

\noindent
The {\bf current density} beween grid points $q$ and $q+1$ per unit energy can be 
written as [eqn. (\ref{eq:expectation_J})],
\begin{eqnarray}
J_{q \rightarrow q+1}(E) &=& \frac{e\hbar}{2mai} 2
      [({\Psi^{(L)}_q}^\dagger \Psi^{(L)}_{q+1} -
        {\Psi^{(L)}_{q+1}}^\dagger \Psi^{(L)}_{q}) f_L(E) +
       ({\Psi^{(R)}_q}^\dagger \Psi^{(R)}_{q+1} -
        {\Psi^{(R)}_{q+1}}^\dagger \Psi^{(R)}_{q}) f_R(E) ]
                      \mbox{ . }        \label{eq:expectation_J1}
\end{eqnarray}
Now, following the derivation for electron density above 
[eqn. \ref{eq:expectation_q3}], it is straight forward to derive that
the current density
\begin{eqnarray}
J_{q \rightarrow q+1} &=& \frac{i e\hbar}{2ma^2} 2 \int \frac{dE}{2\pi} 
                [G_{q,1}(E) \Sigma_L^{in}(E) G^\dagger_{1,q+1}(E)
               + G_{q,n}(E) \Sigma_R^{in}(E) G^\dagger_{n,q+1}(E) \nonumber\\
            & &  \;\;\;\;\;\;\;\;\;\;\;\;\;\;\;\;\;\;\;
               - G_{q+1,1}(E) \Sigma_L^{in}(E) G^\dagger_{1,q}(E)
               - G_{q+1,n}(E) \Sigma_R^{in}(E) G^\dagger_{n,q}(E) ] \mbox.
\end{eqnarray}

\begin{center}
\fcolorbox{red}{white}{
\begin{minipage}{\textwidth}
The current density is given by
\begin{eqnarray}
J_{q \rightarrow q+1} = \frac{i e\hbar}{2ma^2} 2 \int \frac{dE}{2\pi} 
                [\;G(E) \Sigma_{lead}^{in}(E) G^\dagger(E)|_{q,q+1} -
                 G(E) \Sigma_{lead}^{in}(E) G^\dagger(E)|_{q+1,q}\;]
                      \mbox{ .} \label{eq:expectation_J2}
\end{eqnarray}
\end{minipage}
}
\end{center}

\vspace{0.1in}

\noindent
{\bf Electron correlation function:}
More generally, we define the electron correlation function $G^{n}$, which
is the solution to
\begin{eqnarray}
A G^{n} =  \Sigma^{in}_{lead} G^\dagger  \mbox{ .}  \label{eq:G<}
\end{eqnarray}
Noting that $G=A^{-1}$, it is easy to obtain eqn. (\ref{eq:expectation_q31}) $G^{n} =  G \Sigma^{in}_{lead} G^\dagger$.

\begin{center}
\fcolorbox{red}{white}{
\begin{minipage}{\textwidth}
The expressions for electron [eqn. (\ref{eq:expectation_q31})] and
current [eqn. (\ref{eq:expectation_J2})] densities at energy E,
in the phase coherent case at 
finite applied biases can now be written 
as,
\begin{eqnarray}
n_q (E) &=& 2 \frac{G^{n}_{q,q}(E)}{2\pi a} \mbox{ ,} \label{eq:n-general} \\
J_{q \rightarrow q+1} (E) &=& \frac{i e\hbar}{2ma^2} 2 \frac{1}{2\pi} [G^{n}_{q,q+1} (E) - G^{n}_{q+1,q} (E) ]
\mbox{.} \label{eq:JSch-GF} 
\end{eqnarray}
That is, the diagonal and first off-diagonal elements of $G^{n}$ are
related to the electron and current densities respectively. Note that
these equations are entirely equivalent to eqns. (\ref{eq:expectation_q})
and (\ref{eq:expectation_J}) appearing in the Landauer-Buttiker approach.
\end{minipage}
}
\end{center}

\vspace{0.1in}

\noindent
{\bf Hole correlation function:}
In the absence of phase breaking scattering, the  Green's function ($G$)
and the electron correlation function ($G^{n}$) are sufficient for device modeling.
Scattering introduces the need for the hole correlation function ($G^{p}$), 
whose role will become clearer in section \ref{sec:self-energy}. While the less-than
Green's function is directly proportional to the density of occupied states, the 
hole correlation function is proportional to the density of unoccupied states.

The density of unoccupied states at grid point $q$ is also obtained by applying the 
Landauer-Buttiker formalism.
For this, we simply replace the probability of finding an occupied state in the 
contact $f_{L,R}$ by the probability of finding an unoccupied state in the contact, 
$1-f_{L,R}$, in eqn. (\ref{eq:expectation_q}). Then following the derivation leading 
to eqn. (\ref{eq:expectation_q}), we obtain:
\begin{eqnarray}
h_q &=&  2 \int \frac{dE}{2\pi} \left[ G_{q,1}(E) \Sigma_L^{out}(E) G^\dagger_{1,q}(E)  + 
                                        G_{q,n}(E) \Sigma_R^{out}(E) G^\dagger_{n,q}(E) \right]
                                        \cdot \frac{1}{a}
                                  \mbox{, } \label{eq:expectation_q4}  \\
h_q &=&  2 \int \frac{dE}{2\pi} G(E) \Sigma_{lead}^{out} G^\dagger(E) |_{q,q}
\cdot \frac{1}{a} \mbox{, }
\end{eqnarray}
where, the only non zero elements of $\Sigma_{lead}^{out}$ are,
\begin{eqnarray}
{\Sigma_{lead}^{out}}_{_{1,1}}(E)&=& \Sigma_L^{out}(E) 
                  = 2 t_{d,l} \frac{1}{t} sin(k_la) t_{l,d} (1-f_L)  
                  = - 2 Im[\Sigma_L(E)] [1-f_L(E)]
                                                  \label{eq:sigmaL>} \\
{\Sigma_{lead}^{out}}_{_{n,n}}(E)&=& \Sigma_R^{out}(E)
                  = 2 t_{d,r} \frac{1}{t} sin(k_ra) t_{r,d} (1-f_R)
                  = - 2 Im[\Sigma_R(E)] [1-f_R(E)]
                                        \mbox{.}  \label{eq:sigmaR>}
\end{eqnarray}

\begin{center}
\fcolorbox{red}{white}{
\begin{minipage}{\textwidth}
Akin to Eqs. (\ref{eq:G<}) and (\ref{eq:n-general}), the density of unoccupied
states at energy E, at grid point $q$ can be expressed as the diagonal elements of $G^{p}$, 
\begin{eqnarray}
h_q (E) = 2 \frac{G^{p}_{q,q}}{2\pi a} (E) \mbox{ ,} \label{h-general}
\end{eqnarray}
where, $G^{p}$ is in general given by,
\begin{eqnarray}
A G^{p} &=&  \Sigma^{out}_{lead} G^\dagger  \mbox{ .}  \label{eq:G>} 
\end{eqnarray}
\end{minipage}
}
\end{center}

\subsection{\label{sec:self-energy} Electron-phonon scattering }

In section \ref{sect:GF}, we defined the retarded, in-scattering and out-scattering 
self-energies in the device arising from coupling of the device to the external 
contacts. The self-energy $\Sigma^{in}_{L}$ represents in-scattering of 
electrons (in-scattering rate) from the semi-infinite left contact to the device, 
assuming that grid point 1 of the device is empty. A similar statement applies to 
$\Sigma^{in}_{R}$. The in-scattering self-energy of the contacts depend on their 
Fermi distribution functions and surface density of states. 

\begin{figure}
  \begin{center}
    \leavevmode
\includegraphics[height=5cm,angle=-0]{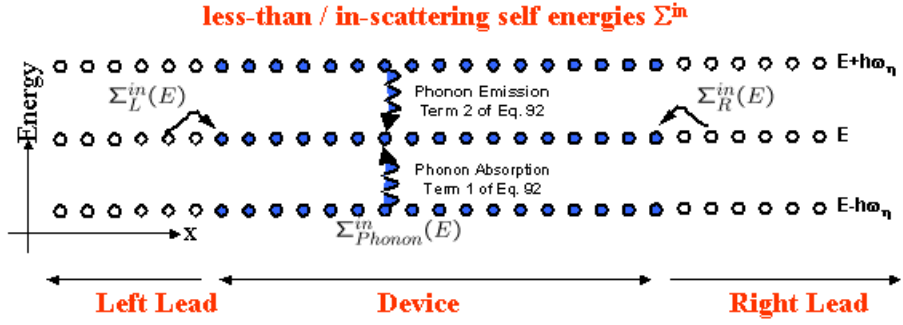}
  \end{center}
\caption{Pictorial representation of the meaning the two in-scattering self-energies that 
appear in this chapter. $\Sigma_L^{in}(E)$ and $\Sigma_R^{in}(E)$ are the self-energies 
of the leads, and are non-zero only at the first and last Device grid points. 
${\Sigma_{Phonon}}^{in}(E)$ is the self-energy due to electron-phonon interaction, 
and is non-zero at all Device grid points.}
\label{fig:in-scatt}
\end{figure}

A second source for in-scattering to grid point $q$ and energy $E$ [$(q,E)$] is electron-phonon 
interaction.  The self-energy at $(q,E)$ has two terms corresponding to in-scattering 
from $(q,E+\hbar \omega_{phonon})$ and $(q,E-\hbar \omega_{phonon})$, 
as shown in Fig. \ref{fig:in-scatt}. Intuitively, the in-scattering self-energy 
(in-scattering rate) at $(x,E)$ should depend on the Bose factor for phonon occupancy, 
the deformation potential for electron-phonon scattering and the availability of 
electrons at energies $E+\hbar \omega_{phonon}$ and $E-\hbar \omega_{phonon}$. 
It follows rigorously, within the Born approximation  that the in-scattering 
self energy into a fully \underline{empty state} at energy $E$ and grid point $q$ is \cite{Mah87}
\begin{eqnarray}
{\Sigma^{in}_{inel}}_{_q} (E) = \sum_{\eta} D_q^{\eta}
\left[n_B(\hbar \omega_{phonon}) G^{n}_q (E-\hbar \omega_{phonon})
  +
  (n_B(\hbar \omega_{phonon})+1) G^{n}_q (E+\hbar \omega_{phonon})
 \right] \mbox{ .} \label{eq:inel_self_en_<0}
\end{eqnarray}
$D_q^{\eta}$ represents the electron-phonon scattering strength
at grid point $q$. The first term of eqn. (\ref{eq:inel_self_en_<0})
represents in-scattering to $E$ from $E-\hbar\omega_{phonon}$ (phonon
absorbtion). $n_B$ is the Bose distribution function for phonons of 
energy $\hbar\omega_{phonon}$ and $G^{n}_q (E-\hbar \omega_{phonon})$ is the 
electron density at $E-\hbar \omega_{phonon}$. 
The first and second terms of 
eqn. (\ref{eq:inel_self_en_<0}) represents in-scattering of electrons from 
$E-\hbar\omega_{phonon}$ (phonon absorbtion) and $E+\hbar\omega_{phonon}$ 
(phonon emission) to $E$ respectively. The in-scattering rate at grid point $q$ is given by
\begin{eqnarray}
\mbox{In-scattering rate at grid point $q$: }\;\; \frac{\hbar}{\tau^{in}_q (E)} = \Sigma^{in}_q(E) 
\mbox{ ,} \label{eq:in-scatt}
\end{eqnarray}
where $\Sigma^{in}_q$ is the sum of all in-scattering self energies at grid point $q$.

\begin{figure}
  \begin{center}
    \leavevmode
\includegraphics[height=5cm,angle=-0]{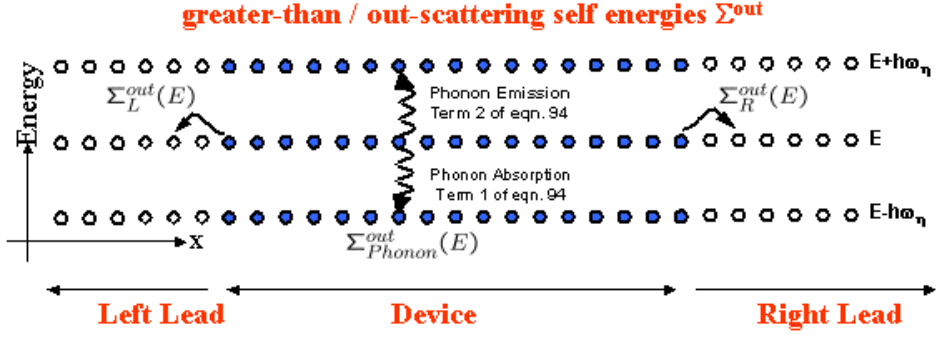}
  \end{center}
\caption{Pictorial representation of the meaning the two out-scattering self-energies
that appear in this chapter. ${\Sigma_{lead}}_{_q}^{out} (E)$ is self-energy
due to leads, which is non-zero only at the first and last Device grid
points ${\Sigma_{Phonon}}_{_q}^{out} (E)$ is self-energy due to 
electron-phonon interaction, which is non-zero at all Device grid points.}
\label{fig:out-scatt}
\end{figure}

The out-scattering self-energy $\Sigma^{out}_{L}$ in eqn. (\ref{eq:sigmaL>}) represents 
out-scattering of electrons from grid point $1$ in the device to the semi-infinite left 
contact, assuming that grid point 1 of the device 
was occupied. The out-scattering self-energy due to the left contact $\Sigma^{out}_{L}$ 
depends on the probability of finding an unoccupied state in the left contact $1-f_L$ 
and the surface density of states of the left contact. A similar statement applies to 
$\Sigma^{out}_{R}$. A second source for out-scattering of electrons from 
an occupied state at $(q,E)$ is electron-phonon interaction, which leads 
to scattering to $(q,E+\hbar \omega_{phonon})$ and $(q,E-\hbar \omega_{phonon})$ 
as represented in Fig. \ref{fig:out-scatt}. Intuitively, the out-scattering self-energy 
(out-scattering rate) at $(q,E)$ should depend on the Bose factor for phonon occupancy, 
the deformation potential for electron-phonon scattering and the availability of 
unoccupied states at enegies $E+\hbar \omega_{phonon}$ and $E-\hbar \omega_{phonon}$. 
It follows rigorously, within the Born approximation that the out-scattering 
self energy from a fully \underline{filled state} at $(q,E)$ is \cite{Mah87}
\begin{eqnarray}
{\Sigma^{out}_{Phonon}}_{_q} (E) = \sum_{\eta} D_q^{\eta}
\left[(n_B(\hbar \omega_{phonon})+1) G^{p}_q (E-\hbar \omega_{phonon})
  +
       n_B(\hbar \omega_{phonon}) G^{p}_q (E+\hbar \omega_{phonon})
 \right] \mbox{ .} \label{eq:inel_self_en_>0}
\end{eqnarray}
In the above equation, $G^{p}_q (E-\hbar \omega_{phonon})$ and $G^{p}_q (E+\hbar \omega_{phonon})$ 
are the \underline{densities of unoccupied states} at $E-\hbar \omega_{phonon}$ and 
$E+\hbar \omega_{phonon}$. So the first and second terms of eqn. (\ref{eq:inel_self_en_>0}) 
represents out-scattering of electrons
from $E$ to $E+\hbar\omega_{phonon}$ (phonon emission) and $E-\hbar\omega_{phonon}$ (phonon absorbtion) respectively.
The out-scattering rate at grid point $q$ is given by
\begin{eqnarray}
\mbox{Out-scattering rate at grid point $q$: }\;\; \frac{\hbar}{\tau^{out}_q (E)} = \Sigma^{out}_q(E) 
\mbox{ ,} \label{eq:out-scatt}
\end{eqnarray}
where $\Sigma^{out}_q$ is the sum of all out-scattering self energies at grid point $q$. 

\begin{figure}
  \begin{center}
    \leavevmode
\includegraphics[height=5cm,angle=-0]{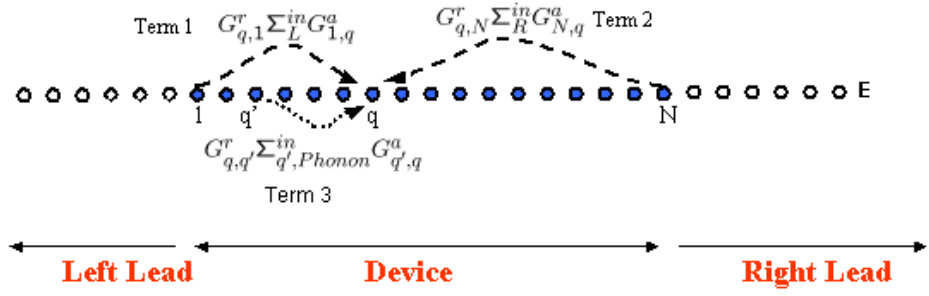}
  \end{center}
\caption{Contributions to electron density from leads and electron-phonon scattering.}
\label{fig:scatt-dens}
\end{figure}

We now discuss how the in-scattering self-energy due to electron-phonon scattering affects 
the expression for electron density. The electron density at grid point $q$ 
in the phase-coherent case (eqn. (\ref{eq:expectation_q31}))
is the sum of two terms,
\begin{eqnarray}
n_q =  
2 \int \frac{dE}{2\pi} \left[G_{q,1}(E) \Sigma^{in}_{L}(E) G^\dagger_{1,q}(E) + G_{q,1}(E) 
\Sigma^{in}_{R}(E) G^\dagger_{1,q}(E) \right]   
\cdot \frac{1}{a}
\mbox{ . } 
\end{eqnarray}
The first term represents in-scattering of electrons from the left contact [($\Sigma^{in}_L(E)$), 
which is propagated to grid point $q$ via the term $G_{q,1}(E) G^\dagger_{1,q}(E)$. 
The interpretation of the second term is similar except that it involves the right contact. 
In the presence of electron-phonon interaction, 
the in-scattering functions $\Sigma^{in}_{Phonon}$ is non zero at all grid points. 
As a result, an electron can scatter from $(q^\prime,E^\prime)$ to $(q^\prime,E)$ and 
then propagate to grid point $(q,E)$ via the term $G_{q,q^\prime} (E) G_{q^\prime,q}^\dagger (E)$. 
The expression for the electron density can be generalized to include such terms,
\begin{eqnarray}
n_q &=&  
2 \int \frac{dE}{2\pi} \left[G_{q,1}(E) \Sigma^{in}_{L}(E) G^\dagger_{1,q}(E)  
+ G_{q,1}(E) \Sigma^{in}_{R}(E) G^\dagger_{1,q}(E) + \sum_{q^\prime} G_{q,q^\prime}(E) 
\Sigma^{in}_{q^\prime,Phonon}(E) G^\dagger_{q^\prime,q}(E) \right]
\cdot \frac{1}{a}
\nonumber   \\
      &=& 2 \int \frac{dE}{2\pi} \left[G(E) \Sigma^{in}_{lead}(E) G^\dagger(E) +  G(E) 
\Sigma^{in}_{Phonon}(E) G^\dagger(E) \right]|_{q,q}
\cdot \frac{1}{a}
\label{eq:expectation_q31-2}  \\
      &=& 2 \int \frac{dE}{2\pi} G^n_{q,q}(E)  
      \cdot \frac{1}{a} \nonumber
\end{eqnarray}
where the third term corresponds to propagation of electrons from grid point 
$q^\prime$ to $q$ after a scattering event at $q^\prime$ as shown in Fig. (\ref{fig:scatt-dens}). 
The in-scattering self energies due to phonon scattering are given by eqn. 
(\ref{eq:inel_self_en_<0}).  More generally, $G^{n}$ is given by,
\begin{eqnarray}
G^{n}(E) &=& G(E) \Sigma^{in} (E) G^\dagger(E)  \\
\left[ E-H-\Sigma(E) \right] G^{n}(E) &=& \Sigma^{in}(E) G^\dagger(E) \mbox{, }
\end{eqnarray}
where $\Sigma^{in}$ is the sum of self-energies due to leads, electron-phonon 
interaction and all other processes. The reader can compare the above two equations to 
Eqs. (\ref{eq:Sigma<lead-1}) and (\ref{eq:G<}), which are valid in the phase coherent limit.

The density of unoccupied states can be written in a manner identical to eqn. (\ref{eq:expectation_q31-2}) as,
\begin{eqnarray}
h_q &=&  
2 \int \frac{dE}{2\pi} \left[G_{q,1}(E) \Sigma^{out}_{L}(E) G^\dagger_{1,q}(E)  + G_{q,1}(E) 
\Sigma^{out}_{R}(E) G^\dagger_{1,q}(E) + \sum_{q^\prime} G_{q,q^\prime}(E) 
\Sigma^{out}_{q^\prime,Phonon}(E) G^\dagger_{q^\prime,q}(E) \right]
\cdot \frac{1}{a}
\nonumber              \\
      &=& 2 \int \frac{dE}{2\pi} \left[G(E) \Sigma^{out}_{lead}(E) G^\dagger(E) +  G(E) 
\Sigma^{out}_{Phonon}(E) G^\dagger(E) \right]|_{q,q}
\cdot \frac{1}{a}
\label{eq:expectation_h31-2}  \\
      &=& 2 \int \frac{dE}{2\pi} G^p_{q,q}(E)
      \cdot \frac{1}{a}
      \nonumber   \mbox{ .}
\end{eqnarray} 
More generally, the $G^{p}$ matrix is given by,
\begin{eqnarray}
G^{p}(E) &=& G(E) \Sigma^{out} (E) G^\dagger(E)  \\
\left[E-H-\Sigma(E)\right] G^{p}(E) &=& \Sigma^{out}(E) G^\dagger(E) \mbox{, }
\end{eqnarray}
where $\Sigma^{out}$ is the sum of self-energies due to leads, electron-phonon 
interaction and all other processes. 

Note that in general $G^{n}$ and $G^{p}$ are full matrices, 
the diagonal elements of which correspond to density of occupied and unoccupied states 
respectively, and the first off diagonal elements of $G^{n}$ and $G^{p}$ correspond to the current density.

The Green's function $G$ in the device region is obtained by solving,
\begin{eqnarray}
\left[E - H - \Sigma_{lead} (E) - \Sigma_{Phonon} (E) \right]G = I    \mbox{ ,}
\end{eqnarray}
which is similar to eqn. (\ref{eq:Gr0}) for the phase coherent case, except for the 
additional self-energy due to phonon scattering.

\section{\label{section:ngf-summary} 
Non-equilibrium Green's Function Equations for Layered Structures}

%%%%%%%%%%%%%%%%%%%%%%%%%%%%%%%%%%%%%%%%%%%%%%%%%%
\begin{figure}
  \begin{center}
    \leavevmode
\includegraphics[height=7cm,angle=-0]{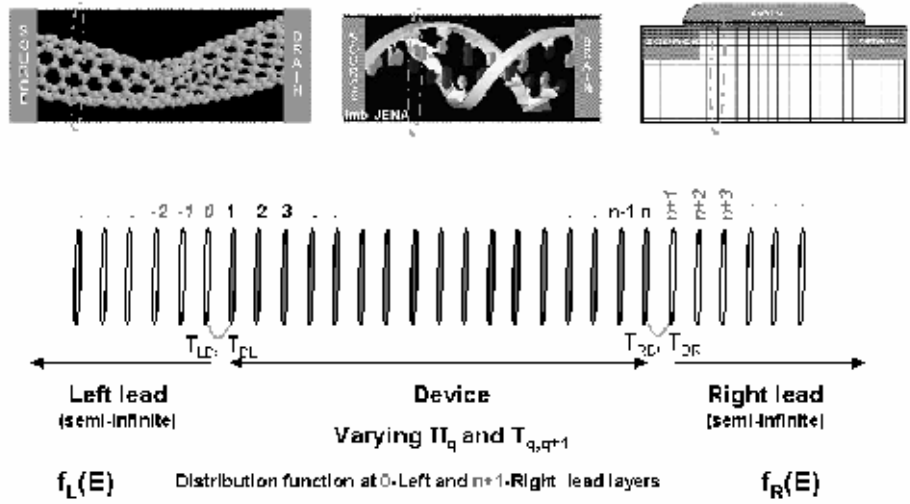}
  \end{center}
\caption{(Top) Examples of the layered structures: carbon nanotube, DNA molecule, MOSFET.
(Bottom) Scheme of simulation domains in the layered structure: device, left and right leads.}
\label{fig:layered}
\end{figure}
%%%%%%%%%%%%%%%%%%%%%%%%%%%%%%%%%%%%%%%%%%%%%%%%%%

The previous section dealt with a simple one dimensional Hamiltonian. In this section, we 
will present the NEGF equations for a family of more realistic structures called 
{\it layered structures}. A layer can be considered to be a generalization of a 
single grid points / orbital (Fig. \ref{fig:1D-chain}) to a set of grid point / orbitals 
(Fig. \ref{fig:layered}). Three examples of layered structures are shown in Fig. \ref{fig:layered}. 
The left most figure is a carbon nanotube, the middle figure is a DNA strand and the right 
figure is a MOSFET. A {\it layer} consists of the atoms / grid points between the dashed 
lines in Fig. \ref{fig:layered}.

A common approximation used to describe the Hamiltonian of layered structures consists 
of interaction only between nearest neighbor layers. That is, each layer $q$ interacts only
with itself and its nearest neighbor layers $q-1$ and $q+1$. Then, the
single particle Hamiltonian of the layered structure is a 
\underline{block tridiagonal matrix}, where diagonal blocks $H_q$
represent the Hamiltonian of layer $q$ and off-diagonal blocks 
$T_{q,q+1}$ represent interaction between layers $q$ and $q+1$:
\begin{eqnarray}
H=\left(
\begin{array}{cccccccccc}
\bullet & \bullet & \bullet & & & & & & & \\
\bullet & H_1 & T_{12} & & & & & & & \\
 & T_{12}^\dagger & H_2 &T_{2,3} & & & & & & \\
 & & \bullet & \bullet & \bullet & & & & & \\
 & & & \bullet & \bullet & \bullet & & & & \\
 & & & & \bullet & \bullet & \bullet & & & \\
 & & & & & T_{n-2,n-1}^\dagger & H_n & T_{n-1,n} & & \\
 & & & & & & T_{n-1,n}^\dagger & H_n & \bullet & \\
 & & & & & & & \bullet & \bullet & \bullet \\
\end{array}
\right) \mbox{ .}
\end{eqnarray}

The Green's function equation in the presence of electron-phonon scattering is
\begin{eqnarray}
[E I - H - \Sigma_{Phonon}] G = I \mbox{ ,} \label{eq:GrP} 
\end{eqnarray}
where, $\Sigma_{Phonon}$ is the self-energy due to electron-phonon scattering. 
We partition the layered structure into Left lead, Device and Right lead as shown in 
Fig. \ref{fig:layered}. The Device corresponds to the region where we solve for the 
nonequilibrium electron density, and the leads are the highly conducting regions 
connected to the nanodevice. While the Device region, where we seek the non equilibrium 
density consists of only $n$ layers, the matrix equation corresponding to eqn. (\ref{eq:GrP}) 
is infinite dimensional due to the semi-infinite leads. We will now show how the influence 
of the semi-infinite leads can be folded into the Device region. In a manner akin to the 
previous section, the influence of the semi-infinite leads is to affect layers 1 and n of 
the Device region. An important difference is that the derivation here includes electron-phonon 
scattering and does not assume a flat potential in the semi-infinite leads.
We first define
\begin{eqnarray}
A^\prime = [E I - H - \Sigma_{Phonon}]  \mbox{ .}
      \label{eq:A-define2}
\end{eqnarray}
Then, eqn. (\ref{eq:GrP}) can be written as,
\begin{eqnarray}
A^\prime G = I \;\;\;\;\;\;\;\;\;\;\;\; \rightarrow
               \;\;\;\;\;\;\;\;\;\;\;\;
\left(
\begin{array}{ccc}
A^\prime_{LL} & A^\prime_{LD}  & O        \\
A^\prime_{DL} & A^\prime_{DD}  & A^\prime_{DR}   \\
O      & A^\prime_{RD}  & A^\prime_{RR}  
\end{array}
\right)
\left(
\begin{array}{ccc}
G_{LL}    & G_{LD}    & G_{LR}        \\
G_{DL}    & G_{DD}    & G_{DR}        \\
G_{RL}    & G_{RD}    & G_{RR}        
\end{array}
\right) 
=
\left(
\begin{array}{ccc}
I      & O    & O    \\
O      & I    & O    \\
O      & O    & I 
\end{array}
\right) \mbox{ ,}  \label{eq:AG1}
\end{eqnarray}
where,
\begin{eqnarray}
A^\prime_{LL} &=& 
\left(
\begin{array}{cccccc}
\bullet&\bullet&\bullet&             &           &             \\
       &\bullet&\bullet&   \bullet  &            &             \\
       &       & -T_{l3}^\dagger  &A^\prime_{l3} &   -T_{l2}     &             \\
       &       &       &   T_{l1}^\dagger     &A^\prime_{l2}&   -T_{l1}        \\
       &       &       &            &     -T_{l1}^\dagger   & A^\prime_{l1}
\end{array}
\right)   \mbox{ corresponds to the left semi-infinite lead, } \label{eq:lead-Amatrices1L} \\
A^\prime_{RR} &=&
\left(
\begin{array}{cccccc}
A^\prime_{r1} &   -T_{r1}      &            &        &       &         \\
   -T_{r1}^\dagger       &A^\prime_{r2} &    -T_{r2}    &        &       &         \\
             &  -T_{r2}^\dagger       &A^\prime_{r3} &   -T_{r3} &       &         \\
             &             &  \bullet   &\bullet &\bullet&         \\
             &             &            &\bullet &\bullet&\bullet
             \end{array}
\right)\mbox{ corresponds to the right semi-infinite lead, and} \label{eq:lead-Amatrices1R}
\end{eqnarray}
\begin{eqnarray}
A^\prime_{DD}=\left(
\begin{array}{ccccccc}
A^\prime_1              &    -T_{12}    &         &         &      &      &    \\
-T_{12}^\dagger  &     A^\prime_2       &-T_{2,3} &         &      &      &    \\
             &    \bullet   & \bullet & \bullet &            & &    \\
             &              & \bullet & \bullet & \bullet    & &    \\
             &              &         & \bullet & \bullet    & \bullet &    \\
             &              &         & &-T_{n-2,n-1}^\dagger& A^\prime_{n-1} &-T_{n-1,n}  \\
             &              &         &         &            & -T_{n-1,n}^\dagger &A^\prime_n \\
\end{array}
\right) \mbox{ corresponds to the Device region.}
\end{eqnarray}
\begin{eqnarray}
A^\prime_{LD} =
\left(
\begin{array}{cccccc}
O      &  O    &     \bullet  &  \bullet    & O        & O \\
O      &  O    &     \bullet  &  \bullet    & O        & O\\
O      &  O    &     \bullet  &  \bullet    & O        & O\\
O      &  O    &     \bullet  &  \bullet    & O        & O\\
-T_{LD}&  O    &     O        &  \bullet    &  \bullet & O 
\end{array}
\right)   \;\;\;\;\;\mbox{and}\;\;\;\;\;
A^\prime_{RD}=
\left(
\begin{array}{cccccc}
-T_{RD}&  O    &     \bullet  &  \bullet    & O        & O\\
O      &  O    &     \bullet  &  \bullet    & O        & O\\
O      &  O    &     \bullet  &  \bullet    & O        & O\\
O      &  O    &     \bullet  &  \bullet    & O        & O\\
O      &  O    &     O        &  \bullet    &  \bullet & O   
\end{array}
\right) \label{eq:lead-Amatrices2}
\end{eqnarray}
\normalsize
are the coupling between the Left and Right leads and Device respectively.
Note that $A^\prime_{DL} = {A^\prime_{LD}}^\dagger  \mbox{ , } A^\prime_{DR} = {A^\prime_{RD}}^\dagger$, and
$A^\prime_{LD}$ and $A^\prime_{DL}$ ($A^\prime_{RD}$, and $A^\prime_{DR}$) are sparse matrices. 
Their only non-zero entry represents coupling between the Left (Right) lead and Device. 
$O$ represents zero matrices. From eqn. (\ref{eq:AG1}), we have:
\begin{eqnarray}
&&G_{LD} = - A^{\prime -1}_{LL}  A^\prime_{LD} G_{DD}  
      \label{eq:GLD}\\
&&G_{RD} = - A^{\prime -1}_{RR}  A^\prime_{RD} G_{DD}  
      \label{eq:GRD} \\
&&A^\prime_{DL} G_{LD} + A^\prime_{DD} G_{DD} + 
     A^\prime_{DR} G_{RD} = I \mbox{ .}
      \label{eq:GDD0}
\end{eqnarray}
Substituting Eqs. (\ref{eq:GLD}) and (\ref{eq:GRD}) in eqn. (\ref{eq:GDD0}),
we obtain a matrix equation with dimension corresponding to total number of 
grid points / orbitals in the $n$ device layers ,
\begin{eqnarray}
[A^\prime_{DD} - A^\prime_{DL} A^{\prime -1}_{LL} A^\prime_{LD} - 
A^\prime_{DR} A^{\prime -1}_{RR} A^\prime_{RD}] G_{DD} = I \mbox{ .}
      \label{eq:GDD}
\end{eqnarray}
The second and third terms of eqn. (\ref{eq:GDD}) are self energies
due to coupling of the Device region to Left and Right leads respectively.

The Green's functions of the
isolated semi-infinite leads by definition are,
\begin{eqnarray}
A^\prime_{LL} g_{L} = I \mbox{ and } A^\prime_{RR} g_{R} = I
\mbox{ .} \label{eq:grL-def}
\end{eqnarray}
The {\bf surface Green's function} of the Left and Right leads are the
Green's function elements corresponding to the edge layers $-1$ and 
$n+1$ respectively,
\begin{eqnarray}
{g_{L}}_{_{-1,-1}}   = {A^{\prime -1}_{LL}}_{_{1,1}} \mbox{ and }
{g_{R}}_{_{n+1,n+1}} = {A^{\prime -1}_{RR}}_{_{1,1}} \mbox{ . }
\label{eq:sg}
\end{eqnarray}

Eq. (\ref{eq:GDD}) can now be rewritten in a form very similar to eqn. (\ref{eq:Gr0}),
\begin{eqnarray}
{[E I - H - \Sigma_{Phonon} - \Sigma_{lead}]} G_{DD} = I  \label{eq:GDD2p}
\end{eqnarray}
where, 
\begin{eqnarray}
{\Sigma_{lead}}_{1,1} &=& T_{DL} {g_L}_{_{-1,-1}} T_{LD} = \Sigma_L \label{eq:self1} \\
{\Sigma_{lead}}_{n,n} &=& T_{DR} {g_R}_{_{n+1,n+1}} T_{RD} = \Sigma_R \mbox{ .} \label{eq:self2}
\end{eqnarray}
All other elements of $\Sigma_{lead}$ are zero. $\Sigma_L$ and $\Sigma_R$ are {\bf self-energies} due 
to the Left and Right leads respectively, and $T_{DL}=T_{LD}^\dagger$ and
$T_{DR}=T_{RD}^\dagger$. Finally, defining,
\begin{eqnarray}
A_{DD} = E I - H - \Sigma_{Phonon} - \Sigma_{lead} \mbox{ ,} \label{eq:ADD1}
\end{eqnarray}
eqn. (\ref{eq:GDD2p}) can be written as
\begin{eqnarray}
A_{DD} G_{DD} = I \label{eq:GDD2} \mbox{ .}
\end{eqnarray}

The main information needed to solve eqn. (\ref{eq:GDD2p}) are the surface Green's functions of $g_L$ and $g_R$.
We will disucss two methods to obtain this surface Green's functions for a constant 
potential  in the Left and Right leads.  When the potential does not vary, $A^\prime_{LL}$ 
and $A^\prime_{RR}$ are semi-infinite periodic 
matrices with all diagonal / off-diagonal blocks being equal:
\begin{eqnarray}
&& A_{l1}=A_{l2}=A_{l3}=...=A_{l} \\
&& T_{l1}=T_{l2}=T_{l3}=...=T_{l} \mbox{ .}
\end{eqnarray}
${g_L}_{_{-1,-1}}$ is obtained by solving the matrix quadratic
equation:
\begin{eqnarray}
[A_l - T_{l}^\dagger {g_L}_{_{-1,-1}} T_{l}] {g_L}_{_{-1,-1}} = I 
     \mbox{ .} \label{eq:gs}
\end{eqnarray}
This equation can be solved iteratively by,
\begin{eqnarray}
[A_l - T_{l}^\dagger {g_L}_{_{-1,-1}}^{<m-1>} T_{l}] {g_L}_{_{-1,-1}}^{<m>}
                         = I \mbox{ ,} \label{eq:gsi}
\end{eqnarray}
where, the superscript of $g_L$ represents the iteration number.
Note that the solution to eqn. (\ref{eq:gs}) is analytic when the 
dimension of $A_l$ is one.
A second simpler solution to obtain $g_L$ involves  
transforming to an eigen mode basis using an unitary transformation 
(S), such that 
\begin{eqnarray}
S^{-1} A_l S = A_{ld} \mbox{ and } S^{-1} T_l S = T_{ld} \mbox{ ,} 
\end{eqnarray}
where, both $A_{ld}$ and $T_{ld}$ are diagonal matrices. The
surface Green's function in this new basis is simply a diagonal matrix,
whose elements are obtained by solving the scalar quadratic version of
eqn. (\ref{eq:gs}). The Green's function in the original basis (in which
$A_l$ is not diagonal) can be obtained using the inverse unitary 
transformation.

\vspace{0.1in}

\noindent
{\bf Electron ($G^{n}$) and Hole ($G^{p}$) Green's Function:}

The electron density is equal to [see the discussion of
electron density in section \ref{sect:GF}],
\begin{eqnarray}
n(\vec{r},E) = 2 \frac{G^{n}(\vec{r},\vec{r},E)}{2\pi} \mbox{ .}\label{eq:elec-dens1}
\end{eqnarray}
The governing equation for $G^{n}$ is
\begin{eqnarray}
[E I - H - \Sigma_{Phonon}] G^{n} = \Sigma^{in}_{Phonon} G^\dagger \mbox{ ,} \label{eq:G<-def1}
\end{eqnarray}
where $G^\dagger$ is the hermitian conjugate Green's function and $\Sigma^{in}_{Phonon}$ 
is the in-scattering self-energy due to phonon scattering. The dimension of 
Eq. (\ref{eq:G<-def1}) is essentially infinite due to the semi-infinite Left and 
Right leads. It can however be converted to a finite dimensional matrix with 
dimension equal to the number of grid points / orbitals corresponding to the 
$n$ device layers. In a manner identical to the derivation of eqn. (\ref{eq:GDD2p}) 
for the retarded Green's function, it can be shown that the role of the leads can be 
folded into layers 1 and n to yield,
\begin{eqnarray}
{[E I - H - \Sigma_{Phonon} - \Sigma_{lead}]} G^{n}_{DD} = \Sigma^{in}_{DD} G^\dagger_{DD} \mbox{ .} \label{eq:G<layerp}
\end{eqnarray}
or
\begin{eqnarray}
A_{DD} G^{n}_{DD} = \Sigma^{in}_{DD} G^\dagger_{DD} \mbox{ ,} \label{eq:G<layer}
\end{eqnarray}
where $A_{DD}$ has been defined in eqn. (\ref{eq:ADD1}). The self-energy $\Sigma^{in}_{DD}$ has 
contributions due to both electron-phonon interaction and leads, 
\begin{eqnarray}
 {\Sigma^{in}_{DD}}_{_{1,1}} &=& {\Sigma^{in}_{Phonon}}_{_{1,1}} + \Sigma^{in}_L \\
 {\Sigma^{in}_{DD}}_{_{n,n}} &=& {\Sigma^{in}_{Phonon}}_{_{n,n}} + \Sigma^{in}_R \\
 {\Sigma^{in}_{DD}}_{_{q,q}} &=& {\Sigma^{in}_{Phonon}}_{_{q,q}}
\mbox{ , where, q=2, 3, 4, ... n-1.} 
\end{eqnarray}
The {\bf in-scattering self-energies} due to the leads, $\Sigma^{in}_L$ and 
$\Sigma^{in}_R$ have forms very similar to eqns. (\ref{eq:sigmaL<-r}) and (\ref{eq:sigmaR<-r})
of section \ref{sect:GF},
\begin{eqnarray}
\Sigma^{in}_L(E) &=& -2 \;\mbox{Im}[\Sigma_L(E)] f_L(E) 
                =  \Gamma_L(E) f_L(E) \label{eq:SigmaL}
\\
\Sigma^{in}_R(E) &=& -2 \;\mbox{Im}[\Sigma_R(E)] f_R(E)
              =  \Gamma_R(E) f_R(E) \mbox{ ,} \label{eq:SigmaR}
\end{eqnarray}
where,
\begin{eqnarray}
\Gamma_L(E) &=& - 2 \;\mbox{Im}[\Sigma_L(E)] \label{eq:GammaL} \\
\Gamma_R(E) &=& - 2 \;\mbox{Im}[\Sigma_R(E)] \mbox{ .} \label{eq:GammaR}
\end{eqnarray}
$f_L$ and $f_R$ are the distribution functions in the Left and
Right leads respectively (Fermi factors at equilibrium). The
self-energies $\Sigma_L(E)$ and $\Sigma_R(E)$ have been defined
in eqn. (\ref{eq:self1}) and (\ref{eq:self2}). 

The diagonal elements of the hole correlation functions $G^{p} (E)$
represents the density of unoccupied states,
\begin{eqnarray}
h(\vec{r},E) = 2 \frac{G^{p}(\vec{r},\vec{r},E)}{2\pi} \mbox{ .}
\end{eqnarray}
The governing equation for $G^{p} (E)$ is,
\begin{eqnarray}
 [E I - H - \Sigma_{Phonon}] G^{p} &=& \Sigma^{out}_{Phonon} G^\dagger \mbox{} \label{eq:G>-def1}
\end{eqnarray}
 or
 \begin{eqnarray}
 A_{DD} G^{p}_{DD} &=&  \Sigma^{out}_{DD} G^\dagger_{DD} \label{eq:G>layer}\\
 {\Sigma^{out}_{DD}}_{_{11}} &=& {\Sigma^{out}_{Phonon}}_{_{11}} + \Sigma^{out}_L \\
 {\Sigma^{out}_{DD}}_{_{nn}}&=& {\Sigma^{out}_{Phonon}}_{_{nn}} + \Sigma^{out}_R \\
 {\Sigma^{out}_{DD}}_{_{ii}} &=& {\Sigma^{out}_{Phonon}}_{_{ii}}
\mbox{ , where, i=2, 3, 4, ... n-1}
\end{eqnarray}
where $\Sigma^{out}_{Phonon}$ is the out-scattering self-energy due to phonon 
scattering.
The {\bf out-scattering self-energies} due to the leads, $\Sigma^{out}_L$ and 
$\Sigma^{out}_R$ have forms very similar to eqns. (\ref{eq:sigmaL>}) and (\ref{eq:sigmaR>})
\begin{eqnarray}
\Sigma^{out}_L(E) &=& - 2 \;\mbox{Im}[\Sigma_L(E)] (1-f_L(E)) 
                 = \Gamma_L (1-f_L(E))     \\
\Sigma^{out}_R(E) &=& - 2 \;\mbox{Im}[\Sigma_R(E)] (1-f_R(E)) 
              = \Gamma_R (1-f_R(E))    \mbox{ ,}
\end{eqnarray}
where $1-f_L(E)$  ($1-f_R(E)$) is the probability of finding an unoccupied 
state in the left (right) contact at energy E.

Eqns. (\ref{eq:G<layer}) and (\ref{eq:G>layer}) for $G^{n}$ and $G^{p}$ can be written as,
\begin{eqnarray}
G^{n}_{DD} &=& A_{DD}^{-1} \Sigma^{in}_{DD} G^\dagger_{DD} = G_{DD} \Sigma^{in}_{DD} G^\dagger_{DD} \label{eq:G<3}\\
G^{p}_{DD} &=& A_{DD}^{-1} \Sigma^{out}_{DD} G^\dagger_{DD} = G_{DD} \Sigma^{out}_{DD} G^\dagger_{DD} 
\mbox{ .} \label{eq:G>3}
\end{eqnarray}
While these equations appear often in literature, we do not suggest using them 
to compute the diagonal elements of $G^{n}$ and $G^{p}$ of layered structures. This is 
because their use requires knowledge of the entire $G_{DD}$ matrix when $\Sigma^{in}_{DD}$ 
is non zero at all grid points. Computation of the entire $G_{DD}$ amounts to inversion of 
$A_{DD}$. Matrix inversion is computationally expensive and scales as $N^{2.7}$ where $N$ 
is the dimension of $A_{DD}$. The diagonal
elements of $G^{n}$ and $G^{p}$ of layered structures can be computed more efficiently 
without calculating the entire $G_{DD}$ matrix using the algorithm discussed in appendix \ref{section:algorithm}.

\vspace{0.1in}

\noindent
{\bf Current Density:} We will now present some expression for current density commonly used in literature.
The current flowing between layers $q$ and $q+1$ is (eqn. (\ref{eq:JSch-GF}) of 
section \ref{sect:GF}):
\begin{eqnarray}
J_{q \rightarrow q+1} =  \frac{i e}{\hbar}
2 \int \frac{dE}{2\pi} \mbox{Tr} \left[
T_{q,q+1} G^{n}_{q+1,q} (E) - T_{q+1,q} G^{n}_{q,q+1} (E) \right ]
\mbox{.}
\label{eq:elec-current}
\end{eqnarray}
Eq. (\ref{eq:elec-current}) frequently appears in the literature in other useful forms that are derived below. 
Expanding both terms of eqn. (\ref{eq:elec-current}) using eqn. (\ref{eq:discreteG<5}) of appendix 
\ref{section:dyson}, we get,
\begin{eqnarray}
J_{L} &=&  \frac{i e}{\hbar}
2 \int \frac{dE}{2\pi} \mbox{Tr}  (
 [T_{LD} G_{1,1}(E) T_{DL} {g^{n}_L}_{_{0,0}}(E) +
  T_{LD} G^{n}_{1,1}(E) T_{DL} {g^\dagger_L}_{_{0,0}}(E)] \nonumber \\
                      & &\;\;\;\;\;\;\;\;\;\;\;\;\;\; \;
-[T_{DL} {G^{n}_L}_{_{0,0}}(E) T_{LD} G^\dagger_{1,1}(E) +
  T_{DL} {g^\dagger_L}_{_{0,0}}(E) T_{LD} G^{n}_{1,1}(E)] )
                                                \label{eq:elec-current1} \\
&=&  \frac{i e}{\hbar}
2 \int \frac{dE}{2\pi} \mbox{Tr} \left(
[G_{1,1}(E)-G^\dagger_{1,1}(E)] T_{DL} {g^{n}_L}_{_{0,0}}(E) T_{LD}
-
G^{n}_{1,1}(E) T_{DL} [{g_L}_{_{0,0}}(E) - {g^\dagger_L}_{_{0,0}}(E)] T_{LD}\right)
                                              \label{eq:elec-current2}
\end{eqnarray}
Using the relationships,
\begin{eqnarray}
 \Sigma^{in}_L &=& T_{DL} {g^{n}_L}_{_{0,0}} T_{LD} \\
-i\Gamma_L &=& T_{DL} [{g_L}_{_{0,0}} - {g^\dagger_L}_{_{0,0}}] T_{LD} \mbox{ ,}
\end{eqnarray}
Eq. (\ref{eq:elec-current2}) can be written as,
\begin{eqnarray}
J_{L} &=& \frac{e}{\hbar} 2 \int \frac{dE}{2\pi} \mbox{Tr}
(i [G_{1,1}(E)-G^\dagger_{1,1}(E)] \Sigma^{in}_L(E) - G^{n}_{1,1}(E) \Gamma_L(E))
                                                \label{eq:elec-current31}
\end{eqnarray}
Eqns. (\ref{eq:elec-current}) and (\ref{eq:elec-current31}) 
are both general expression for current density valid in the presence 
of electron-phonon scattering in the device \cite{Mei92}. The advantage of using eqn. 
(\ref{eq:elec-current}) is that the current density can be calculated 
at every layer in the device. This expression is useful in understanding 
how the current density is energetically redistributed along the length 
of the device as a result of scattering. 

In the phase coherent limit, $\Sigma_{Phonon} = 0$ and the only non 
zero self-energies are in layers 1 and n due to the contacts. 
We define matrices, $\tilde{\Gamma}_L$ and $\tilde{\Gamma}_R$, 
which consist of $n$ blocks corresponding to the $n$ Device layers 
(dimension of $A_{DD}$ matrix) and with the following non zero elements:
\begin{eqnarray}
{\tilde{\Gamma}_{L}}|_{_{1,1}} = \Gamma_L \mbox{, }
\;\;\;
{\tilde{\Gamma}_{R}}|_{_{n,n}} = \Gamma_R \mbox{ .}
\end{eqnarray}
Now left multiplying eqn. (\ref{eq:GDD2}) by $G^\dagger$ and right multiplying
the Hermitian conjugate of eqn. (\ref{eq:GDD2}) by $G$, and
subtracting the resulting two equations, we have,
\begin{eqnarray}
G - G^\dagger = G^\dagger (\Sigma - \Sigma^\dagger) G \mbox{ ,} \label{eq:Ga-Gr}
\end{eqnarray}
where $\Sigma$ is the total self-energy due to phonons and the
leads. The Hermitean conjugate Green's functions and self-energies are
$G^\dagger$ and $\Sigma^\dagger$. 
In the absence of phonon scattering, the self energies only have components due to 
the leads and so eqn. (\ref{eq:Ga-Gr}) can be written as,
\begin{eqnarray}
i [G - G^\dagger] = G^\dagger ( \tilde{\Gamma}_{L} + \tilde{\Gamma}_{R} )
               G \mbox{ , where } \label{eq:Gr-Ga-pc}
\end{eqnarray}
eqns. (\ref{eq:GammaL}) and (\ref{eq:GammaR}) have been used. It also 
follows from Eqs. (\ref{eq:SigmaL}), (\ref{eq:SigmaR}) and 
(\ref{eq:G<3}) that
\begin{eqnarray}
G^{n} = G\;(\tilde{\Gamma}_{L} f_L + \tilde{\Gamma}_{R} f_R) G^\dagger
				    \label{eq:G<-pc}
\mbox{ .}
\end{eqnarray}
Now using eqns. (\ref{eq:Gr-Ga-pc}) and (\ref{eq:G<-pc}) in eqn. 
(\ref{eq:elec-current31}), the current in the phase coherent limit is,
\begin{eqnarray}
J_{L} = \frac{e}{\hbar} 2 \int \frac{dE}{2\pi} T(E) [f_L(E) - f_R(E) ]
                                        \mbox{ .}
                                                \label{eq:elec-current5}
\end{eqnarray}
The total transmission at energy $E$ is identified from eqn.
(\ref{eq:elec-current5}) to be
\begin{eqnarray}
T (E) =  Tr[ \tilde{\Gamma}_L (E) G (E) \tilde{\Gamma}_{R} (E) G^\dagger (E) ] 
				\mbox{ .} \label{eq:transmission}
\end{eqnarray}
Note that to compute the total transmission using 
eqn. (\ref{eq:transmission}), only the elements of $G$ connecting 
layers $1$ and $n$ are required because $\tilde{\Gamma}_L$ and
$\tilde{\Gamma}_R$ are non zero only in layers $1$ and $n$ respectively.

\subsection{\label{section:crib-Sheet} Crib Sheet}

%%%%%%%%%%%%%%%%%%%%%%%%%%%%%%%%%%%%%%%%%%%%%%%%%%
\begin{figure}
\begin{center}
    \leavevmode
\includegraphics[height=12cm,angle=-0]{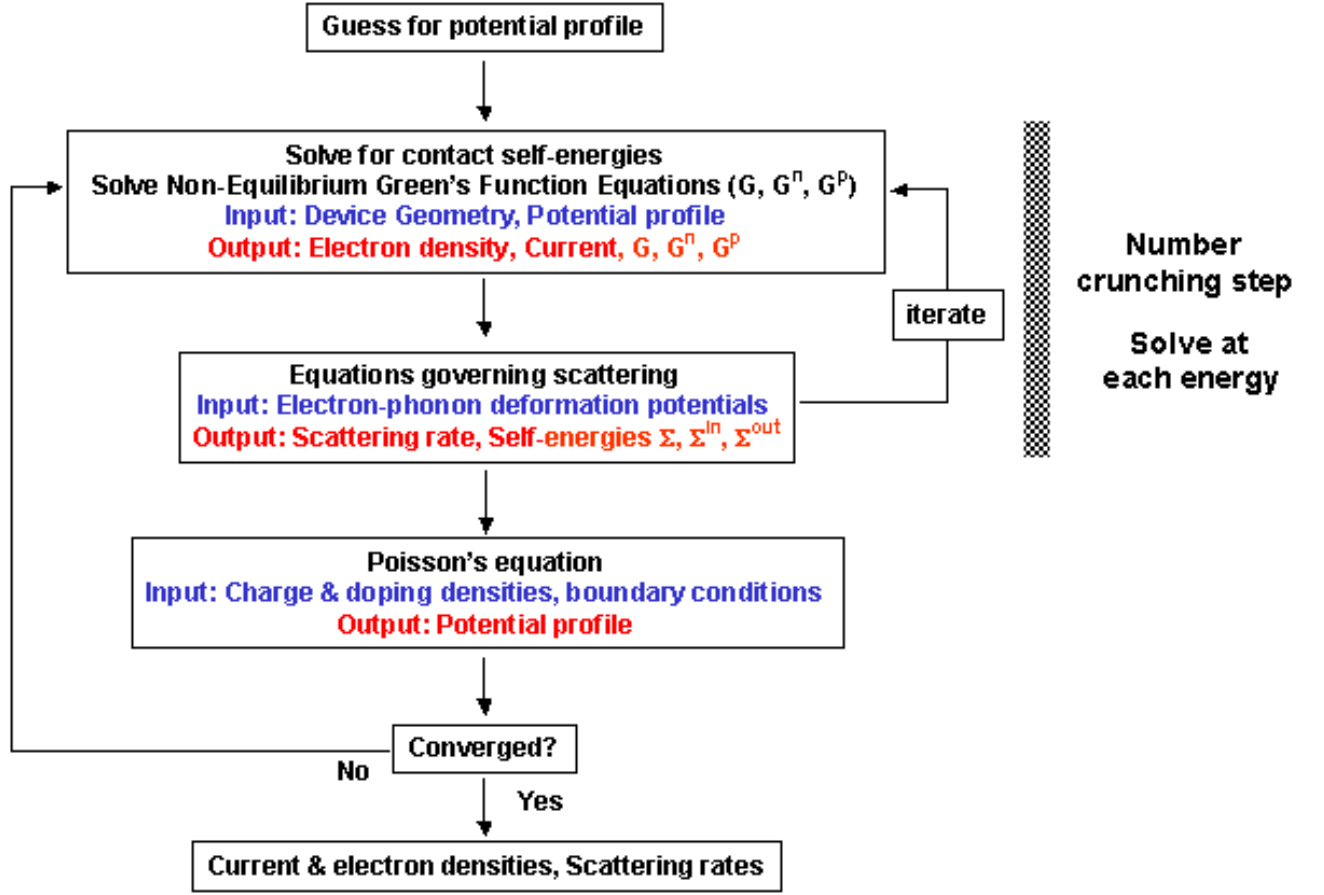}
  \end{center}
\caption{Flowchart of a typical simulation involved in modeling of a nanodevice.}
\label{fig:flowchart}
\end{figure}
%%%%%%%%%%%%%%%%%%%%%%%%%%%%%%%%%%%%%%%%%%%%%%%%%%

The algorithmic flow in modeling nanodevices using the non equilibrium Green's function consists of 
the following steps (Fig. \ref{fig:flowchart}). We first find a guess for the electrostatic 
potential $V(\vec{r})$ and calculate the self energies due to the contacts 
(eqns. (\ref{eq:sig-cs}) - (\ref{eq:sig-out-cs})). The self-energies due to 
electron-phonon scattering are set to zero. The non equilibrium Green's function 
equations for $G$, $G^n$ and $G^p$ (eqs. (\ref{eq:Gr-def}) - eqs(\ref{eq:G>-def})) 
are then solved. Following this, the self energies due to electron phonon scattering 
and contacts (eqns. (\ref{eq:sig-cs}) - (\ref{eq:sig-out-cs}))are calculated. 
As the equations governing the Green's functions depend on the self energies, 
we iteratively solve for the Green's function and self-energies, as indicated 
by the inner loop of Fig. \ref{fig:flowchart}. Then, the electron density 
(diagonal elements of $G^n$) is used in Poisson's equation to obtain a new 
potential profile. We use this updated electrostatic potential profile as an 
input to solve for updated non equilibrium Green's functions, and continue the 
above process iteratively until convergence is achieved (outer loop of Fig. \ref{fig:flowchart}). 
A number of equations that are repeatedly used in nanodevice modeling are listed below.
 
\vspace{0.2in}

{\bf Physical Quantities:}

\noindent

\begin{eqnarray}
\mbox{Scattering Rate: } 
\frac{\hbar}{\tau (E)} = {-2\;\mbox{Im}[\Sigma(E)]} = \Gamma (E)
                   \mbox{}
      \label{eq:scatt-rate}\\
\end{eqnarray}

\noindent
\begin{eqnarray}
\mbox{Density of States at $(\vec{r}, E)$: }
&&N(\vec{r},E) = - \frac{1}{\pi} \mbox{Im} (G (\vec{r},\vec{r},E))
       \mbox{}
      \label{eq:dos}\\
&& \mbox{{Use recursive algorithm to calculate DOS (Do not invert $A$).}}
\nonumber
\end{eqnarray}

\begin{eqnarray}
\mbox{Electron / Occupied density at $\vec{r}$: }
&&n(\vec{r})=  2 \int \frac{dE}{2\pi} G^{n} (\vec{r},\vec{r},E)
      \mbox{}
      \label{eq:elec-dens} \\
&&\mbox{{Use recursive algorithm to calculate $n$ (Do not use
$G^{n}=G \Sigma^{in} G^\dagger$).}}
\nonumber
\end{eqnarray}

\begin{eqnarray}
\mbox{Unoccupied Density at $\vec{r}$: }
&&h(\vec{r})=  2 \int \frac{dE}{2\pi} G^{p} (\vec{r},\vec{r},E)
      \mbox{}
      \label{eq:hole-dens} \\
&&\mbox{{Use recursive algorithm to calculate $h$ (Do not use
$G^{n}=G \Sigma^{in} G^\dagger$).}}
\nonumber
\end{eqnarray}

Current density flowing between layers $q$ and $q+1$ (valid with scattering in device):
\begin{eqnarray}
J_{q \rightarrow q+1} =  \frac{i e}{\hbar}
2 \int \frac{dE}{2\pi} \mbox{Tr} \left[
T_{q,q+1} G^{n}_{q+1,q} (E) - T_{q+1,q} G^{n}_{q,q+1} (E) \right ] \mbox{}
                                                \label{eq:elec-current01}
\end{eqnarray}
Current density flowing from the Left lead into layer 1 of Device
(valid with scattering in device):
\begin{eqnarray}
J_{L} & = & \frac{e}{\hbar} 2 \int \frac{dE}{2\pi} \mbox{Tr}
\{i[G_{1,1}(E)-G^\dagger_{1,1}(E)] \Sigma^{in}_L(E) - G^{n}_{1,1}(E) \Gamma_L(E) \}
                                                \label{eq:elec-current34} \\
& = & \frac{e}{\hbar} 2 \int \frac{dE}{2\pi} \mbox{Tr}
\{G^p_{1,1}(E) \Sigma^{in}_L(E) - G^{n}_{1,1}(E) \Sigma^{out}_L(E) 
\}
\end{eqnarray}
Current density flowing from the Left lead into layer 1 of Device
(valid only in the phase coherent limit):
\begin{eqnarray}
J_{L} = \frac{e}{\hbar} 2 \int \frac{dE}{2\pi} T(E) \; [f_L(E) - f_R(E) ]
                                        \mbox{,}
                                                \label{eq:elec-current51}
\end{eqnarray}
where the total transmission from the Left to Right lead at energy $E$ is given by
\begin{eqnarray}
&& T (E) =  Tr[ \tilde{\Gamma}_L (E) G (E) \tilde{\Gamma}_{R} (E) G^\dagger (E) ]
                                \mbox{.} \label{eq:transmission1} \\
&&\mbox{{Only elements of $G$ connecting layers $1$ and $n$ are
necessary.}} \nonumber
\end{eqnarray}

{\bf Equations Solved:}
\begin{eqnarray}
\mbox{ Green's Function: } &&
[EI - H - \Sigma] G (E) = I \rightarrow A G = I 
      \label{eq:Gr-def}   \\
\mbox{Hermitean conjugate Green's Function: } &&
G^\dagger (E) [EI - H - \Sigma^\dagger] = I 
      \label{eq:Ga-def}   \\
\mbox{electron correlation Function: } &&
[EI - H - \Sigma] G^{n} (E) = \Sigma^{in} (E) G^\dagger(E) 
 \rightarrow A G^{n} = \Sigma^{in} G^\dagger 
      \label{eq:G<-def} \\
\mbox{hole correlation Function: } &&
[EI - H - \Sigma] G^{p} (E) = \Sigma^{out} (E) G^\dagger(E)
 \rightarrow A G^{p} = \Sigma^{out} G^\dagger 
      \label{eq:G>-def}
\end{eqnarray}
\begin{eqnarray}
\Sigma^\alpha(E)&=&\Sigma^\alpha_{lead} (E) + \Sigma^\alpha_{Phonon} (E)
            \mbox{,} \;\;\mbox{where $\alpha \in \;in,\;out$}  \label{eq:sig-cs} \\
{\Sigma^\alpha_{lead}}_{_{1,1}} &=& \Sigma^\alpha_L(E) = T_{DL} {g^\alpha_L}_{_{0,0}} T_{LD} \\
{\Sigma^\alpha_{lead}}_{_{n,n}} &=& \Sigma^\alpha_R(E) = T_{DR} {g^\alpha_R}_{_{n+1,n+1}} T_{RD} \\
{\Sigma^\alpha_{lead}}_{_{i,i}} &=& 0 \;\;\;\;\;\forall \;\;\;\; i \neq {1,n}
\end{eqnarray}
\begin{eqnarray}
\Gamma(E)     &=&  - 2\;\mbox{Im}[\Sigma(E)]   \\ 
\Gamma_L(E)   &=& - 2\;\mbox{Im}[\Sigma_L(E)] \\ 
\Gamma_R(E)   &=&  - 2\;\mbox{Im}[\Sigma_R(E)] \\
\Sigma^{in}_L(E) &=& \Gamma_L(E) f_L(E) \\
\Sigma^{in}_R(E) &=& \Gamma_R(E) f_R(E) \mbox{}  \\
\Sigma^{out}_L(E) &=& \Gamma_L [1-f_L(E)]  \\
\Sigma^{out}_R(E) &=& \Gamma_R [1-f_R(E)] \mbox{} \label{eq:sig-out-cs}
\end{eqnarray}
The diagonal and nearest neighbor off-diagonal elements of $G$ and
$G^{n}$ are computed repeatedly as they correspond to physical quantities
such as the density of states, electron density and current. Non local scattering
mechanisms, which requires calculation of further off-diagonal elements, are not
discussed here.

\vspace{0.1in}

\noindent
{\bf Useful Relationships:}
\begin{eqnarray}
i[G - G^\dagger] &=& G^{p} + G^{n} = {\rm DOS}
      \label{eq:Grprop2} \\
i [\Sigma - \Sigma^\dagger] &=& \Sigma^{out} + \Sigma^{in} = \Gamma
      \label{eq:Grprop3} \\
G - G^\dagger &=& G [\Sigma^\dagger - \Sigma] G^\dagger  
	\label{eq:Grprop4} \\
          &=& G^\dagger [\Sigma^\dagger - \Sigma] G \\
            &=& - i G \Gamma G^\dagger = - i G^\dagger \Gamma G
      \label{eq:Grprop5} \\
{G^{n}}^\dagger &=& {G^{n}} 
      \label{eq:Grprop6} \\
{G^{p}}^\dagger &=& {G^{p}}
	\label{eq:Grprop7}
\end{eqnarray}

\section{\label{section:2D-ballistic-mosfet}
Application to a ballistic nanotransistor}

Quantum mechanics is playing an increasingly important role in modeling 
transistors with channel lengths in the ten nanometer regime for many
reasons: 
(i) {\bf Tunneling} from gate to channel and source to drain determine the off current \cite{Svi02}. 
(ii) {\bf Ballistic} flow of electrons in the channel is 
    important as the channel length becomes comparable to the electron 
    mean free path.
(iii) Classically, the electron distribution in the inversion layer is
    a sheet charge at the Si-SiO$_2$ interface. Quantum mechanically,
    the inversion layer charge is distributed over a few nanometers 
    perpendicular to Si-SiO$_2$ interface due to {\bf quantum confinement}.
Methods based on the drift-diffusion and Boltzmann equations do 
    not apriori capture the above mentioned quantum mechanical features.
    
In this section, we will discuss the equations involved in the two dimensional 
modeling of nanotransistors within the effective mass frame work \cite{Ren03}. 
The quantum mechanical and semiclassical results are compared to illustrate their differences.
The discretized matrix equations that we solve are discussed in section \ref{subsection:DiscretizedEquations}.
The application of the quantum mechanical method to illustrate the role of polysilicon gate depletion, 
and the slopes of
$I_d$ versus gate ($V_g$) and drain ($V_d$) voltages are discussed in section \ref{subsection:pot}.

\vspace{0.1in}

\begin{center}
\begin{tabular}{|l|p{3in}|}
\hline
\multicolumn{2}{|c|}{Table 1. List of Abbreviations: Length Scales}
\\ \hline \hline
$t_{ox}$&oxide thickness \\ \hline
$L_P$&polysilicon gate thickness in x-direction \\ \hline
$L_B$&boundary of substrate region in x-direction \\ \hline
$L_y$&Poisson's and NEGF equations are solved from $-L_y/2$ to
                         $+L_y/2$ \\ \hline
$L_g$&length of  polysilicon gate region in $y$-direction \\ \hline
\end{tabular}
\end{center}

\subsection{\label{subsection:DiscretizedEquations} Discretized equations}

\begin{figure}[htbp]
  \begin{center}
    \leavevmode
\includegraphics[height=10cm,angle=-0]{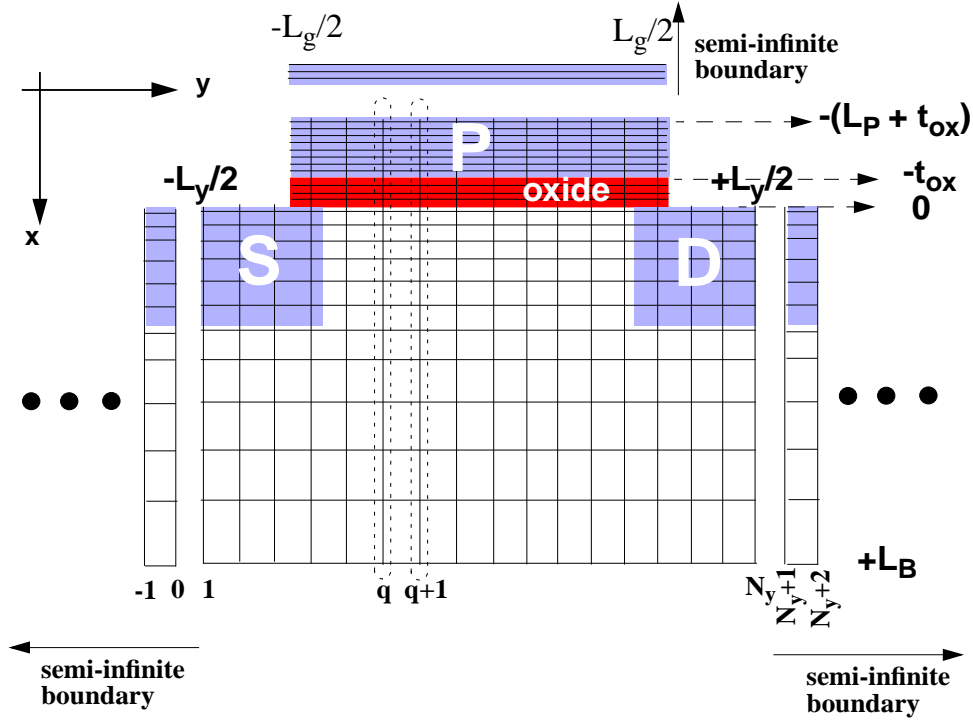}
 \end{center}
 \caption{The equations are solved in a two dimensional non 
uniform spatial grid, with semi-infinite boundaries as shown. Each 
column $q$ corresponds to the diagonal blocks of the Green's function
equations.}
\label{fig:grid}
\end{figure}

We consider $N_b$ independent valleys for electrons within the
effective mass approximation. The Hamiltonian of valley $b$ is
\begin{eqnarray}
H_b(\vec{r}) = -\frac{\hbar^2}{2} [\frac{d}{dx} \left(\frac{1}{m^b_x}
\frac{d}{dx}\right) + \frac{d}{dy} \left(\frac{1}{m^b_y}\frac{d}{dy}
\right) + \frac{d}{dz} \left( \frac{1}{m^b_z}\frac{d}{dz} \right) ]
+ V(\vec{r}) \mbox{,} \label{eq:H-eff-mass}
\end{eqnarray}
where $(m^b_x,m^b_y,m^b_z)$ are components of effective mass in
valley $b$. The equation of motion for the Green's function ($G$) and 
electron correlation function ($G^{n}$) are
\begin{eqnarray}
[E - H_{b_1} (\vec{r}_1)] G_{b_1,b_2} (\vec{r}_1,\vec{r}_2,E)
- \int d\vec{r} \; \Sigma_{b_1,b^\prime} (\vec{r}_1,\vec{r},E)
            G_{b^\prime,b_2} (\vec{r},\vec{r}_2,E) =
            \delta_{b_1,b_2}  \delta(\vec{r}_1-\vec{r}_2) \label{eq:Gr1}
\end{eqnarray}
and
\begin{eqnarray}
& & [E - H_{b_1} (\vec{r}_1)] G^{n(p)}_{b_1,b_2} (\vec{r}_1,\vec{r}_2,E)
- \int d\vec{r} \; \Sigma_{b_1,b^\prime} (\vec{r}_1,\vec{r},E)
      G^{n(p)}_{b^\prime,b_2} (\vec{r},\vec{r}_2,E) =  \nonumber  \\
& &
\hspace{3in}
\int d\vec{r} \; \Sigma^{in(out)}_{b_1,b^\prime} (\vec{r}_1,\vec{r},E)
          G^\dagger_{b^\prime,b_2} (\vec{r},\vec{r}_2,E) \label{eq:G<1}
\mbox{.}
\end{eqnarray}
The coordinate spans only the device in Eqs. (\ref{eq:Gr1}) and
(\ref{eq:G<1}). The influence of the semi-infinite source (S), drain 
(D) and polysilicon gate (P) leads, and electron-phonon interaction
are included via self-energy terms $\Sigma_{b_1,b^\prime}$ and
$\Sigma^{in(out)}_{b_1,b^\prime}$ as discussed in section 
\ref{section:ngf-summary}. The contact self-energies are diagonal in 
the band index, $\Sigma^\alpha_{b_1,b_2,C} = \Sigma^\alpha_{b_1,C} 
\;\delta_{b_1,b_2}$ ($C$ represents contacts). 

The electrostatic potential varies in the $(x,y)$ plane of 
Fig. \ref{fig:grid}, and the system
is translationally invariant along the z-axis.  So, all quantities
$A(\vec{r}_1,\vec{r}_2,E)$ depend only on the difference coordinate
$z_1-z_2$. Using the relation
\begin{eqnarray}
A(\vec{r}_1,\vec{r}_2,E) = \int \frac{d k_z}{2\pi} e^{ik_z (z_1 - z_2)}
                                    A(x_1,y_1,x_2,y_2,k_z,E) \mbox{ ,}
\end{eqnarray}
the equations of motion for $G$ and $G^{n(p)}$ simplify to
\begin{eqnarray}
[E - \frac{\hbar^2 k_z^2}{2m_z} - H_b (\vec{r}_1)]
                  G_{b} (\vec{r}_1,\vec{r}_2,k_z,E)
- \int d\vec{r} \; \Sigma_{b} (\vec{r}_1,\vec{r},k_z,E)
            G_{b} (\vec{r},\vec{r}_2,k_z,E) =
                     \delta(\vec{r}_1-\vec{r}_2) \label{eq:Gr2}
\end{eqnarray}
and
\begin{eqnarray}
[E - \frac{\hbar^2 k_z^2}{2m_z} - H_b (\vec{r}_1)]
                  G_{b} (\vec{r}_1,\vec{r}_2,k_z,E)
- \int d\vec{r} \; \Sigma_{b} (\vec{r}_1,\vec{r},k_z,E)
            G^{n(p)}_{b} (\vec{r},\vec{r}_2,k_z,E) =
\;\;\;\;\;\;\;\;\; && \nonumber \\
\int d\vec{r} \; \Sigma^{in(out)}_{b} (\vec{r}_1,\vec{r},k_z,E)
        G^\dagger_{b} (\vec{r},\vec{r}_2,k_z,E)\mbox{,}  &&
\;\;\;\;\;\;\;
\label{eq:G<2}
\end{eqnarray}
where $Z_b = Z_{b,b}$, and $\vec{r} \rightarrow (x,y)$ for the
remainder of this section.

Eqs. (\ref{eq:Gr2}) and (\ref{eq:G<2}) can be written in matrix 
form as,
\begin{eqnarray}
A^\prime G &=& \lambda \mbox{ and } \label{eq:discreteGr1} \\
\mbox{ and } \;\;\;\;\;\;\;\
A^\prime G^{n(p)} &=& \Sigma^{in(out)} G^\dagger \mbox{ . } \label{eq:discreteG<1}
\end{eqnarray}
The self-energies due to S, D and P are non zero only along the lines
$y=y_S=y_1$, $y=y_D=y_{N_y}$ and $x=-(L_P+t_{ox})$ respectively of Fig. 
\ref{fig:grid}. The $A^\prime$ matrix is ordered such that all grid 
points at a y-coordinate (layer) correspond to a diagonal block of
dimension $N_x$, and there are $N_y$ such blocks. In the notation
adopted $A^\prime_{j_1,j_2}(i,i^\prime)$ is the
off-diagonal entry corresponding to grid points $(x_i,y_{j_1})$ and
$(x_i^\prime,y_{j_2})$. The non zero elements of the diagonal blocks of
the $A^\prime$ matrix are
\begin{eqnarray}
A^\prime_{j,j}(i,i) \!\! &=& \!\! E^\prime - V_{i,j}  - T_{j,j} (i+1,i)
                - T_{j,j} (i-1,i) - T_{j+1,j} (i,i) - T_{j-1,j} (i,i)
                                                        \nonumber\\
           \!\!  & & \!\!
- \Sigma_S(x_i,x_i) \delta_{j,1} - \Sigma_D(x_i,x_i) \delta_{j,N_y}
- \Sigma_P(y_j,y_j) \delta_{i,1} -\Sigma (x_i,y_j;x_i,y_j)  \\
A^\prime_{j,j}(i\pm1,i) \!\! &=& \!\! T_{j,j} (i\pm1,i)
- \Sigma_S(x_{i\pm1},x_i) \delta_{j,1} - \Sigma_D(x_{i\pm1},x_i)
  \delta_{j,N_y} \nonumber \\
             & & \;\;\;\;\;\;\;\;\;\;\;\;\;\;\;\;\;\;\;\;\;\;\;\;\;\;\;
\;\;\;\;\;\;\;\;\;\;\;\;\;\;\;\;\;\;\;\;\;\;\;\;\;\;\;\;\;\;\;\;\;\;\;\;
-\Sigma (x_{i\pm1},y_j;x_i,y_j) \\
A^\prime_{j,j}(i,i^\prime) \!\! &=& \!\! - \Sigma_S(x_i,x_{i^\prime})
            \delta_{j,1} - \Sigma_D(x_i,x_{i^\prime}) \delta_{j,N_y}
\mbox{, for $i^\prime \neq i,\;i\pm1$ ,}
\end{eqnarray}
where  $E^\prime = E - \hbar^2 k_z^2 / 2m_z$ and $V_{i,j}=V(x_i,y_j)$.
The off-diagonal blocks are
\begin{eqnarray}
A^\prime_{j\pm1,j}(i,i) &=& T_{j\pm1,j} (i,i)
               - \Sigma_P(y_j,y_{j\pm1}) \delta_{i,1} \nonumber \\
A^\prime_{j,j^\prime} (i,i^\prime) &=& 0 \mbox{, for $j^\prime \neq j,\;
                                                j\pm1$,}
\end{eqnarray}
and the non zero elements of the $T$ matrix are
\begin{eqnarray}
T_{j,j} (i\pm1,i) & = & \frac{\hbar^2}{2m^{\pm x}}
                  \frac{2}{x_{i+1}-x_{i-1}} \frac{1}{|x_{i\pm1}-x_i|}
                        \label{eq:T1}  \\
T_{j\pm1,j} (i,i) & = & \frac{\hbar^2}{2m^{\pm y}}
                  \frac{2}{y_{j+1}-y_{j-1}} \frac{1}{|y_{j\pm1}-y_j|}
                                                        \mbox{,}
                        \label{eq:T2}
\end{eqnarray}
where $m^{\pm x}=\frac{m_{i\pm 1,j} + m_{i,j}}{2}$ and
$m^{\pm y}=\frac{m_{i,j\pm 1} + m_{i,j}}{2}$.
Non zero elements of $\Sigma_P(y_j,y_j^\prime)$, where $j^\prime \neq
j, j\pm1$ are neglected to ensure that $A^\prime$ is block tridiagonal; The
algorithm to calculate $G$ and $G^{n}$ relies on the block tridiagonal
form of $A^\prime$. The $\lambda$ appearing in eqn.
(\ref{eq:discreteGr1}) corresponds to the delta function in eqn.
(\ref{eq:Gr2}). $\lambda$ is a diagonal matrix whose elements are
given by
\begin{eqnarray}
\lambda_{i,j;i,j} = \frac{4}{(x_{i+1}-x_{i-1}) (y_{i+1}-y_{i-1})}
\mbox{ .}
\end{eqnarray}

\subsection{\label{subsection:pot} Results}

\begin{figure}[htbp]
\begin{center}
\leavevmode
\includegraphics[height=6.5cm,width=6.5cm,angle=-0]{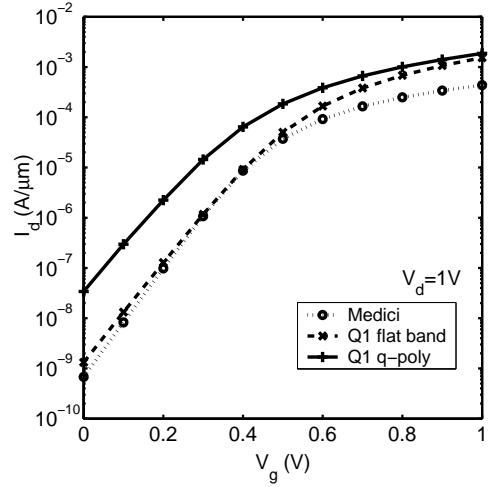}
\end{center}
\caption{Plot of drain current versus gate voltage from the quantum mechanical
calculations and Medici, at $V_d=1$V. At small gate voltages, the
drain current from Medici are comparable to the
'Q1 flat band' results. The drain current from 'Q1 q-poly' is however
significantly different at large gate voltages.}
\label{fig:Id1vsVg_quantum_medici}
\end{figure}

We will now discuss quantum mechanical aspects of transport in a two dimensional ballistic nanotransistor.
 We do so by comparing the current-voltage characteristics from our quantum and 
drift-diffusion simulations as shown in
Fig. \ref{fig:Id1vsVg_quantum_medici}. 
The important features are a higher off-current, threshold voltage 
shift, smaller subthreshold slope and much higher on-current, in the quantum case. 

\begin{figure}[htbp]
  \begin{center}
    \leavevmode
\includegraphics[height=6.5cm,width=6.5cm,angle=-0]{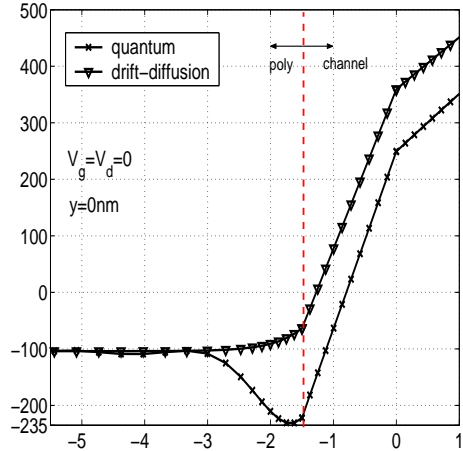}
  \end{center}
  \caption{Potential profile at the y=0 slice of MIT25, calculated using 
  quantum and drift-diffusion methods by assuming flat band in the 
  polysilicon gate.}
  \label{fig:poly-effect}
\end{figure}

The change in threshold voltage results from the very different 
quantum mechanically calculated potential profile in the polysilicon gate.  
The quantum mechanical band bending is opposite to the drift-diffusion case as 
shown in Fig. \ref{fig:poly-effect}. In the quantum case, the conduction band 
at the polysilicon-SiO$_2$ is lower by approximately 130 meV. 
The physical reason for the qualitatively different quantum potential 
profile arises due to the tiny quantum mechanical probability density for
electron occupancy close to the barrier. As a consequence, the electronic
charge density is smaller than the uniform background doping density, near
the SiO$_2$ barrier in the polysilicon region. This causes the conduction
band in the polysilicon gate to bend in a direction opposite to that
computed classically.

%%%%%%%%%%%%%%%%%%%%%%%%%%%%%%%%%%%%%%%%%%%%%
%%%%%%%%%%%%%%%%%%%%%%%%%%%%%%%%%%%%%%%%%%%%%

\begin{figure}[htbp]
  \begin{center}
    \leavevmode
\includegraphics[height=6.5cm,width=6.5cm,angle=-0]{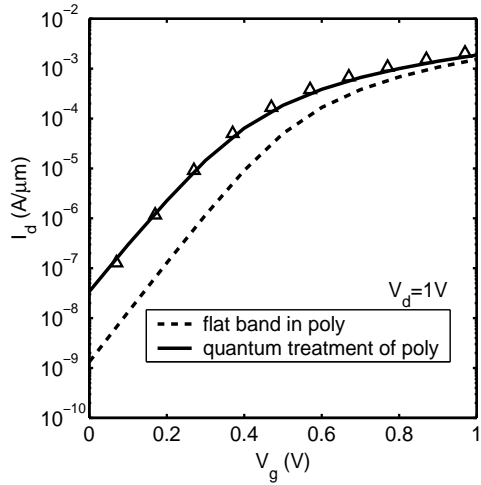}
  \end{center}
  \caption{Drain current versus gate voltage for $V_d=1$ V. Quantum mechanical
treatment of the polysilicon gate results in much higher current (solid line). 
The triangles correspond to the $I_d(V_g)$ calculated using a flat band in the
polysilicon region shifted by the equilibrium built-in potential of 130 meV
in the polysilicon region.}
\label{fig:IdvsVg_1B_1BP}
\end{figure}

The quantum mechanically calculated $I_d$ versus $V_g$ with and with out the
quantum mechanical band bending in the gate region is shown in Fig. 
\ref{fig:IdvsVg_1B_1BP}. The gate voltage shift is approximately
equal to the band bending in the polysilicon gate.
A shift in $I_d (V_g)$ from the flat band result by the equilibrium
1D built-in potential does a reasonable job (triangles in Fig. \ref{fig:IdvsVg_1B_1BP}) of 
reproducing the results
obtained by quantum mechanical treatment of the polysilicon region.
This is especially true at small gate voltages and becomes progressively
poorer with increase in gate voltage.

%%%%%%%%%%%%%%%%%%%%%%%%%%%%%%%%%%%%
%%%%%%%%%%%%%%%%%%%%%%%%%%%%%%%%%%%%

\begin{figure}[htbp]
  \begin{center}
    \leavevmode
\includegraphics[height=6.5cm,width=6cm,angle=-0]{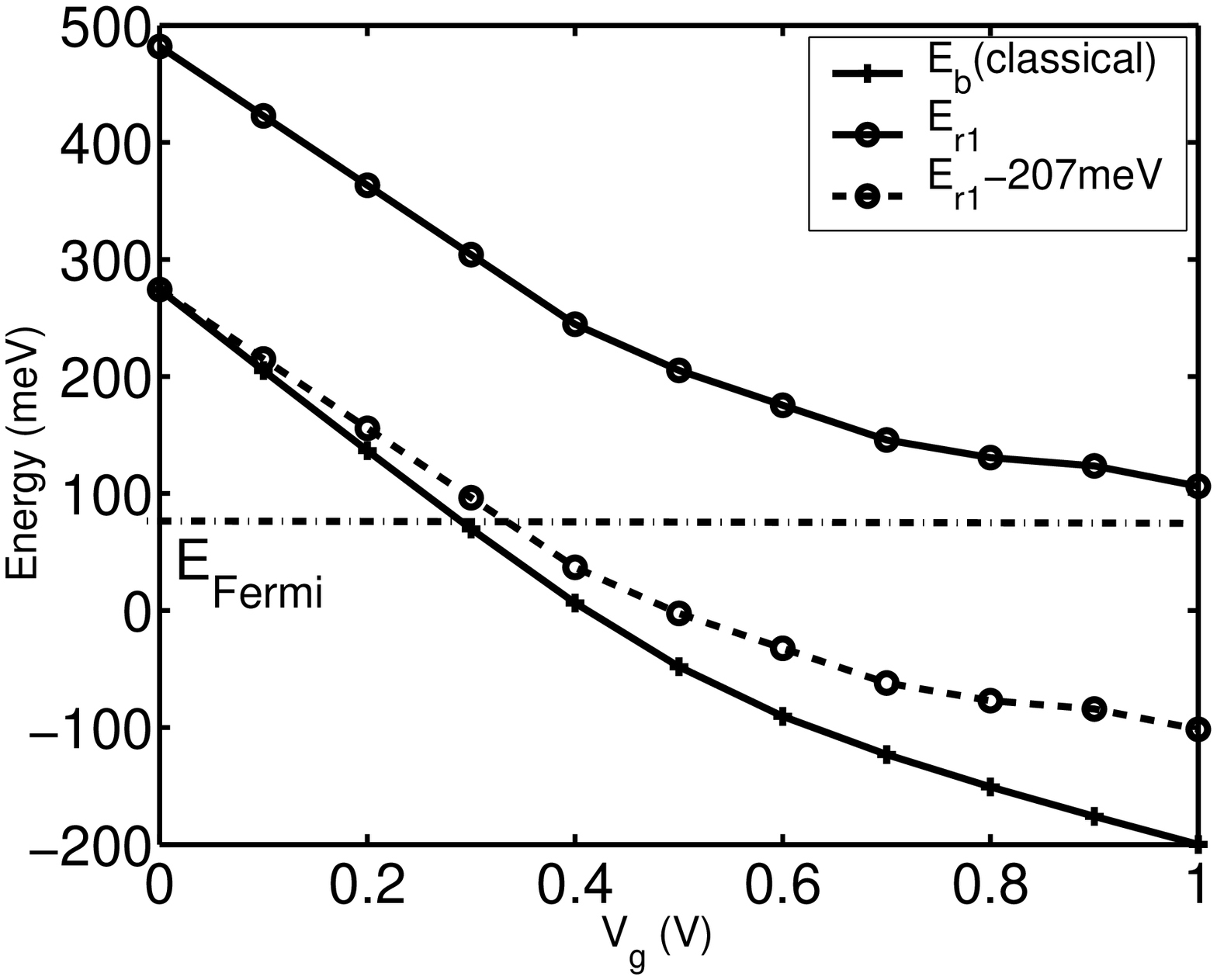}
\hspace{0.5in}
\includegraphics[height=6.cm,width=6cm,angle=-0]{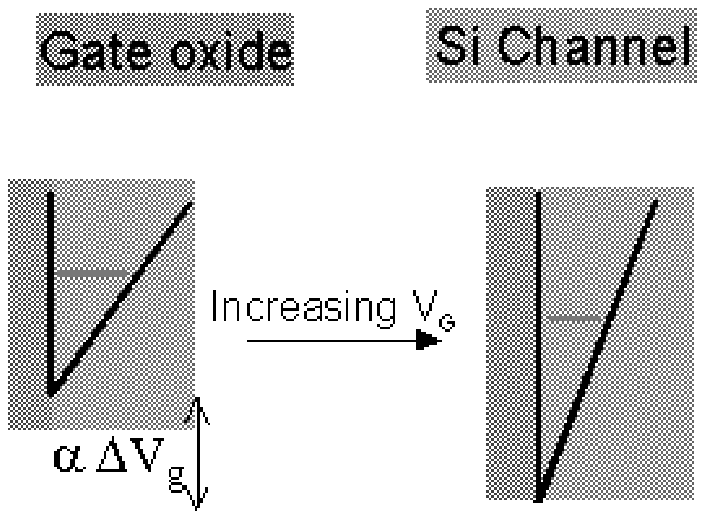}
  \end{center}
  \caption{{\it Left:} Location of the first resonant level $E_{r1}$ versus gate voltage
and the classical source injection barrier $E_b(classical)$. Note
that $E_{r1}$ decreases slower than $E_b(classical)$ with gate
voltage due to narrowing of channel potential well. {\it Right}: Narrowing of the triangular 
well in the channel with increase in gate bias. $E_b(classical)$ is the bottom of the 
triangular well and the resonant level is shown by the horizontal line.
}
\label{fig:EbERvsVg}
\end{figure}

The subthreshold slope $d[log(I_d)]/dV_g$ is smaller in the quantum 
case compared to the drift-diffusion case. To understand the reason for this, we first note that
the subthreshold current resulting from the simple intuitive expression
\begin{eqnarray}
I = I_{q0} \; e^{\frac{-E_{r1}}{kT}} \label{eq:q-curr}
%\;\;\;\; \mbox{ and } \;\;\;\;
%I = I_{c0} \; e^{\frac{E_f-E_{b}}{kT}} \label{eq:c-curr} \mbox{ ,}
\end{eqnarray}
matches the quantum result quite accurately. $I_{q0}$ is a prefactor
chosen to match the current at $V_g=0$. $E_{r1}$ is the energy of the source injection barrier
due to the lowest resonant level in the channel, which varies with gate bias. 
The higher resonant levels do not carry appreciable current. The difference in 
slope between the classical and quantum
results for $I_d(V_g)$ is because of the slower variation of $E_{r1}$
in comparison to the drift-diffusion barrier height ($E_b(classical)$) as a function of $V_g$ (Fig.
\ref{fig:EbERvsVg}). We also find that the decrease of $E_{r1}$
with increase in gate voltage is slower than the barrier height
determined from the quantum potential profiles. The slower variation of $E_{r1}$ 
arises due quantum confinement in the triangular well in the channel that becomes 
progressively narrower with increase in gate voltage (Fig. \ref{fig:EbERvsVg}). 
This change in confinement is a non issue in the classical case.

%%%%%%%%%%%%%%%%%%%%%%%%%%%%%%%%%%%%%%%%%%%%
%%%%%%%%%%%%%%%%%%%%%%%%%%%%%%%%%%%%%%%%%%%%

\begin{figure}[htbp]
  \begin{center}
    \leavevmode
\includegraphics[height=6.5cm,width=6.5cm,angle=-0]{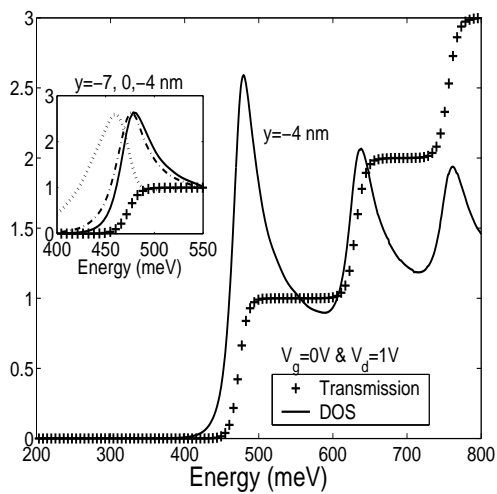}
  \end{center}
  \caption{Transmission (+) and density of states (solid) versus energy at a
spatial location close to the source injection barrier, at $V_g=0$V
and $V_d=1$V. The peaks in the density of states represent the
resonant levels in the channel.
Inset: The density of states at three different y-locations and total transmission (+).
The points y = -7 and 0 nm are to the left and right of the location
where the source injection barrier is largest (close to y = -4 nm).}
\label{fig:DOSandTvsEbw}
\end{figure}

We will now address the value of total transimssion in a ballistic MOSFET. 
The transmission is related to drain current by (eqn. (\ref{eq:elec-current5})),
\begin{eqnarray}
I_d = \frac{2e}{h} \int dE \; T (E) \left[f_S (E) - f_D (E) \right] \mbox{ ,}
\end{eqnarray} 
where $T$ is the total transmission from source to drain. $f_S$ and $f_D$ are the 
Fermi factors in the source and drain, and the factor of 2 accounts for spin. 
The main factors that determine the transmission probability are tunneling and 
scattering in the two dimensional potential profile, as an electron transits 
from source to drain.  Traditionally, simple theories of  ballistic nanotransistors 
have assumed that the transmission from the source to drain to be integers 
in the above expression for current. The quantum mechanically computed transmission 
versus energy is shown in Fig. \ref{fig:DOSandTvsEbw}. The transmission increases 
in a step-wise manner, with the integer values at plateaus equal to the number of 
conducting modes in the channel. The steps turn on at an energy determined by an 
effective 'subband dependent' source injection barrier ($E_{r1}$). That is, the 
maximum subband energy between the source and drain due to quantization perpendicular 
to the gate plane (x-direction of Fig. \ref{fig:grid}).
The total transmission assumes integer values at an energy slightly above the 
maximum in 2D density of states as shown in the inset of Fig. \ref{fig:DOSandTvsEbw}. 
Further, the transmission steps develop over a 50 meV energy window. 
In the case of MIT25 the current is predominantly carried at energies where the transmission is not an integer.

%%%%%%%%%%%%%%%%%%%%%%%%%%%%%%%%%%%%%%%%%%%%
%%%%%%%%%%%%%%%%%%%%%%%%%%%%%%%%%%%%%%%%%%%%

\section{\label{section:mode-space}
Application to nanotransistors with electron-phonon scattering}

The channel, scattering and screening lengths become comparable in 
transistors with diminishing channel lengths. Carrier transport is
however not fully ballistic. Realsitic nanodevice modeling will involve
phase-breaking scattering such that transport is between the ballitic and 
diffusive limits. In this regime, carriers are not 
thermally relaxed in the drain-end of the transistor, in contrast to long 
channel devices. The reflection of hot 
carriers from the drain-end towards the source-end, both above and below 
the source injection barrier should be explicitly included in models to 
compute the drive current. It is in this intermediate regime that the
NEGF method has a distinct advantage over solving Schrodinger and Poisson 
equations self-consistently.

\begin{figure}
  \begin{center}
    \leavevmode
\includegraphics[height=5cm,angle=-0]{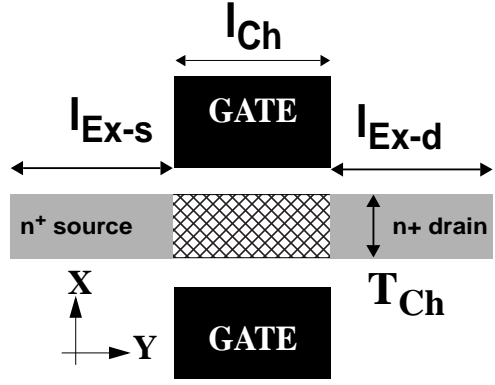}
  \end{center}
\caption{
Schematic of a Dual Gate MOSFET (DG MOSFET). Ex-s and Ex-d are the
extension regions and the hatched region is the channel. The white
region between the source / drain / channel and gate is the oxide. The
device dimension normal to the page is infinite in extent.}
\label{fig:dg}
\end{figure}

To illustrate modeling of electron-phonon scattering in nanotransistors,
we consider a prototype dual gate MOSFET (DGMOSFET) with a channel 
length of 25 nm and channel thickness ($t_{ch}$) of 1.5 nm. The quantized
energy levels in the channel due to quantization perpendicular to the 
gate (x-direction of Fig. \ref{fig:dg}) are,
\begin{eqnarray}
E_n = \frac{n^2 \pi^2 \hbar^2}{2 m t_{ch}^2} \mbox{ .} \label{eq:E_n}
\end{eqnarray}
When $t_{ch}$ is small, the energy level separation is large and very few
subbands are occupied in the highly doped extension regions. The lowest three
quantized energy levels [eqn. (\ref{eq:E_n})] are at 173,
691 and 891 meV above the bulk conduction band. The Fermi energy at the
contact doping of 1E20 cm$^{-3}$ is 60 meV above the bulk conduction 
band. As a result, electrons are injected only into one subband from the
source-end at the operating voltage of $V_d=V_g=0.6 V$. In this section,
we discuss modeling of such MOSFETs using the mode space approach. The
mode space approach consists of solving a one dimensional Schrodinger's 
equation for each subband. Inter subband scattering which can arise due
to a change in electrostatic potential profile is neglected. 
 Electron-phonon scattering
between different subbands is included within the Born approximation.

The three dimensional effective mass Hamiltonian is the same as 
eqn. (\ref{eq:H-eff-mass}). As the $z$ direction is infinite, the
wave function can be expanded as, 
$\Phi(x,y,z) = e^{ik_zz} \phi(x,y)$.
Schrodinger's equation is then,
\begin{eqnarray}
 \left[ E - \frac{\hbar^2 k_z^2}{2m_z^b} - \left(-\frac{\hbar^2}{2}
\frac{d}{dy} \left(\frac{1}{m^b_y}\frac{d}{dy} \right)
                                         - \frac{\hbar^2}{2}
\frac{d}{dx} \left(\frac{1}{m^b_x}\frac{d}{dx} \right) + V(x,y) \right)
\right] \phi(x,y)  = 0 \mbox{ .}
\end{eqnarray}
In the first step of the mode space approach, the eigen values
[$E_n(y)$] at every cross section $y$ along the source-drain direction is 
computed by solving for a subband dependent eigen value that varies with $y$,
\begin{eqnarray}
\left[-\frac{\hbar^2}{2 m_x^b} \frac{d}{dx}
\left(\frac{1}{m^b_x}\frac{d}{dx} \right) + V(x,y) \right] \;
\Psi_n(x,y) = E_n (y) \Psi_n(x,y) \mbox{ .} \label{eq:Sch}
\end{eqnarray}
$n=\{\nu, b\}$, where $\nu$ is the quantum number due to quantization
in the X-direction and $b=1\mbox{, }2 \mbox{ and }3$ are the valley
indices. 

In the second step, the Green's function equations for $G$, $G^{n}$
and $G^{p}$ are solved for each subband $n$:
\begin{eqnarray}
&&
\left[ E - \frac{\hbar^2 k_z^2}{2m_z^n} - \left(-\frac{\hbar^2}{2}
\frac{d}{dy} \left(\frac{1}{m^n_y}\frac{d}{dy} \right) + E_n(y) \right)
\right] G_{n} (y,y^\prime,k_z,E)  \nonumber \\
&&\;\;\;\;\;\;\;\;\;\;\;\;\;\;\;\;\;\;\;\;\;\;\;\;\;\;\;\;\;\;\;\;\;\;\;\;\;\;\;\;\;
\;\;\;\;\;\;\;\;\;\;\;\;\;\;\;\;\;\;\;\;\;\;\;\;\;\;\;
- \int dy_1 \; \Sigma_{n} (y,y_1,k_z,E) G_{n} (y_1,y^\prime,k_z,E) =
                 \delta(y-y^\prime)  \mbox{ , and} \label{eq:Gr_scatt}
\end{eqnarray}
\begin{eqnarray}
&&
 \left[ E - \frac{\hbar^2 k_z^2}{2m_z^n} - \left(-\frac{\hbar^2}{2}
\frac{d}{dy} \left(\frac{1}{m^n_y}\frac{d}{dy} \right) + E_n(y) \right)
\right] G^{n(p)}_{n} (y,y^\prime,k_z,E)  \nonumber \\
&&\;\;\;\;\;\;\;\;\;\;\;\;\;\;\;\;\;\;\;\;\;\;\;\;\;\;\;\;\;\;\;\;\;\;
- \int dy_1 \; \Sigma_{n} (y,y_1,k_z,E)
G^{n(p)}_{n} (y_1,y^\prime,k_z,E) =
\int dy \; \Sigma^{in(out)}_{n} (y,y_1,k_z,E)
           G^\dagger_{n} (y_1,y^\prime,k_z,E) \mbox{ ,} \label{eq:Gl_scatt}
\end{eqnarray}
$m_y^n$ and $m_z^n$ are the effective masses
of silicon in the $y$ and $z$ directions that give rise to subband
index $n$. Note that in Eqs. (\ref{eq:Gr_scatt}) and 
(\ref{eq:Gl_scatt}), $E_n(y)$ is effectively an electrostatic potential
for electrons in subband $n$.

We designate the Green's/correlation functions $G^{\alpha}_{n}$
with $\alpha \in \text{(empty)}, \; p, \; n$
and, respectively, self-energies $\Sigma^{\alpha}_{n}$
with $\alpha \in \text{(empty)}, \; out, \; in$. 
They can be written as,
\begin{eqnarray}
\Sigma^\alpha_{n} &=& \Sigma^\alpha_{n,C} + \Sigma^\alpha_{n,P}
\mbox{, where} \\
\Sigma^\alpha_{n,P} &=& \Sigma^\alpha_{n,el} +
                                              \Sigma^\alpha_{n,inel}
\mbox{ .}
\end{eqnarray}
$\Sigma^\alpha_{n,C}$ is the self-energy due to the leads. The phonon
self-energy $\Sigma^\alpha_{n,P}$ consists of two terms,
$\Sigma^\alpha_{n,el}$ due to elastic and $\Sigma^\alpha_{n,inel}$ due
to inelastic scattering. Scattering between the lowest three subbands
due to electron-phonon interaction is included and all other inter 
subband scattering is neglected. The self-energy due to leads is non 
zero only at the first (source) and last (drain) grid points.

Assuming isotropic scattering, an equilibrium phonon bath and the
self-consistent Born approximation, the self-energies due to
electron-phonon scattering at grid point $y_i$ are \cite{Mah87},
\begin{eqnarray}
\Sigma^\alpha_{el,n} (y_i,E) &=& \sum_{n^\prime}
D_{n,n^\prime}^{el}
\frac{\sqrt{m_z^{n^\prime}}}{\pi \hbar \sqrt{2}} \int dE_z
\frac{1}{\sqrt{E_z}} G^\alpha_{n^\prime}(y_i,E_z,E) \mbox{ ,}
\label{eq:el_self_en_r}
\end{eqnarray}
\begin{eqnarray}
&&\Sigma^{in}_{inel,n} (y_i,E) = \sum_{n^\prime,\eta}
D_{n,n^\prime}^{i,\eta} \frac{\sqrt{m_{z}^{n^\prime}}}{\pi \hbar \sqrt{2}}
\int dE_z \frac{1}{\sqrt{E_z}}  \nonumber \\
&& \;\;\;\;\;\;\left[n_B(\hbar \omega_\eta)
G^{n}_{n^\prime}(y_i,E_z,E-\hbar \omega_\eta)  +
(n_B(\hbar \omega_\eta)+1)
G^{n}_{n^\prime}(y_i,E_z,E+\hbar \omega_\eta) \right] \mbox{ ,}
\label{eq:inel_self_en_<}
\end{eqnarray}
and
\begin{eqnarray}
&&\Sigma^{out}_{inel,n} (y_i,E) = \sum_{n^\prime,\eta}
D_{n,n^\prime}^{i,\eta} \frac{\sqrt{m_{z}^{n^\prime}}}{\pi \hbar \sqrt{2}}
\int dE_z \frac{1}{\sqrt{E_z}}  \nonumber \\
&& \;\;\;\;\;\;\left[n_B(\hbar \omega_\eta)
G^{p}_{n^\prime}(y_i,E_z,E+\hbar \omega_\eta)  +
(n_B(\hbar \omega_\eta)+1)
G^{p}_{n^\prime}(y_i,E_z,E-\hbar \omega_\eta) \right] \mbox{ .}
\label{eq:inel_self_en_>}
\end{eqnarray}
Here $\eta$ represents
the phonon modes, and the square of the matrix elements for phonon
scattering are given by,
\begin{eqnarray}
D_{n,n^\prime}^{el} &=& (\delta_{\nu,\nu^\prime} + \frac{1}{2})
\delta_{b,b^\prime} \frac{D_A^2 kT}{\rho v^2} \label{eq:def1} \\
D_{n,n^\prime}^{i,\eta} &=& (\delta_{\nu,\nu^\prime} + \frac{1}{2})
\left[\delta_{b,b^\prime} \frac{D_{g\eta}^2 \hbar}{2\rho \omega_{g\eta}}
+ (1-\delta_{b,b^\prime}) \frac{D_{f\eta}^2 \hbar}{\rho \omega_{f\eta}}
\right] \label{eq:def2}
\end{eqnarray}
The contribution to elastic scattering is only from acoustic phonon
scattering. The values of the deformation potential, $D_A$, $D_{g\eta}$
and $D_{f\eta}$, and phonon frequencies $\omega_{g\eta}$ and
$\omega_{f\eta}$ are taken from \cite{Lun00}. $\rho$ is the
mass density, $k$ is the Boltzmann constant, $T$ is the temperature and
$v$ is the velocity of sound. $b$ and $b^\prime$ are indices
representing the valley. The following scattering processes are
included:
acoustic phonon scattering in the elastic approximation and g-type
intervalley scattering with phonon energies of 12, 19 and 62 meV. The
imaginary part of the electron-phonon self-energy which is responsible
for scattering induced broadening of energy levels and energetic 
redistribution of carriers are included but the real part is set to 
zero.

In the numerical solution, $N$ uniformly spaced grid points in the 
$Y$-direction with the grid spacing equal to $\Delta y$ are considered.
The discretized form of Eqs. (\ref{eq:Gr_scatt}) and (\ref{eq:Gl_scatt})
are then:
\begin{eqnarray}
 A_{i,i}   G_{n} (y_i,y_i^\prime,k_z,E)
 + A_{i,i+1} G_{n} (y_{i+1},y_i^\prime,k_z,E)
 + A_{i,i-1} G_{n} (y_{i-1},y_i^\prime,k_z,E) =
\frac{\delta_{i,i^\prime}}{\Delta y}
              \mbox{ , and} \label{eq:Gr-scatt-discrete}
\end{eqnarray}
\begin{eqnarray}
&& \!\!\!\!\!\!\!\!\!\!\!\!\!\!\!\!\!\!\!\!\!\!\!\!\!\!\!\!\!\!\!\!
A_{i,i}   G^\alpha_{n} (y_i,y_i^\prime,k_z,E)
 + A_{i,i+1} G^\alpha_{n} (y_{i+1},y_i^\prime,k_z,E)
 + A_{i,i-1} G^\alpha_{n} (y_{i-1},y_i^\prime,k_z,E) = \nonumber \\
&& \;\;\;\;\;\;\;\;\;\;\;\;\;\;\;\;\;\;\;\;\;\;\;\;
 \;\;\;\;\;\;\;\;\;\;\;\;\;\;\;\;\;\;\;\;\;\;\;\;
 \;\;\;\;\;\;\;\;\;\;\;\;\;\;\;\;\;\;\;\;\;\;\;\;
\Sigma_n^\alpha (y_i,E) G^\dagger_{n} (y_i,y_i^\prime,k_z,E)
              \mbox{ , } \label{eq:G<-scatt-discrete}
\end{eqnarray}
where,
\begin{eqnarray}
A_{i,i} &=& E - \frac{\hbar^2 k_z^2}{2m_z^n} -
                \frac{\hbar^2}{m^n_y \Delta y^2} -
                 E_n(y_i) - \Sigma_n (y_i,k_z,E) \mbox{ and }\\
A_{i\pm1,i} &=& + \frac{\hbar^2}{2m_z^n \Delta y^2}
\end{eqnarray}

The contact self-energies are:
\begin{eqnarray}
\Sigma_{n,C}(y_1,k_z,E) &=& (\frac{\hbar^2}{2m_z^n \Delta y^2})^2
g_s(k_z,E) \\
\Sigma_{n,C}(y_N,k_z,E) &=& (\frac{\hbar^2}{2m_z^n \Delta y^2})^2
g_d(k_z,E) \\
\Sigma^{in}_{n,C}(y_1,k_z,E) &=& -2 \mbox{Im}(\Sigma_{n,C}(y_1,k_z,E))
f_{s}(E) \\
\Sigma^{in}_{n,C}(y_N,k_z,E) &=& -2 \mbox{Im}(\Sigma_{n,C}(y_N,k_z,E))
f_{d}(E) \\
\Sigma^{out}_{n,C}(y_1,k_z,E) &=&  -2 \mbox{Im}(\Sigma_{n,C}(y_1,k_z,E))
[1-f_{s}(E)] \\
\Sigma^{out}_{n,C}(y_N,k_z,E) &=& -2 \mbox{Im}(\Sigma_{n,C}(y_N,k_z,E))
[1-f_{d}(E)] \mbox{ ,}
\end{eqnarray}
where $y_1$ an $y_N$ are the left (source-end) and right
(drain-end) most grid points respectively, $g_s(k_z,E)$ and
$g_d(k_z,E)$ are the surface Green's functions of the source and drain
leads respectively, and $f_{s}$ and $f_{d}$ are the Fermi functions in
the source and drain contacts respectively.

The electron and current densities per energy given by 
Eqs. (\ref{eq:elec-dens1}) and (\ref{eq:elec-current}) can be simplified to 
\begin{eqnarray}
n_n(y_i,k_z,E) &=& G_n^{n}(y_i,y_i,k_z,E) \label{eq:dens}  \\
J_n(y_i,k_z,E) &=& \frac{i e}{\hbar} \sum_n
                 \frac{\hbar^2}{2m_y^n \Delta y^2}
      [ G_n^{n}(y_i,y_{i+1},k_z,E) - G_n^{n}(y_{i+1},y_i,k_z,E)] \mbox{ .}
   \label{eq:current}
\end{eqnarray}

The total electron and current densities at grid point $y_i$ are given by,
\begin{eqnarray}
n(y_i) &=& 4 \sum_n \frac{\sqrt{m_z^{n^\prime}}}{\pi \hbar \sqrt{2}}
\int \frac{dE}{2\pi} \int dE_z \frac{1}{\sqrt{E_z}} n_n(y_i,E_z,E) \\
J(y_i) &=& 4 \sum_n \frac{\sqrt{m_z^{n^\prime}}}{\pi \hbar \sqrt{2}}
\int \frac{dE}{2\pi} \int dE_z \frac{1}{\sqrt{E_z}}
J_n(y_i,E_z,E) \mbox{ ,}
\end{eqnarray}
where the prefactor of 4 in the above two equations account for two
fold spin and valley degeneracies.
The non equilibrium electron and current densities are calculated in
both the channel and extension regions using the algorithm for $G^{n}$
presented in section \ref{subsect:recursive-G<}.

In solving the the Green's function and Poisson's equation, the
applied bias corresponds to a difference in the Fermi levels used
in the source and drain regions. The electrostatic potential at the
left and right most grid points of the source and drain extension
regions are calculated self consistently using the floating boundary 
condition that $dV(y)/dy=0$. Poisson's equation is solved
in two dimensions and the electron density is computed from 
Eqs. (\ref{eq:Sch}) and (\ref{eq:dens}) using,
\begin{eqnarray}
n(x_i,y_i,k_z,E) = n_n(y_i,k_z,E) |\Psi_n(x_i,y_i)|^2 \mbox{ .}
\end{eqnarray}

When does the mode-space approach fail? Ref. \cite{Ven03} has extensively
analysed the regime of validity of the mode space approach. The
mode space approach is valid when the wave function 
$\Psi_n(x,y)$ at various $y$ cross sections in eqn.
(\ref{eq:Sch}) satisfies $\frac{\Psi_n(x,y)}{dy} \sim 0$. That is
the shape of the wave function at each cross section should not
change significantly along the transport direction, which means
that intersubband scattering due to changes in potential profile
is absent. This approximation seems to be valid until a 
channel thickness of 5 nm for silicon.

\subsection{Results}

Using the equations presented in the previous section, we show
results illustrating the role of scattering along the channel
length of a nanotransistor.
First, we show that in devices where the scattering length is
comparable to channel length, the nanotransistor drive current
is affected by scattering at all points in the channel. Second,
we show that when hot electrons enter the drain extension region
of a nanotransistor, the drain extension region cannot be 
modeled as a series resistance. Instead, the drain extension
should be included as part of the non equilibrium simulation
region.  

To illustrate the role of scattering along the channel, we calculate
the drain current as a function of the {\it right boundary of
scattering} ($Y_{R-Scatt}$) \cite{Svi03}. Scattering is included only from the
source end of the channel (-5 nm) to $Y_{R-Scatt}$ by setting the 
deformation potential in Eqs. (\ref{eq:def1}) and (\ref{eq:def2})
to zero to the right of $Y_{R-Scatt}$. The scattering lengths are
decreased by a factor $\alpha$ by modifying the deformation 
potential in Eqs. (\ref{eq:def1}) and (\ref{eq:def2}) by an overall
multiplicative factor of $\sqrt{\alpha}$. 

\begin{figure}
  \begin{center}
    \leavevmode
\includegraphics[height=6.5cm,width=6.5cm,angle=-0]{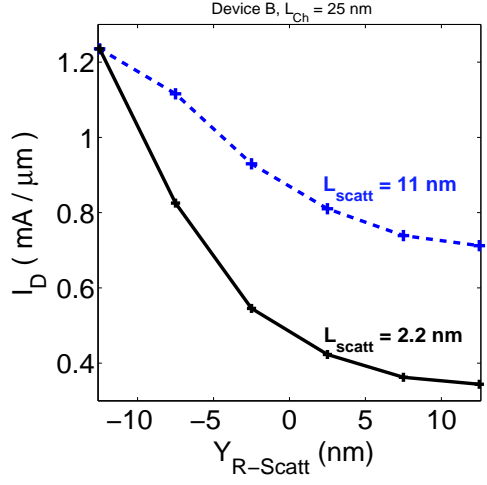}
  \end{center}
\caption{Drain current versus $Y_{R-Scatt}$ for two different scattering lengths.
The channel length is 25 nm.}
\label{fig:dgscatt2}
\end{figure}

\begin{figure}[htbp]
  \begin{center}
    \leavevmode
\includegraphics[height=6cm,angle=-0]{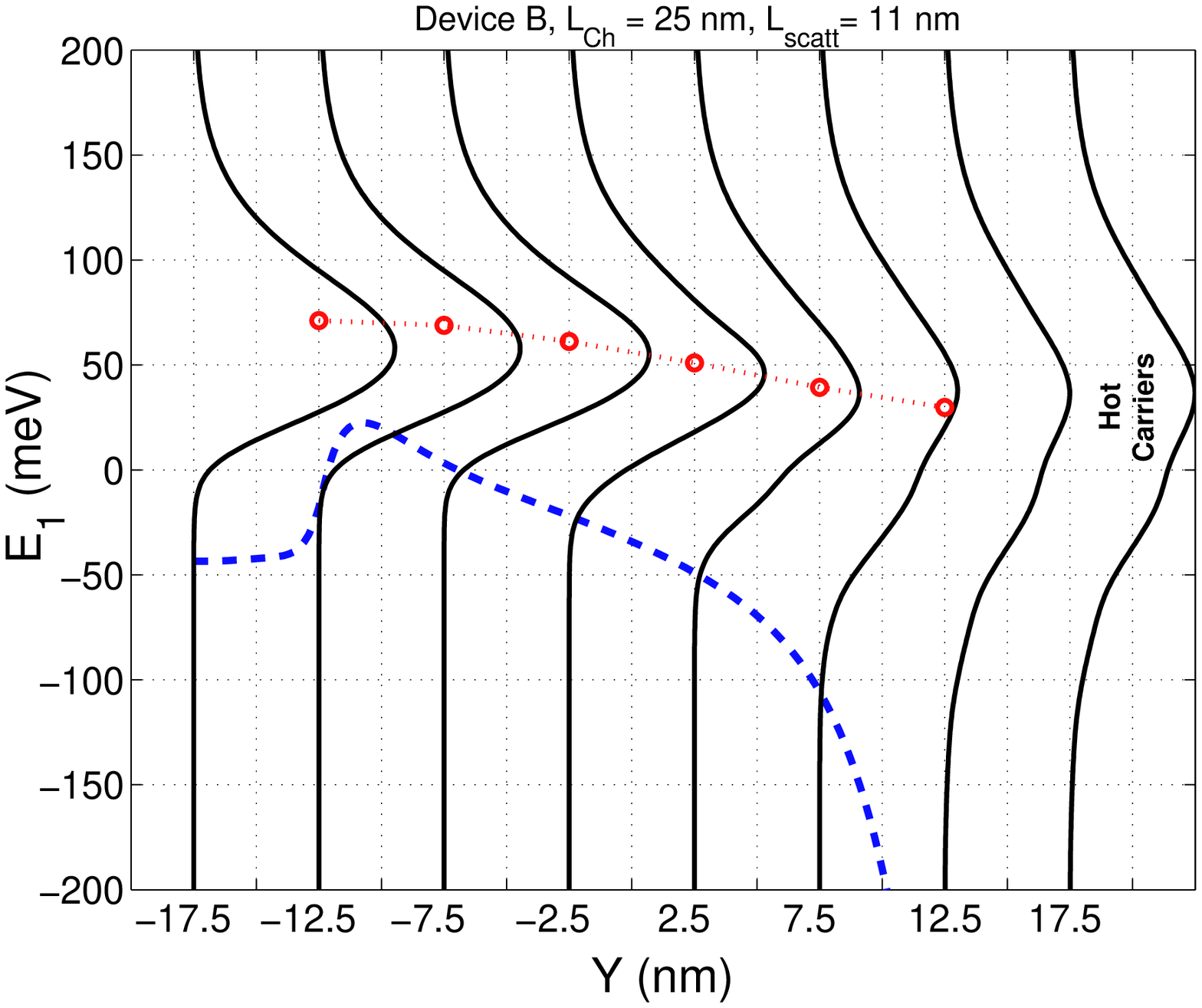}
\includegraphics[height=6cm,angle=-0]{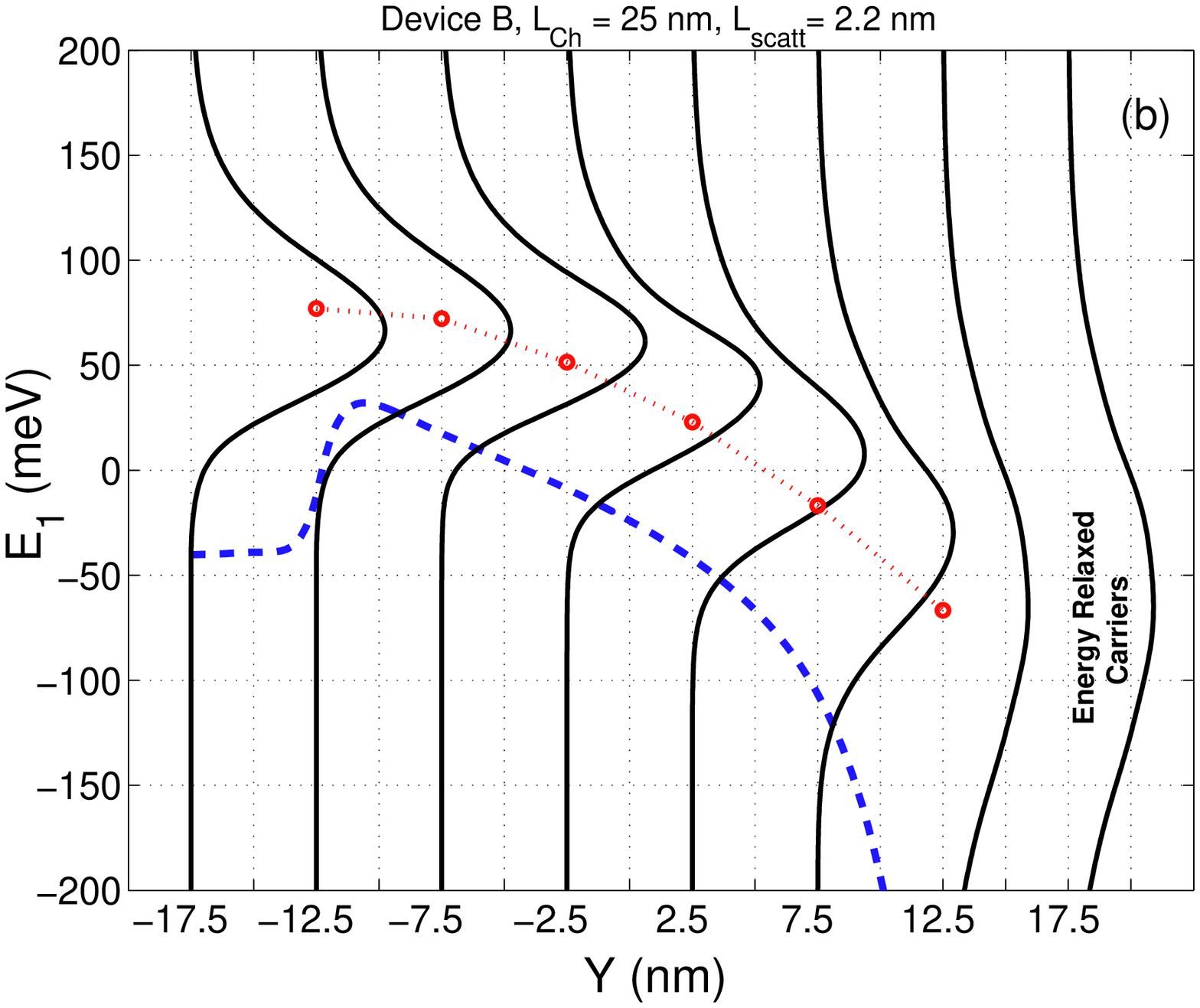}
  \end{center}
\caption{Solid lines represent $J(Y,E)$ for Y equal to -17.5, -12.5 -7.5, -2.5, 2.5, 7.5, 
12.5, 17.5 nm respectively, when scattering is included every where in the channel. 
The dashed lines are the first resonant level ($E_1$) along the channel. The dotted 
lines represent the first moment of energy with respect to the current distribution function, 
$\int dE E J(Y,E)$. (a) and (b) correspond to $L_{scatt} = 11$ and $2.2$ nm respectively.}
\label{fig:dgscatt3}
\end{figure}

The device considered has a channel length of 25 nm, body thickness
of 1.5 nm, oxide thickness of 1.5 nm, doping of 1 E+20 cm$^{-3}$ in the
source and drain extension regions and an intrinsic channel. The 
scattering length due to electron-phonon interaction is 11 nm. We see
from Fig. \ref{fig:dgscatt2} that scattering in the right half of the
channel (0 to 12.5 nm) is important and that the drive current 
degradation due to scattering in the right half of the channel is 
30\%. To understand the large reduction in drain current due to 
scattering in the right half of the channel, we plot the current 
density as a function of energy at various cross sections, $J(Y,E)$.
$J(Y,E)$ shows the energetic redistribution of carriers along the
channel. When the channel length is comparable to the scattering
length, $J(Y,E)$ in the right half of the channel is peaked in 
energy above the source injection barrier as shown in Fig. 
\ref{fig:dgscatt3} (a). Scattering causes reflection of these
energetic electrons toward the source. These reflected electrons lead
to an increase in the channel electron density (classical MOSFET
electrostatics). As the charge in the channel should be approximately
$C_{ox}(V_G-V_S)$, the source injection barrier floats to a higher 
energ toy compensate for the reflected electrons. The increase in 
source injection barrier and reflection leads to the decrease in drain 
current \cite{Lun97}. 

To gain further insight into the role of carrier relaxation, we
now discuss the case when the scattering length is 2.2 nm, which
is five times smaller than in the previous discussion. We see from
Fig. \ref{fig:dgscatt2} that scattering in the right half of the
channel now decreases the drive current by a smaller amount of 15\%,
when compared to the case with $L_{scatt}=11$ nm. As the channel
length of 25 nm is much larger than 2.2 nm, multiple scattering 
events now lead to an energy distribution of current that is peaked
well below the source injection barrier in the right half of the 
channel as shown in Fig. \ref{fig:dgscatt3} (b). The first moment 
of energy with respect to the current distribution function,
which is defined as the ratio of $\int dE E J(Y,E)$ and total current
is also shown in Fig. \ref{fig:dgscatt3}. When the 
 scattering length is much smaller than the
channel length, the carriers relax classically such that the first 
moment ($\int dE E J(Y,E)$) closely tracks the potential profile as seen in Fig. \ref{fig:dgscatt3} (b).
Thermalized carriers that are reflected in the right half of the channel can no longer
reach the source injection barrier due to the
large barrier to the left, and so contribute less significantly to
the charge density. Thus, explaining the diminished influence
of scattering in the right half of the channel relative to the left
half of the channel.

\begin{figure}
  \begin{center}
    \leavevmode
\includegraphics[height=6.5cm,width=6.5cm,angle=-0]{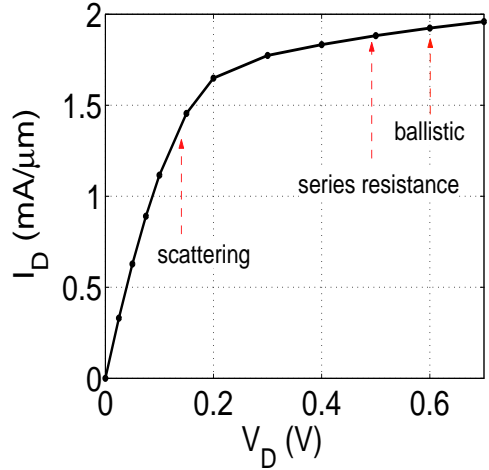}
  \end{center}
\caption{Ballistic $I_D(V_D)$ with the drive currents obtained in the ballistic limit, 
with the series resistance picture and NEGF calculations marked.}
\label{fig:dgscatt4}
\end{figure}

To further demonstrate the use of NEGF simulations, we study the role of assuming that the 
extension regions can be modeled as a classical series resistance. Within the classical 
series resistance picture the current with scattering ($I_D^{scatt}$) can be related to 
the current without scattering ($I_D^{noscatt}$) by \cite{Tau98}, 
\begin{eqnarray}
I_D^{scatt} (V_D) \sim I_D^{noscatt} (V_D - \delta V_D) \mbox{ ,}
\end{eqnarray}
where we have assumed that the source extension region and device do not experience scattering. 
The potential drop in the drain region within the classical series resistance picture is 
$\delta V_D = I_D^{scatt} (V_D) R_D$. In Fig. \ref{fig:dgscatt4}, the values of the drain 
current in the ballistic limit, using the classical series resistance picture and with the 
NEGF method are marked. Clearly, the series resistance picture underestimates the deterimental 
nature of scattering in the drain end. The physics of the large reduction in drain current was 
discussed in the context of Fig. \ref{fig:dgscatt3}: When scattering in the channel does not effectively thermalize
carriers, the current distribution is peaked at energies
above the source injection barrier, upon carriers exiting the channel. Scattering in the
drain extension region then causes reflection of electrons toward 
the source-end. As a result, the source injection increases so as to keep the electron
density in the channel approximately at $C_{ox}(V_G-V_S)$. The
drain current decreases dramatically as a result of the increase in source injection barrier height.

\section{\label{section:Discussion} Discussion and Summary}

Our objectives in this chapter has been to: (i) review the underlying assumptions of the traditional, 
semiclasssical treatment of carrier transport in semiconductor devices, (ii) describe how the semiclassical 
approach can be applied to ballistic transport, (iii) discuss the Landauer-Buttiker approach to quantum 
transport in the phase coherent limit, (iv) introduce important elements of the non-equilibrium 
Green's function approach using Schrodinger's equation as a starting point and (v) finally demonstrate 
the application of the NEGF method to the MOSFET in the ballistic limit and with electron-phonon scattering. 

It is appropriate to make a few comments about the computational burden of the various transport models.  
One reason that device engineers continue to use drift-diffusion simulations rather that the more rigorous 
Monte Carlo simulations is the enormous difference in computational burden.  For semiclassical transport,
 the fundamental quantity is the carrier distribution function, $f(\vec{r},\vec{k},t)$. 
To find $f(\vec{r},\vec{k})$, we must solve the BTE, which is a six-dimensional equation.  
The difficulty of solving this six-dimensional equation is one reason that engineers continue 
to rely on simplified models.  For quantum transport, we can take the Green's function 
$G^{n}(\vec{r},\vec{r}^\prime,E)$   as the fundamental quantity.  
The Green's function is a correlation function that describes the phase relationship between 
the wavefunction at $\vec{r}$ and $\vec{r}^\prime$ for an electron injected at energy E. 
The quantum transport problem is seven dimensional, which makes it much harder than the semiclassical problem.  
We can think of $\vec{r}$ and $\vec{r}^\prime$ as analogous to $\vec{r}$ and $\vec{k}$ in 
the semiclassical approach, but there is no E in the semiclassical approach. 
The reason is that for a bulk semiconductor or in a device in which the potential 
changes slowly, there is a relation between E and $\vec{k}$, as determined by the 
semiconductor bandstructure $E(\vec{k})$.  When the potential varies rapidly, however, 
there is no $E(\vec{k})$, and energy becomes a separate dimension. Analysis of electronic 
devices by quantum simulation is however becoming practical because device dimensions are 
shrinking, which reduces the size of the problem. Quantum simulations are also essential 
to accurately model devices whose dimensions are comparable to the phase breaking length,
 and rely on tunneling and wave interference for operation. The resonant tunneling diode 
is the most successful example in this category.

\section{\label{section:ack} Acknowledgements}

MPA is grateful to T. R. Govindan and Alexei Svizhenko for collaboration and discussions over the 
last five years. MPA and DEN would also like to thank Supriyo Datta for many useful discussions.

\appendix
\section{\label{sect:appendix1}Derivation of 
eqn. (\ref{eq:Gr0})} 

Eq. (\ref{eq:genGr}) can be expanded as,
\begin{eqnarray}
&&G_{LD} = - A^{\prime -1}_{LL}  A^\prime_{LD} G_{DD}
      \label{eq:GLD0}\\
&&G_{RD} = - A^{\prime -1}_{RR}  A^\prime_{RD} G_{DD}
      \label{eq:GRD0} \\
&&A^\prime_{DL} G_{LD} + A^\prime_{DD} G_{DD} +
     A^\prime_{DR} G_{RD} = I \mbox{ .}
      \label{eq:GDD00}
\end{eqnarray}
Substituting Eqs. (\ref{eq:GLD0}) and (\ref{eq:GRD0}) in
eqn. (\ref{eq:GDD00}), we have,
\begin{eqnarray}
[A^\prime_{DD} - A^\prime_{DL} A^{\prime -1}_{LL} A^\prime_{LD} -
A^\prime_{DR} A^{\prime -1}_{RR} A^\prime_{RD}] G_{DD} = I \mbox{ .}
      \label{eq:GDD0_app}
\end{eqnarray}
Noting the sparsity of $A^\prime_{LD}$ and $A^\prime_{RD}$, it follows
that only the {\bf surface Green's functions} of the Left and Right
leads,
\begin{eqnarray}
{g_{L}}_{_{-1,-1}} = {A^{\prime -1}_{LL}}_{_{1,1}} \mbox{ and }
{g_{R}}_{_{n+1,n+1}} = {A^{\prime -1}_{RR}}_{_{1,1}} \label{eq:sg_app}
\end{eqnarray}
are required in eqn. (\ref{eq:GDD0_app}).
Then, we can rewrite eqn. (\ref{eq:GDD0_app}) in terms of lead
self-energies as,
\begin{eqnarray}
A G_{DD} = [EI - H_D - \Sigma_{lead}] G_{DD} = I \mbox{,}
\end{eqnarray}
where $\Sigma_{lead}$ has been defined in eqn. (\ref{eq:sigmaLr})
and (\ref{eq:sigmaRr}). The reader can verify that
${g_{L}}_{_{-1,-1}}=e^{+ik_la} t_l^{-1}$ and
${g_{R}}_{_{n+1,n+1}}=e^{+ik_ra} t_r^{-1}$.

\section{\label{section:dyson} Dyson's equation for layered structures} 

\begin{center}
({\bf Note: }Matrix $A$ of this section is equivalent to Matrix $A_{DD}$ of
section \ref{section:ngf-summary})
\end{center}

\vspace{0.1in}

\begin{figure}
  \begin{center}
    \leavevmode
\includegraphics[height=5cm,angle=-0]{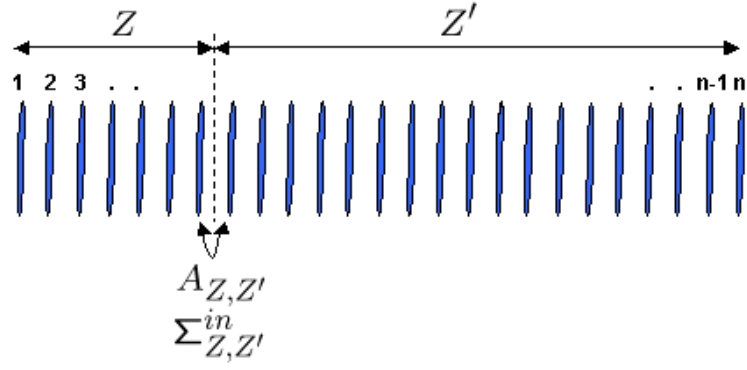}
  \end{center}
\caption{Scheme of device for application of Dyson's equation by splitting the device in two parts.}
\label{fig:dyson}
\end{figure}

Partition the device layers into two regions $Z$ and $Z^\prime$ as shown in Fig. \ref{fig:dyson}.
Dyson's equation is a very useful method that relates the Green's function
of the full system $Z+Z^\prime$ in terms of the subsystems
$Z$, $Z^\prime$ and the coupling between $Z$ and 
$Z^\prime$. We will see below that from a computational point of view,
Dyson's equation provides us with a systematic framework to calculate
the diagonal blocks of $G$ and $G^{n}$ without full inversion of the
$A$ matrix. The reader should note that Dyson's equation has a 
significantly more general validity than implied in our application
here \cite{Mah90}.

\subsection{\label{subsect:Gr} Dyson's equation for $G$}

The  Green's function equation over the 
device layers [eqn. (\ref{eq:GDD2})],
\begin{eqnarray}
A G = I
\label{eq:G_def_viaA}
\end{eqnarray}
can be written as,
\begin{eqnarray}
{
\left(
\begin{array}{cc}
A_{Z,Z}    & A_{Z,Z^\prime}  \\
A_{Z^\prime,Z} & A_{Z^\prime,Z^\prime}
\end{array}
\right)
\left(
\begin{array}{cc}
G_{Z,Z} & G_{Z,Z^\prime} \\
G_{Z^\prime,Z} & G_{Z^\prime,Z^\prime}
\end{array}
\right)
=
\left(
\begin{array}{cc}
I  & O         \\
O  & I
\end{array}
\right)
} \mbox{.} \label{eq:Gr}
\end{eqnarray}
The solution of eqn. (\ref{eq:G_def_viaA}is,
\begin{eqnarray}
G &=& G^{0} + G^{0} U G  \label{eq:dysonGr2} \\
    &=& G^{0} + G U G^{0} \mbox{  ,} \label{eq:dysonGr3}
\end{eqnarray}
where,
\begin{eqnarray}
G =
{ \left(
\begin{array}{cc}
G_{Z,Z}        & G_{Z,Z^\prime}         \\
G_{Z^\prime,Z} & G_{Z^\prime,Z^\prime}
\end{array}
\right)}
\mbox{, }
G^{0}=
{
{ \left(
\begin{array}{cc}
G^{0}_{Z,Z}  & O         \\
O  & G^{0}_{Z^\prime,Z^\prime}
\end{array}
\right)}
=
{ \left(
\begin{array}{cc}
A_{Z,Z}^{-1}  & O         \\
O  & A_{Z^\prime,Z^\prime}^{-1}
\end{array}
\right) }}
\mbox{ and }
U=
{ \left(
\begin{array}{cc}
O    & - A_{Z,Z^\prime}  \\
- A_{Z^\prime,Z} & O
\end{array}
\right) }
\mbox{ .} \label{eq:GrandU}
\end{eqnarray}
The Hermitean conjugate Green's function ($G^\dagger$) is by definition related to $G$
by
\begin{eqnarray}
G^\dagger &=& G^{\dagger 0} + G^{\dagger 0} U^\dagger G^\dagger 
            \label{eq:Ga1} \\
                      &=& G^{\dagger 0} + G^{\dagger} U^\dagger G^{\dagger 0}
\mbox{  .} 
\label{eq:Ga2}
\end{eqnarray}
Eq. (\ref{eq:dysonGr2}) is Dyson's equation for the  Green's 
function.

\subsection{\label{subsect:G<} Dyson's equation for $G^{n}$}

The electron correlation function equation over the device layers
[eqn. (\ref{eq:G<layer})],
\begin{eqnarray}
A G^{n} = \Sigma^{in} G^\dagger
\label{eq:Gn_def_viaA} 
\end{eqnarray}
can be written as,
\begin{eqnarray}
{
\left(
\begin{array}{cc}
A_{Z,Z}    & A_{Z,Z^\prime}  \\
A_{Z^\prime,Z} & A_{Z^\prime,Z^\prime}
\end{array}
\right)
\left(
\begin{array}{cc}
G^{n}_{Z,Z}         & G^{n}_{Z,Z^\prime} \\
G^{n}_{Z^\prime,Z}  & G^{n}_{Z^\prime,Z^\prime}
\end{array}
\right)
=
\left(
\begin{array}{cc}
\Sigma^{in}_{Z,Z}         & \Sigma^{in}_{Z,Z^\prime}  \\
\Sigma^{in}_{Z^\prime,Z}  & \Sigma^{in}_{Z^\prime,Z^\prime}
\end{array}
\right)
\left(
\begin{array}{cc}
G^\dagger_{Z,Z}         & G^\dagger_{Z,Z^\prime} \\
G^\dagger_{Z^\prime,Z}  & G^\dagger_{Z^\prime,Z^\prime}
\end{array}
\right)
} \mbox{ .} 
\label{eq:discreteG<3}
\end{eqnarray}
The solution of eqn. (\ref{eq:Gn_def_viaA}) is,
\begin{eqnarray}
G^{n} &=& G^{0} U G^{n} + G^{0} \Sigma^{in} G^\dagger \mbox{,} \label{eq:dysonG<1}
\end{eqnarray}
where $G^{0}$ and $U$ have been defined in Eqs. (\ref{eq:GrandU}).
Functions $G^{n}$ and $G^\dagger$ are readily defined by
eqns. (\ref{eq:Gn_def_viaA}) and (\ref{eq:Ga2}), respectively. 
Using $G^\dagger=G^{\dagger 0} + G^{\dagger 0} U^\dagger G^{\dagger}$,
eqn. (\ref{eq:dysonG<1}) can be written as
\begin{eqnarray}
G^{n} &=& G^{n 0} + G^{n 0} U^\dagger G^\dagger + G^{0} U G^{n}
                                        \label{eq:dysonG<3}\\
    &=& G^{n 0} + G U G^{n 0} + G^{n} U^\dagger G^{\dagger 0}
                                        \label{eq:dysonG<4}
                                                     \mbox{, } \\
 \mbox{where  }
G^{n 0} &=& G^{0} \Sigma^{in} G^{\dagger 0}  \mbox{ .}  \label{eq:G<0}
\end{eqnarray}

\section{\label{section:algorithm} Algorithm to calculate $G$ and 
$G^{n}$}

\begin{center}
({\bf Note: }Matrix $A$ of this section is equivalent to Matrix $A_{DD}$ of
section \ref{section:ngf-summary})
\end{center}

\vspace{0.1in}

\noindent
{\bf Why algorithm:} A typical simulation of a nanoelectronic 
device consists of solving Poisson's
equation self-consistently with the Green's function equations. 
The input to Poisson's equation is the charge density, 
which is obtained via integrating over energies the 
elements, $G^{n}_{q,q}(E)$, from the main diagonal of the electron correlation function.
The index $q$ here runs over the layers of the device.
In order to calculate the current density one requires the elements,
$G^{n}_{q,q+1}(E)$, from the diagonals adjacent to the main diagonal.
That is, we {\it do not} require the 
entire $G^{n}$ matrix in most situations. 
Same goes for the hole correlation function $G^{p}$ and the Green's function
$G$.

Provided that $N_x$ is the dimension of the Hamiltonian of each layer and
$N_y$ is the total number of layers, the size of the matrix $A$ equals $N_xN_y$ 
The operation count for the full matrix inversion $G=A^{-1}$ is proportional to
$(N_xN_y)^3$.
The computational cost of
obtaining the diagonal elements of the $G^{n}$ matrix at each energy is
approximately $N_x^3 N_y^3$ operations if $G^{n}=G \Sigma^{in} G^\dagger$ is
used.
Therefore it is highly desirable
to find less expensive algorithms that avoid full inversion of matrix
$A$ and take advantage of the fact the diagonal elements of Green's functions.
Another reason to prefer such algorithms is the memory storage.
If one had had to retain the whole matrix $G$ in the memory, it might had 
required using the RAM or the hard drive instead of on-chip cache.
That would have significantly slowed down the calculations.

One such algorithm which is valid for the block tridiagonal form
of matrix $A$ is presented in this section. The operation count of this
algorithm scales approximately as $N_x^3 N_y$. The dependence on 
$N_x^3$ arises because matrices of dimension of the sub Hamiltonian 
of layers should be inverted, and the dependence on $N_y$ corresponds
to one such inversion for each of the $N_y$ layers.

The algorithm consists of two steps. In the first step, the diagonal
blocks of the left connected and full  Green's function are 
evaluated (section \ref{subsect:recursive-Gr}). In the second step, 
these results are used to evaluate the diagonal blocks of the less-than
Green's function (section \ref{subsect:recursive-G<}).

\subsection{\label{subsect:recursive-Gr} Recursive algorithm for $G$}
\begin{figure}
  \begin{center}
    \leavevmode
\includegraphics[height=6.5cm,angle=-0]{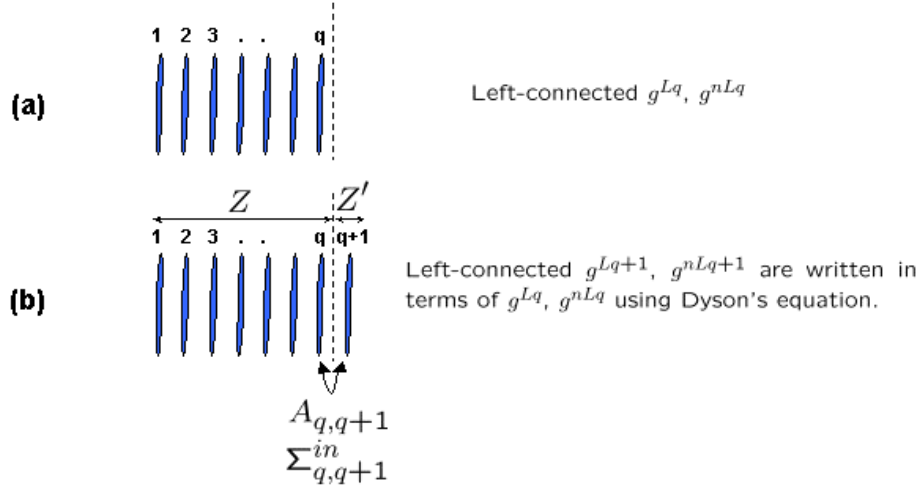}
  \end{center}
\caption{Illustration for the relation between the left-connected Green's functions for
adjacent layers.}
\label{fig:left-connected}
\end{figure}

(i) \underline{Left-connected  Green's function} (Fig.
\ref{fig:left-connected}):

The left-connected (superscript $L$)  Green's
function $g^{Lq}$ is defined by the first $q$ blocks of 
eqn. (\ref{eq:GDD2}) by
\begin{eqnarray}
A_{1:q,1:q} \; g^{Lq} = I_{q,q}. 
\end{eqnarray}
where we introduce a shorthand $I_{q,q}=I_{1:q,1:q}$
The matrix $g^{Lq+1}$ is defined in a manner identical to $g^{Lq}$ except that
the left-connected system is comprised of the first $q+1$ blocks of
eqn. (\ref{eq:GDD2}). In terms of eqn. (\ref{eq:Gr}), the
equation governing $g^{Lq+1}$ can be expressed via the solution $g^{Lq}$ by setting $Z=1:q$ and 
$Z^\prime=q+1$.
Using Dyson's equation [eqn. (\ref{eq:dysonGr2})], we obtain
\begin{eqnarray}
g^{Lq+1}_{q+1,q+1}= \left(A_{q+1,q+1} -
             A_{q+1,q} g^{Lq}_{q,q}
          A_{q,q+1} \right)^{-1} \mbox{ .} 
\label{eq:dysongrL2}
\end{eqnarray}
Note that the last element of this progression $g^{LN}$ is equal to the fully
connected Green's function $G$, which is the solution to
eqn. (\ref{eq:GDD2}).

(ii) \underline{Full Green's function in terms of the left-connected
Green's function}:

Consider the special case of eqn. (\ref{eq:Gr}) in which $A_{Z,Z}=A_{1:q,1:q}$,
$A_{Z^\prime,Z^\prime}=A_{q+1:N,q+1:N}$ and
$A_{Z,Z^\prime}=A_{1:q,q+1:N}$. Noting that the only nonzero element
of $A_{1:q,q+1:N}$ is $A_{q,q+1}$ and using eqn. (\ref{eq:dysonGr2}),
we obtain
\begin{eqnarray}
G_{q,q} &=& g^{Lq}_{q,q} + g^{Lq}_{q,q}
       \left(A_{q,q+1} G_{q+1,q+1} A_{q+1,q} \right) g^{Lq}_{q,q} 
\label{eq:discreteGr3} \\
          &=& g^{Lq}_{q,q} - g^{Lq}_{q,q} A_{q,q+1} G_{q+1,q}
\label{eq:discreteGr4}
\mbox{ .}
\end{eqnarray}
The equations for the adjacent diagonals are obtained similarly
\begin{eqnarray}
G_{q+1,q} &=& - G_{q+1,q+1} A_{q+1,q} g^{Lq}_{q,q},  
\label{eq:discreteGl_off} \\
G_{q,q+1} &=& - g^{Lq}_{q,q} A_{q,q+1} G_{q+1,q+1}
\label{eq:discreteGu_off}
\mbox{ .}
\end{eqnarray}

Both $G_{q,q}$ and $G_{q+1,q}$ are used in the algorithm for
electron density, and so storing both sets of matrices is
necessary.

Making use of the above equations, the algorithm to obtain the three diagonals of $G$ is

\begin{center}
\begin{enumerate}
\item $g^{L1}_{11} = A_{11}^{-1}$.
\item For $q=1,2, ...,N-1$, compute eqn. (\ref{eq:dysongrL2}).
\item For $q=1,2, ...,N$,  compute $\left( g^{Lq}_{qq} \right) ^\dagger$.
\item $G_{N,N} = g^{Lq}_{q,q}.$ 
\item For $q=N-1, N-2, ..., 1$, compute eqns. (\ref{eq:discreteGl_off}), 
(\ref{eq:discreteGu_off}) and (\ref{eq:discreteGr4}) (in this order).
\item For $q=1,2, ...,N$,  compute $\left( G_{q,q+1} \right) ^\dagger$
and $\left( G_{q+1,q} \right) ^\dagger$.
\end{enumerate}
\end{center}

\subsection{\label{subsect:recursive-G<} Recursive algorithm for $G^{n}$}
(i) \underline{Left-connected $G^{n}$} (Fig.  \ref{fig:left-connected}):

The function $g^{nLq}$ is the counterpart of $g^{Lq}$, and is defined by the first
$q$ blocks of eqn. (\ref{eq:G<layer}):
\begin{eqnarray}
A_{1:q,1:q} \; g^{nLq} = 
\Sigma^{in}_{1:q,1:q} \; g^{\dagger Lq}_{1:q,1:q} \mbox{ .}
\end{eqnarray}
$g^{nLq+1}$ is defined in a manner identical to $g^{nLq}$ except that
the left-connected system is comprised of the first $q+1$ blocks of
eqn. (\ref{eq:G<layer}). 
 
The equation governing $g^{nLq+1}$ follows 
from eqn. (\ref{eq:discreteG<3})
by setting $Z=1:q$ and
$Z^\prime=q+1$. Using the Dyson's equations for $G$ 
[eqn. (\ref{eq:dysonGr2})] and $G^{n}$ [eqn. (\ref{eq:dysonG<3})],
$g^{nLq+1}_{q+1,q+1}$ is recursively obtained as

\begin{eqnarray}
g^{nLq+1}_{q+1,q+1}
=
g^{Lq+1}_{q+1,q+1} \left[\Sigma^{in}_{q+1,q+1}
+ \sigma^{in}_{q+1} \right] g^{Lq+1 \dagger}_{q+1,q+1}
\mbox{ ,} \label{eq:g<L}
\end{eqnarray}

where $\sigma^{in}_{q+1} = A_{q+1,q} g^{nLq}_{q,q} A_{q,q+1}^\dagger$.
Eqn. (\ref{eq:g<L}) has the physical meaning that $g^{nLq+1}_{q+1,q+1}$ has contributions 
due to an effective self-energy due to the left-connected structure that ends at $q$, 
which is represented by $\sigma^{in}_{q+1}$ and  the diagonal self-energy component at 
grid point $q+1$ ($\Sigma^{in}_{DD}$ of eqn. (\ref{eq:G<layer})).

\noindent
(ii) \underline{Full electron correlation function in terms of
left-connected Green's function}:
Consider eqn. (\ref{eq:discreteG<3}) such that $A_Z=A_{1:q,1:q}$,
$A_Z^\prime=A_{q+1:N,q+1:N}$ and $A_{Z,Z^\prime}=A_{1:q,q+1:N}$.
Noting that the only nonzero element of $A_{1:q,q+1:N}$ is 
$A_{q,q+1}$ and using eqn. (\ref{eq:dysonG<3}), we obtain
\begin{eqnarray}
G^{n}_{q,q} = g^{nLq}_{q,q} 
                 - g^{nLq}_{q,q} A_{q,q+1}^\dagger G^\dagger_{q+1,q}   
                 - g^{Lq}_{q,q} A_{q,q+1} G^{n}_{q+1,q}
                         \mbox{ .} \label{eq:discreteG<4}
\end{eqnarray}
Using eqn. (\ref{eq:dysonG<4}), $G^{n}_{q+1,q}$ can be written in terms 
of $G^{n}_{q+1,q+1}$ and other known Green's functions as
\begin{eqnarray}
G^{n}_{q+1,q} = 
                   - G_{q+1,q+1} A_{q+1,q} g^{nLq}_{q,q}
           - G^{n}_{q+1,q+1} A^\dagger_{q+1,q} g^{\dagger Lq}_{q,q}
                    \mbox{ .}   \label{eq:discreteG<5}
\end{eqnarray}
Substituting eqn. (\ref{eq:discreteG<5}) in eqn. (\ref{eq:discreteG<4})
and using Eqs. (\ref{eq:dysonGr2}) and (\ref{eq:dysonGr3}), we obtain
\begin{eqnarray}
G^{n}_{q,q}&=&g^{nLq}_{q,q} + 
            g^{Lq}_{q,q} \left(A_{q,q+1} G^{n}_{q+1,q+1} 
             A_{q+1,q}^\dagger \right) g^{\dagger Lq}_{q,q} -
            \left[g^{nLq}_{q,q} A_{q,q+1}^\dagger G^\dagger_{q+1,q} +
             G_{q,q+1} A_{q+1,q} g^{nLq}_{q,q} \right] \mbox{.} 
\label{eq:discreteG<6}
\end{eqnarray}
The terms inside the square brackets of eqn. (\ref{eq:discreteG<6})
are Hermitian conjugates of each other.
In view of the above equations, the algorithm to compute the diagonal
blocks of $G^{n}$ and $G^{p}$ is given by the following steps:

\begin{center}
\begin{enumerate}
\item $g^{nL1}_{11} = g^{L1}_{11} \Sigma^{in}_{11} g^{L 1 \dagger }_{11}$.
\item For $q=1,2, ...,N-1$, compute eqn. (\ref{eq:g<L}).
\item $G^n_{N,N} = g^{nLN}_{NN}$.
\item For $q=N-1, N-2, ..., 1$, compute eqns. (\ref{eq:discreteG<6}) and 
(\ref{eq:discreteG<5}).
\item Use $G^n_{q,q+1} = \left( G^n_{q+1,q} \right)^\dagger $.
\item Use $G^{p} = i \left( G-G^\dagger \right) - G^{n}$.
\end{enumerate}
\end{center}

The above algorithm is illustrated by a Matlab code in Section \ref{se:matlab_recurse}.

\noindent
{\bf Challenging problem:} The algorithm presented here solves for
the three block diagonals of $G$, $G^n$, and $G^p$.
Each of $n$ blocks on the main diagonal corresponding to a layer of the device.
All blocks in the three diagonals is treated as a full matrix.
It is highly desirable to find a more efficient algorithm that finds only
the diagonal elements within each block rather than complete blocks.

\section{Code of the Recursive Algorithm \label{se:matlab_recurse} }

The listing of a Matlab code recursealg3d.m is provided here for illustration of the 
algorithm described in Section \ref{section:algorithm}.
\begin{verbatim}
function [Grl,Grd,Gru,Gnl,Gnd,Gnu,Gpl,Gpd,Gpu,grL,ginL] = recursealg3d(Np,Al,Ad,Au,Sigin,Sigout)
%%%%%%%%%%%%%%%%%%%%%%%%%%%%%%%%%%%%%%%%%%%%%%%%%%%%%%%%%%%%%%%%
% function [Grl,Grd,Gru,Gnl,Gnd,Gnu,Gpl,Gpd,Gpu] = recursealg3d(Np,Al,Ad,Au,Sigin,Sigout)
% recursive algorithm to solve for the diagonal elements of 
% the Non-equilibrium Green's function
% "l" = lower diagonal = [G(2,1); ...; G(end,end-1); 0]
% "d" = main diagonal = [G(1,1); ...; G(end,end)]
% "u" = upper diagonal = [0; G(1,2); ....; G(end-1,end)]
% Grl,Grd,Gru = retarded Green's function
% Gnl,Gnd,Gnu = electron Green's function
% Gpl,Gpd,Gpu = hole Green's function
% Np = size of the matrices
% Al,Ad,Au = matrix of coefficients
% Sigin = matrix of in-scattering self-energies (diagonal)
% Sigout = matrix of out-scattering self-energies (diagonal)
%%%%%%%%%%%%%%%%%%%%%%%%%%%%%%%%%%%%%%%%%%%%%%%%%%%%%%%%%%%%%%%%
% Dmitri Nikonov, Intel Corp. and Siyu Koswatta, Purdue University, 2004
%%%%%%%%%%%%%%%%%%%%%%%%%%%%%%%%%%%%%%%%%%%%%%%%%%%%%%%%%%%%%%%%
flag_Gp = 'no';
Al_cr = conj(Au);
Ad_cr = conj(Ad);                      % Hermitean conjugate of the coefficient matrix
Au_cr = conj(Al);
grL = zeros(1,Np);                     % initialize left-connected function
ginL = zeros(1,Np);                    % initialize left-connected in-scattering function
gipL = zeros(1,Np);                    % initialize left-connected out-scattering function
Grl = zeros(1,Np-1);
Grd = zeros(1,Np);                     % initialize the Green's function
Gru = zeros(1,Np-1);
Gnl = zeros(1,Np-1);
Gnd = zeros(1,Np);                     % initialize the electron coherence function
Gnu = zeros(1,Np-1);
Gpl = zeros(1,Np-1);
Gpd = zeros(1,Np);                     % initialize the hole coherence function
Gpu = zeros(1,Np-1);
grL(1)=1/Ad(1);                        % step 1
for q=2:Np                             % obtain the left-connected function
    grL(q)=1/(Ad(q)-Al(q-1)*grL(q-1)*Au(q-1));
end
gaL = conj(grL);                       % advanced left-connected function
Grd(Np)=grL(Np);                       % step 2
for q=(Np-1):-1:1
    Grl(q)=-Grd(q+1)*Al(q)*grL(q);     % obtain the sub-diagonal of the Green's function
    Gru(q)=-grL(q)*Au(q)*Grd(q+1);     % obtain the super-diagonal of the Green's function
    Grd(q)=grL(q)-grL(q)*Au(q)*Grl(q); % obtain the diagonal of the Green's function
end
Gal = conj(Gru);
Gad = conj(Grd);                       % advanced Green's function
Gau = conj(Gal);
ginL(1)=grL(1)*Sigin(1)*gaL(1);        % step 3  
for q=2:Np
    sla2 = Al(q-1)*ginL(q-1)*Au_cr(q-1);
    prom = Sigin(q) + sla2;
    ginL(q)=grL(q)*prom*gaL(q);        % left-connected in-scattering function
end
Gnd(Np)=ginL(Np);                      % step 4
Gnd = real(Gnd);
for q=(Np-1):-1:1
    Gnl(q) = - Grd(q+1)*Al(q)*ginL(q) - Gnd(q+1)*Al_cr(q)*gaL(q);                 
    % obtain the lower diagonal of the electron Green's function
    Gnd(q) = ginL(q) + grL(q)*Au(q)*Gnd(q+1)*Al_cr(q)*gaL(q) ...
    - ( ginL(q)*Au_cr(q)*Gal(q) + Gru(q)*Al(q)*ginL(q) );
end
Gnu = conj(Gnl);                       % upper diagonal of the electron function
switch flag_Gp
    case 'yes'
gipL(1)=grL(1)*Sigout(1)*gaL(1);       % step 3  
for q=2:Np
    sla2 = Al(q-1)*gipL(q-1)*Au_cr(q-1);
    prom = Sigout(q) + sla2;
    gipL(q)=grL(q)*prom*gaL(q);        % left-connected in-scattering function
end
Gpd(Np)=gipL(Np);                      % step 4
Gpd = real(Gpd);
for q=(Np-1):-1:1
    Gpl(q) = - Grd(q+1)*Al(q)*gipL(q) - Gnd(q+1)*Al_cr(q)*gaL(q);                 
    % obtain the lower diagonal of the hole Green's function
    Gpd(q) = gipL(q) + grL(q)*Au(q)*Gpd(q+1)*Al_cr(q)*gaL(q) ...
    - ( gipL(q)*Au_cr(q)*Gal(q) + Gru(q)*Al(q)*gipL(q) );
end
Gpu = conj(Gpl);                       % upper diagonal of the hole function
    case 'no'
Gpl = i*(Grl-Gal) - Gnl;
Gpd = real(i*(Grd-Gad) - Gnd);         % hole Green's function
Gpu = i*(Gru-Gau) - Gnu;
end
jnzer = find(Gnd<0);
Gnd(jnzer) = 0;
jpzer = find(Gpd<0);
Gpd(jpzer) = 0;
%%%%%%%%%%%%%%%%%%%%%%%%%%%%%%%%%%%%%%%%%%%%%%%%%%%%%%%%%%%%
\end{verbatim}

\end{document}